\documentclass[12pt]{iopart}
\pdfoutput=1
\usepackage[utf8]{inputenc}
    \usepackage[pdftex]{graphicx}  % note the x at the end
      \usepackage[pdftex]{hyperref}
\usepackage{pslatex}
\usepackage{mathptm}    % to use math fonts
\usepackage{mathptmx}   % to use math fonts
\usepackage{dcolumn}% Align table columns on decimal point
\usepackage{color}
\usepackage{booktabs,tabulary}
\usepackage{soul}
\usepackage[export]{adjustbox}
\usepackage{epstopdf}
\usepackage{subcaption}
\usepackage{cite}

\usepackage{amssymb} % A package for extra mathematical symbols.
\usepackage{bm} % A package for bold mathematical symbols.

\newcommand{\eqref}[1]{(\ref{#1})}

\usepackage[colorinlistoftodos,shadow,textwidth=18mm]{todonotes}

\newenvironment{cond}
	{\left\{
    \begin{array}{l l}
    }
    { 
    \end{array} 
    \right.
    }

\begin{document}

\title{Nearest-Neighbor Functions for Disordered Stealthy Hyperuniform Many-Particle Systems}

\author{Timothy M Middlemas$^1$ and Salvatore Torquato$^{1,2,3,4}$}
\address{$^1$ Department of Chemistry, Princeton University, Princeton, New Jersey 08544, USA}
\address{$^2$ Department of Physics, Princeton University, Princeton, New Jersey 08544, USA}
\address{$^3$ Princeton Institute for the Science and Technology of Materials, Princeton University, Princeton, New Jersey 08544, USA}
\address{$^4$ Program in Applied and Computational Mathematics, Princeton University, Princeton, New Jersey 08544, USA}

\ead{torquato@electron.princeton.edu}
\vspace{10pt}
\date{\today}

\begin{abstract}
    
    Disordered stealthy many-particle systems in $d$-dimensional
    Euclidean space $\mathbb{R}^d$ are exotic amorphous states of
    matter that suppress any single scattering events for a finite range of
    wavenumbers around the origin in reciprocal space. They are currently the
    subject of intense fundamental and practical interest.  We derive
    analytical formulas for the nearest-neighbor functions of disordered
    stealthy many-particle systems. First, we analyze asymptotic small-$r$
    approximations and expansions of the nearest-neighbor functions based on
    the pseudo-hard-sphere ansatz. We then consider the problem of
    determining how many of the standard $n$-point correlation functions are
    needed to determine the nearest neighbor functions, and find that a finite
    number suffice. Via theoretical and computational methods, we are able to
    compare the large-$r$ behavior of these functions for disordered stealthy
    systems to those belonging to crystalline lattices. Such ordered and
    disordered stealthy systems have bounded hole sizes, and thus compact
    support for their nearest-neighbor functions. However, we find that the
    approach to the critical-hole size can be quantitatively different,
    emphasizing the importance of hole statistics in distinguishing ordered and
    disordered stealthy configurations. We argue that the probability of
    finding a hole close to the critical-hole size should decrease as a power
    law with an exponent only dependent on the space dimension
    $d$ for ordered systems, but that this probability decays asymptotically
    faster for disordered systems, with either an increase in the exponent of
    the power law or a crossover into a decay faster than any power law. This
    implies that holes close to the critical-hole size are rarer in disordered
    systems. The rarity of observing large holes in disordered systems creates
    substantial numerical difficulties in sampling the nearest neighbor
    distributions near the critical-hole size. This motivates both the need for
    new computational methods for efficient sampling and the development of
    novel theoretical methods for ascertaining the behavior of holes close to
    the critical-hole size. We also devise a simple analytical formula that
    accurately describes these systems in the underconstrained regime for all
    $r$. These results provide a theoretical foundation for the analytical
    description of the nearest-neighbor functions of stealthy systems in the
    disordered, underconstrained regime, and can serve as a basis for
    analytical theories of material and transport properties of these systems.

\end{abstract}
\vspace{2pc}
%\submitto{\JSTAT}

\section{Introduction}

In the study of disordered many-body systems, a large body of recent work (see
Ref. \cite{torquato_hyperuniform_2018} and references therein) has promoted the
concept of hyperuniformity \cite{torquato_local_2003} as a useful principle for
identifying exotic disordered systems with novel physical properties
\cite{torquato_hyperuniform_2018, zhang_perfect_2016, batten_classical_2008,
zhang_can_2017, florescu_designer_2009, man_photonic_2013, man_isotropic_2013,
florescu_optical_2013, froufe-perez_band_2017, milosevic_hyperuniform_2019,
zhang_transport_2016, torquato_multifunctional_2018, batten_novel_2009,
batten_interactions_2009}. Hyperuniformity refers to systems with an anomalous
suppression of long-range density fluctuations. More specifically, given a
$d$-dimensional point process, one considers the variance $\sigma^2_N
(R)$ of the number of particles within a spherical window of radius $R$ as one
uniformly varies the location of the window or averages over an enemble. Quantitatively, a
hyperuniform system is one in which \cite{torquato_local_2003} 
\begin{equation}
    \lim_{R\to\infty} \frac{\sigma^2_N(R)}{v_1(R)} = 0,
\end{equation}
where $v_1(R) = \pi^{d/2} R^d/\Gamma(1+d/2)$ is the volume of a $d$-dimensional
sphere of radius $R$. For typical disordered systems, $\sigma_N^2(R)$
grows as $R^d$, so the above ratio tends to a positive constant.  Thus,
hyperuniformity is defined by an asympotically slow growth of the number
variance, which is a key measure of the density fluctuations associated to a
given scale in the system. Equivalently, one can also identify hyperuniformity
through the following condition on the structure factor $S({\bi k})$
(obtainable through the scattering intensity) associated with the point process
\cite{torquato_local_2003}:
\begin{equation}
    \lim_{|{\bi k}|\to0} S({\bi k}) = 0.\label{hypdef}
\end{equation}
Note that this definition excludes the forward scattering contribution
in the scattering pattern. The structure factor is related to the
widely-used total correlation function $h({\bi r}) = g_2({\bi r}) - 1$, where
$g_2({\bi r})$ is the pair correlation function, through a Fourier transform
\cite{jean_pierre_hansen_theory_1986}:
\begin{equation}
    S({\bi k}) = 1 + \rho \int_{\mathbb{R}^d} e^{-i{\bi k}\cdot{\bi r}} h({\bi r})\,\rmd {\bi r}.
\end{equation}
Thus, Eq. (\ref{hypdef}) amounts to the following sum rule on the
two-point statistics of the point process \cite{torquato_hyperuniform_2018}:
\begin{equation}
    \int_{\mathbb{R}^d} h({\bi r})\,\rmd {\bi r} = -1.
\end{equation}

There are many examples of hyperuniform systems, both ordered and disordered. In
the ordered case, we have trivially that all perfect crystals are
hyperuniform, due to the presence of a Bragg-peak spectrum. As a less trivial
ordered example, we have that perfect quasicrystals are also hyperuniform
\cite{zachary_hyperuniformity_2009, oguz_hyperuniformity_2017,
lin_hyperuniformity_2017}. Disordered hyperuniform systems are considerably more
exotic, since typical disordered systems such as liquids and gases have $S(k\to 0)
\neq 0$ \cite{torquato_local_2003}. Examples include avian photoreceptor
patterns \cite{jiao_avian_2014}, perfect glasses \cite{zhang_perfect_2016},
maximally random jammed packings \cite{donev_unexpected_2005,
zachary_hyperuniform_2011, jiao_maximally_2011, chen_equilibrium_2014,
berthier_suppressed_2011, kurita_incompressibility_2011}, density fluctuations
in the large-scale structure of the Universe \cite{peebles_principles_1993,
gabrielli_statistical_2005, gabrielli_glass-like_2002,
gabrielli_generation_2003}, fermionic point processes
\cite{torquato_point_2008, scardicchio_statistical_2009}, and superfluid helium
\cite{feynman_energy_1956, reatto_phonons_1967}.

\begin{figure}
    \begin{center}
    \begin{subfigure}{.5\textwidth}
        \includegraphics[width=\linewidth]{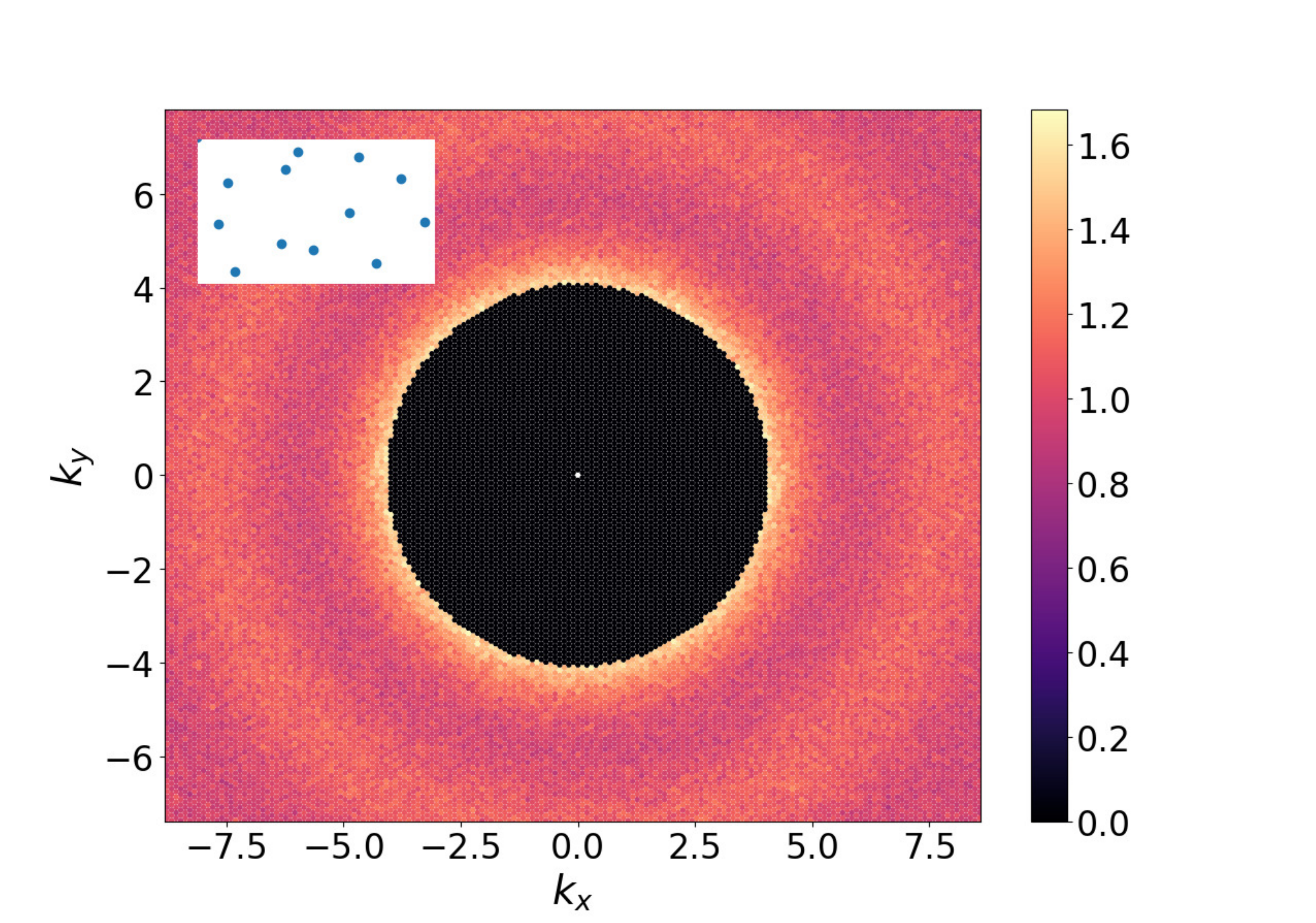}
        \caption{}
    \end{subfigure}%
    \begin{subfigure}{0.5\textwidth}
        \includegraphics[width=\linewidth]{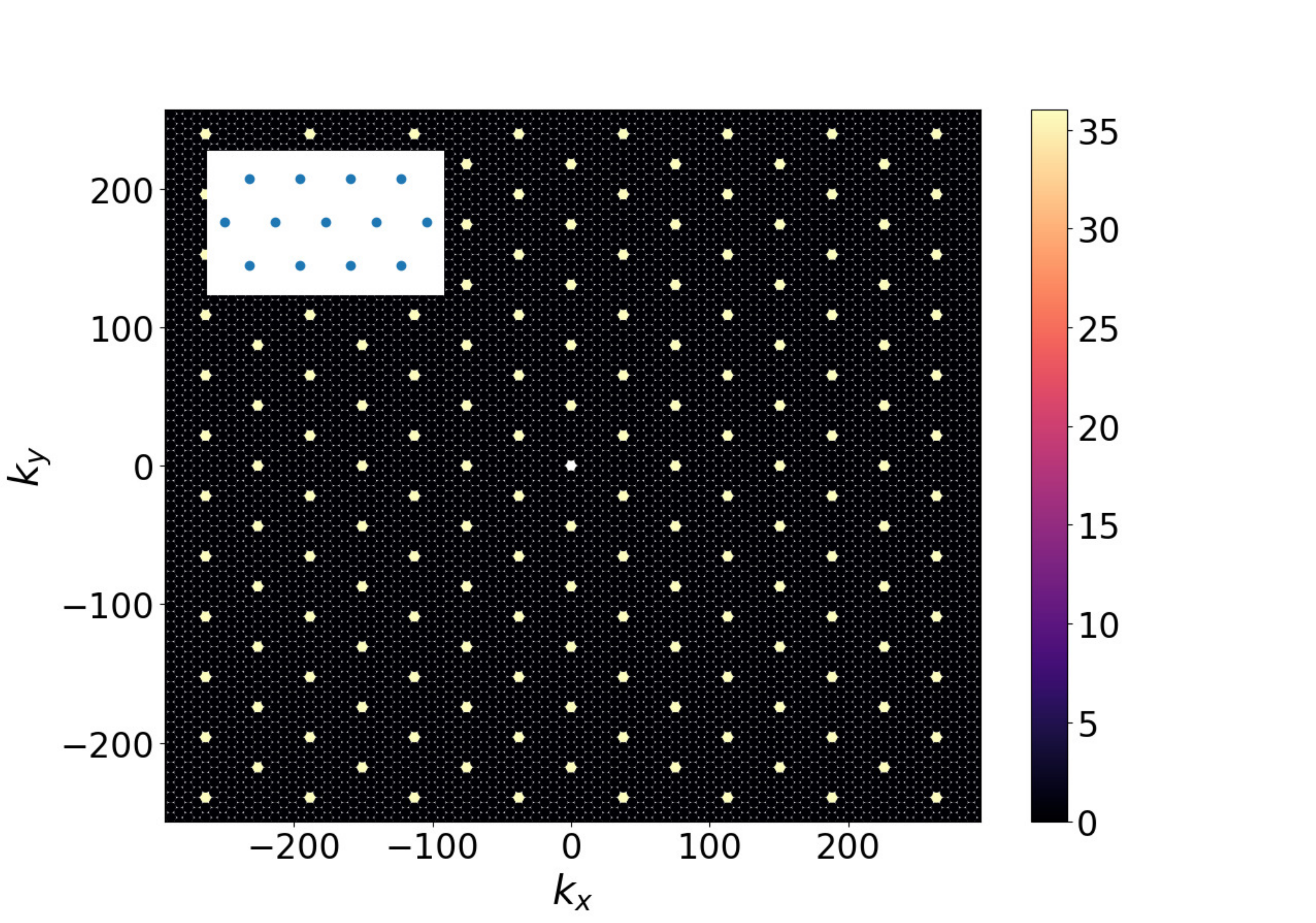}
        \caption{}
    \end{subfigure}
    \end{center}
    \caption{Scattering
    patterns (structure factor) for two 2D stealthy hyperuniform point
    processes and small corresponding
    representative samples of the underlying real-space point processes
    (inset). (a) The scattering pattern of a triangular
    lattice (see inset). (b) The scattering pattern of a disordered stealthy
    hyperuniform system (see inset). Excluding the forward
    scattering contribution, both structure factors exhibit the signature
    exclusion region around the origin in which there are no
    single-scattering events, implying a suppression of density fluctuations
    from infinite down to finite wavelengths. However, the disordered pattern
    lacks sharp Bragg peaks, with the diffuse behavior of the
    scattering pattern away from the origin being closer to that of a liquid.
    Note that while the stealthy disordered pattern possesses
    short-range order more typical of a disordered liquid or gas (see inset),
    it has a bounded hole size \cite{zhang_can_2017, ghosh_generalized_2018}.
    }\label{illustration}
\end{figure}

In this article, we will focus on an important subset of hyperuniformity
known as stealthy hyperuniformity \cite{batten_classical_2008}. Stealthy
hyperuniformity further generalizes the notion of mimicking an aspect of a
crystal's long wavelength behavior while maintaining local disorder. A stealthy
hyperuniform system is one in which the structure factor vanishes in an entire
range of wavelengths near the origin \cite{batten_classical_2008}:
\begin{equation}
    S({\bi k}) = 0,\quad 0 < |{\bi k}| < K.\label{sfdef}
\end{equation}
Crystals, due to their Bragg peaks, trivially satisfy this condition.
Interestingly, one can also find disordered systems that obey stealthy
hyperuniformity \cite{uche_constraints_2004, torquato_ensemble_2015,
zhang_transport_2016}. An example of the scattering pattern for a
stealthy disordered system is compared to a stealthy ordered system in Fig.
\ref{illustration}.  While both the ordered crystal and the disordered pattern
exhibit a spherical exclusion region with no scattering, the disordered pattern
exhibits the continuous scattering usually associated with liquids and gases
everywhere else in the domain \cite{torquato_hyperuniform_2018,
torquato_ensemble_2015}.

One of the most powerful techniques for studying stealthy
hyperuniform systems is a collective coordinate optimization procedure
\cite{fan_constraints_1991, uche_constraints_2004, uche_collective_2006,
batten_classical_2008, batten_novel_2009, batten_interactions_2009,
suto_crystalline_2005, torquato_ensemble_2015, zhang_ground_2015,
zhang_ground_2015-1} that involves finding the ground states of a
class of bounded pair potentials with compact support in Fourier space
\cite{fan_constraints_1991, uche_constraints_2004, suto_crystalline_2005, batten_classical_2008, torquato_ensemble_2015}.
The highly degenerate ground states of such potentials are stealthy
hyperuniform by construction. This technique suggests the utility of defining a
{\it control parameter} $\chi$, which is a dimensionless measure of the ratio of
constrained degrees of freedom to the total degrees of freedom in such an
optimization procedure. In the thermodynamic limit, this control parameter can
be written \cite{batten_classical_2008, torquato_ensemble_2015}
\begin{equation} 
    \chi = \frac{v_1(K)}{2\rho d (2\pi)^d},\label{chieq}
\end{equation}
where $\rho$ is the number density of the point process.
Since this formula involves only the general properties of a stealthy system,
such as the cutoff wavevector $K$, it can be used to classify stealthy systems
even beyond the collective coordinate framework. A system with a small
$\chi$ (relatively unconstrained) is disordered, and as $\chi$ increases, the short-range
order increases within a disordered regime \cite{fan_constraints_1991,
uche_constraints_2004, uche_collective_2006, batten_classical_2008,
batten_novel_2009, batten_interactions_2009, suto_crystalline_2005,
torquato_ensemble_2015, zhang_ground_2015, zhang_ground_2015-1}. Upon reaching a critical value of $\chi$,
there is a phase transition to predominantly crystalline ground states \cite{fan_constraints_1991,
uche_constraints_2004, uche_collective_2006, batten_classical_2008,
batten_novel_2009, batten_interactions_2009, suto_crystalline_2005,
torquato_ensemble_2015, zhang_ground_2015, zhang_ground_2015-1}.

While the stealthiness of crystals is a trivial outcome of Bragg
scattering, disordered stealthy systems display
highly unusual statistical geometric properties. For example, all stealthy
hyperuniform systems have a bounded hole size \cite{zhang_can_2017,
ghosh_generalized_2018}, meaning that one cannot find a sphere devoid of
particles above a certan radius, and an ``anti-concentration'' property that
strictly bounds the density from above in a large enough subset of the system
\cite{ghosh_generalized_2018}. As a result of these crystal-like geometric
properties and fluid-like short-range order, the disordered variants exhibit
novel physical properties with implications for materials discovery.
In particular, the isotropy of these disordered phases generates
direction-independent physical properties, in stark contrast to typical
crystalline systems. For example, disordered stealthy point processes, which
can be mapped to cellular dielectric networks, led to the first discovery of a
complete isotropic photonic band gap \cite{florescu_designer_2009,
florescu_optical_2013, man_photonic_2013, man_isotropic_2013,
froufe-perez_band_2017, milosevic_hyperuniform_2019}, which
enables the construction of free-form waveguides
\cite{milosevic_hyperuniform_2019, man_isotropic_2013, florescu_optical_2013}.
In addition, they possess certain nearly optimal transport properties (while
remaining isotropic) when used to model both inclusion-based and cellular composites
\cite{zhang_transport_2016, torquato_multifunctional_2018}, which emphasizes
the importance of the underlying point process geometry. The link between the
unique structural properties of stealthy disordered processes and
their obvious utility for materials design is still not fully understood, but
it has been conjectured that the bounded hole size property plays a key role in
producing their novel thermodynamic and physical properties, including their
desirable band gap, optical, and transport behaviors
\cite{torquato_hyperuniform_2018}.

However, there is still much we do not know about the fundamental structural
properties of disordered stealthy processes. One such type of fundamental
question involves determining the analytical functional forms for the
nearest-neighbor functions of a given particle or void point in the system
\cite{torquato_nearest-neighbour_1990, torquato_nearest-neighbor_1990}. These
functions encode the statistical distribution of intuitive geometric concepts
such as the size of holes in a system, making them good candidates for capturing
the statistical properties of stealthy disordered processes,
which possess bounded hole sizes. These functions come in two general
varieties: the void nearest-neighbor functions, which identifies the nearest
neighbor of an arbitrary spatial point in the system, and the particle
nearest-neighbor functions, which identify the nearest neighbor of an arbitrary
particle in the system. While these varieties are generally distinct, they can
sometimes be related to each other for specific point processes, such as
equilibrium hard spheres \cite{torquato_nearest-neighbour_1990,
torquato_nearest-neighbor_1990}.

The nearest-neighbor functions and
variants have played a key role in investigating problems in a
variety of scientific fields.
These include the application of the Wigner surmise in nuclear
physics \cite{madan_lal_mehta_random_1991,
torquato_point_2008}, their fundamental appearance in the theory of liquids and
other amorphous systems \cite{rintoul_nearest-neighbor_1996, reiss_statistical_1959,
torquato_nearest-neighbour_1990, torquato_nearest-neighbor_1990,
torquato_computer_1990, torquato_mean_1995, torquato_nearest-neighbor_1995,
finney_random_1970, berryman_random_1983, zachary_anomalous_2011,
zachary_hyperuniformity_2011, rintoul_algorithm_1995, sastry_statistical_1997,
maiti_characterization_2013, klatt_characterization_2016,
viot_nearest-neighbor_1998, macdonald_nearest-neighbor_1992}, the study of
astrophysical dynamics \cite{chandrasekhar_stochastic_1943}, the
characterization of membranes in cells \cite{cornell_modeling_1981}, and the
modeling of granular flows \cite{michele_larcher_segregation_2013}.  They have
also been applied to the study of fundamental problems in the mathematical
discipline of discrete geometry, including the covering and quantizer problems
\cite{torquato_reformulation_2010}.

In addition to their utility in describing systems of
fundamental scientific and mathematical interest, one can use them to derive
statistics to characterize the microstructure of complex materials. One example of
such a derived quantity is the distribution of pore sizes in a heterogeneous material
\cite{torquato_random_2002, sorichetti_determining_2020}. They can also be used to estimate transport
properties, such as the rate of a diffusion-controlled reaction
\cite{keller_extremum_1967, rubinstein_diffusioncontrolled_1988,
torquato_diffusioncontrolled_1989, torquato_random_2002}. Determining accurate
formulas for the nearest-neighbor functions of a system can thus aid in
materials discovery.

Based on strong theoretical and computational evidence, Zhang, Stillinger,
and Torquato \cite{zhang_can_2017} formulated the surprising conjecture that
any stealthy system has the aforementioned bounded hole size property, which
was subsequently proven by Ghosh and Lebowitz \cite{ghosh_generalized_2018}. It
is important to note that the converse is not true; there exist systems such as
random sequential addition at the {\it saturation} state that
have bounded holes by construction but are not stealthy
\cite{torquato_random_2006, zhang_precise_2013}. The nearest-neighbor
functions of disordered stealthy systems have also been
studied computationally in light of their connection with transport properties
\cite{zhang_transport_2016}, and a few results are known based on analytical
approximations we will use later in this article \cite{torquato_ensemble_2015}.
However, to date, there has not been a systematic theoretical investigation of
their nearest-neighbor statistics, and little is known about their asymptotics
as the critical-hole radius (i.e. radius of the largest possible hole) $r_c$ is
approached.

In this article, we obtain accurate theoretical expressions for these functions
for disordered stealthy hyperuniformity. The accuracy of our formulas is verified
through simulations presented in Refs. \cite{batten_classical_2008,
torquato_ensemble_2015, zhang_ground_2015, zhang_ground_2015-1,
zhang_can_2017}. We pay particular attention to the small-$r$ behavior of the
functions and asymptotics on approach to the critical-hole size.

In the small-$r$ regime, we are able to obtain a variety of
approximations and bounds due to the pseudo-hard-sphere ansatz
\cite{torquato_ensemble_2015}, which is valid when considering stealthy point
processes with low to intermediate $\chi$. In particular, we are able to derive
small-$r$ expansions that can provide useful approximations, even outside the
small-$\chi$ limit. We also provide supporting evidence for a new conjecture on
the validity of two upper bounds. Going beyond the methods based on the
pseudo-hard-sphere ansatz, we demonstrate that
the nearest-neighbor functions can be determined by a finite number of
$g_n({\bi r}^n)$, in contrast with the general case, which requires an infinite
number of $g_n({\bi r}^n)$. In the large-$r$ regime, we consider the
scaling behavior of these functions as they approach the critical-hole size. We
compare their behavior to that of ordered point configurations through theoretical
arguments and the analysis of simulation data. We encounter substantial numerical
difficulty due to the rarity of finding holes close to the critical-hole radius, which
we argue is exacerbated in disordered systems due to the expectation that
the hole probability vanishes more quickly in the presence of disorder. This
difficulty points to the need for the development of more efficient simulation
methods for these exotic potentials as well as further
research into theoretical methods for determining the behavior of holes
near the critical-hole radius. We also discuss a useful prescription for
linking the small-$r$ and near-$r_c$ regime into an approximation accurate
over all $r$, as validated by comparison to simulations. Finally, we
comment on the large-$r$ asymptotic behavior of the nearest-neighbor functions
of stealthy systems at positive temperature, where they lose their strict
stealthiness property, and show that they are also expected to lose their
bounded holes property.

Section \ref{prelim} covers the basic theory of the nearest-neighbor functions
and stealthy hyperuniform point processes. In Sec. \ref{pseudohs}, we provide analytical
bounds and approximations obtained through the pseudo-hard-sphere approximation valid at
small-$r$. We consider the problem of determining how many of the $g_n({\bi r}^n)$ are
needed to determine the nearest-neighbor functions of stealthy systems in
Section \ref{higherorder}. Section \ref{tail} presents a description of the asymptotic behavior
near the critical-hole size of the nearest-neighbor functions.
Section \ref{domain} discusses the problem of linking the small and large-$r$
regimes to obtain expressions for the nearest-neighbor functions over all $r$.
Section \ref{finitetemp} describes positive temperature results. In Sec.
\ref{conclusion}, we summarize our findings and makes some concluding remarks.

\section{Preliminaries}\label{prelim}
\subsection{Definitions for Nearest-Neighbor Functions}\label{nnintro}

\subsubsection{``Void'' Quantities}

The nearest-neighbor functions are special cases of the general $n$-point
canonical function and thus obey the same mathematical properties,
such as the rigorous bounds described below \cite{torquato_microstructure_1986}. We
will begin by defining the void nearest-neighbor probability density function
$H_V(r)$ as in \cite{torquato_nearest-neighbor_1990}:
\begin{eqnarray}
    \fl    
    H_V(r)\,\rmd r = \textrm{probability that at an arbitrary located point in the system, the nearest point}\nonumber\\
        \textrm{in the point process lies between } r \textrm{ and } r + dr.
\end{eqnarray}
This probability density is also closely related to the pore-size probability
density function of the two-phase system that forms when the points are
decorated with spheres of radius $R$ \cite{torquato_random_2002}. Under this
assumption, the pore-size function becomes
\cite{torquato_random_2002}
\begin{equation}
    P(\delta) = \frac{H_V(\delta + R)}{\phi_1},
\end{equation}
where $\phi_1$ is the volume fraction of the void phase.

The associated complementary cumulative distribution function, called
the void exclusion probability function, is given by 
\cite{torquato_nearest-neighbor_1990}
\begin{equation}
    E_V(r) = 1 - \int_0^r H_V(r')\,\rmd r'.\label{evhv}
\end{equation}
This has the following interpretation \cite{torquato_nearest-neighbor_1990}:
\begin{eqnarray}
    \fl
    E_V(r) = \textrm{probability that given an arbitrary location in the void, a ball of radius } r\nonumber\\
        \textrm{centered at that location is devoid of points}.
\end{eqnarray}
This definition is often given succintly as the probability of finding a hole of
radius $r$.

We can define a third nearest-neighbor function by expressing $H_V(r)$ in terms
of a conditional probability density $G_V(r)$
\cite{torquato_nearest-neighbor_1990}
\begin{equation}
    H_V(r) = \rho s_1(r) G_V(r) E_V(r),\label{evhvgv}
\end{equation}
where $s_1(r)$ is the surface area of a $d$-dimensional sphere of radius $r$.
Thus, $G_V(r)$ has the interpretation \cite{torquato_nearest-neighbor_1990}:
\begin{eqnarray}
    \fl
    \rho s_1 G_V(r)\,\rmd r = \textrm{probability of finding a particle between } r \textrm{ and } r+ dr \textrm{ given that}\nonumber\\
        \textrm{one has found a hole of radius } r.
\end{eqnarray}

The asymptotic behavior of the function $G_V(r)$ is intimately related to the
work required to create a cavity of radius $r$ in an equilibrium system at
positive temperature \cite{reiss_statistical_1959}. This enables one to
relate the long-range behavior to the ratio of the pressure and temperature of
a system \cite{reiss_statistical_1959}:
\begin{equation}
    G_V(r\to\infty) = \frac{p}{\rho k_B T}. \label{thermo}
\end{equation}

To assist in building intuition for the behavior of these functions, we note that
the Poisson point process has a void exclusion probability function of \cite{hertz_uber_1909}
\begin{equation}
    E_V(r) = \exp\left(-\rho v_1(r)\right).
\end{equation}
One of the key features of the  nearest-neighbor functions of
stealthy systems is their limited support due to their bounded hole size
\cite{zhang_can_2017, ghosh_generalized_2018}, in contrast to the infinite
support of many disordered point processes, including the Poisson
distribution. A more detailed comparison of the void nearest-neighbor
functions for several different physical systems is described in the Supplementary Material
\cite{t_middlemas_supplemental_nodate}.

The $n$th moments $\langle r^n\rangle$ of the functions $H_V(r)$ and $E_V(r)$ are important summary
statistics for a point process, and are defined by
\cite{torquato_reformulation_2010}
\begin{equation}
    \langle r^n \rangle = \int_0^\infty r^n H_V(r)\,\rmd r = n\int_0^\infty r^{n-1} E_V(r)\,\rmd r\qquad n\in \mathbb{Z}^+.
\end{equation}
In particular, the first moment of $H_V(r)$ gives the mean distance $l_V$ 
from an arbitrary location in the void to the nearest point of the process:
\begin{equation}
    l_V \equiv \langle r \rangle = \int_0^\infty r H_V(r)\,\rmd r = \int_0^\infty E_V(r)\,\rmd r. 
\end{equation}

\subsubsection{``Particle'' Quantities}

One can also define a corresponding set of functions that measure the nearest-neighbor
statistics with respect to an arbitrary particle rather than a void point. The
particle nearest-neighbor distribution function is defined \cite{torquato_nearest-neighbor_1990}:
\begin{eqnarray}
    \fl
    H_P(r)\,\rmd r = \textrm{probability that the nearest point to a point of the point}\nonumber\\
        \textrm{process lies between } r \textrm{ and } r+dr.
\end{eqnarray}
We can define $E_P(r)$ and $G_P(r)$ in the same manner as for the
void functions \cite{torquato_nearest-neighbor_1990}:
\begin{eqnarray}
    E_P(r) &=& 1 - \int_0^r H_P(r)\,\rmd r\\
    H_P(r) &=& \rho s_1(r) E_P(r) G_P(r).
\end{eqnarray}

In general, the particle functions differ from the void functions for a given
system, but can sometimes be related to them for special systems. For example,
the expression for $E_P(r)$ for a Poisson point process is
\cite{torquato_nearest-neighbor_1990}
\begin{equation}
    E_P(r) = \exp\left(-\rho v_1(r)\right),
\end{equation}
which is the same as the expression for $E_V(r)$. In addition, the particle
nearest-neigbor functions can be determined from the void variants for
hard-sphere systems \cite{torquato_nearest-neighbor_1990}. We note in passing that the relation
between the void and particle variants of a given statistical quantity are
studied in the subject of Palm theory in stochastic geometry
\cite{daley_introduction_1998, s_n_chiu_stochastic_2013}. The interested
reader can refer to the Supplementary Material \cite{t_middlemas_supplemental_nodate} for a comparison
of the particle nearest-neighbor functions for a variety of physical systems.

The moments of $H_P(r)$ and $E_P(r)$ can be related to each in a similar
manner to that of the void quantities \cite{torquato_reformulation_2010}:
\begin{equation}
    \int_0^\infty r^n H_P(r)\,\rmd r = n\int_0^\infty r^{n-1}E_P(r)\,\rmd r\qquad n\in\mathbb{Z}^+.
\end{equation}
The mean nearest-neighbor distance is the first moment of $H_P(r)$ or
the integral over $E_P(r)$
\cite{torquato_nearest-neighbour_1990, torquato_reformulation_2010}:
\begin{equation}
        l_P = \int_0^\infty r H_P(r)\,\rmd r = \int_0^\infty E_P(r)\,\rmd r.
\end{equation}

\subsubsection{Series and Bounds}\label{introbounds}
Importantly, the nearest-neighbor functions can be represented as a
series expansion involving functionals of the standard $n$-point correlation
functions $g_n(r)$ \cite{torquato_nearest-neighbor_1990}. For example, in the
case of a translationally invariant point process, the void exclusion
probability can be written \cite{torquato_nearest-neighbor_1990}:
\begin{equation}
    E_V(r) = 1 + \sum_{k=1}^\infty (-1)^k \frac{\rho^k}{k!} \int g_k({\bi R}^k)
    \prod_{j=1}^k \Theta(r - |{\bi x} - {\bi R}_j |)\,\rmd {\bi
    R}_j,\label{evexpansion}
\end{equation}
where the value of ${\bi x}$ can be chosen arbitrarily. Note that this implies
that the nearest-neighbor functions incorporate partial information from
higher-order distribution functions.

\begin{figure}
    \centering
    \includegraphics[width=0.6\linewidth]{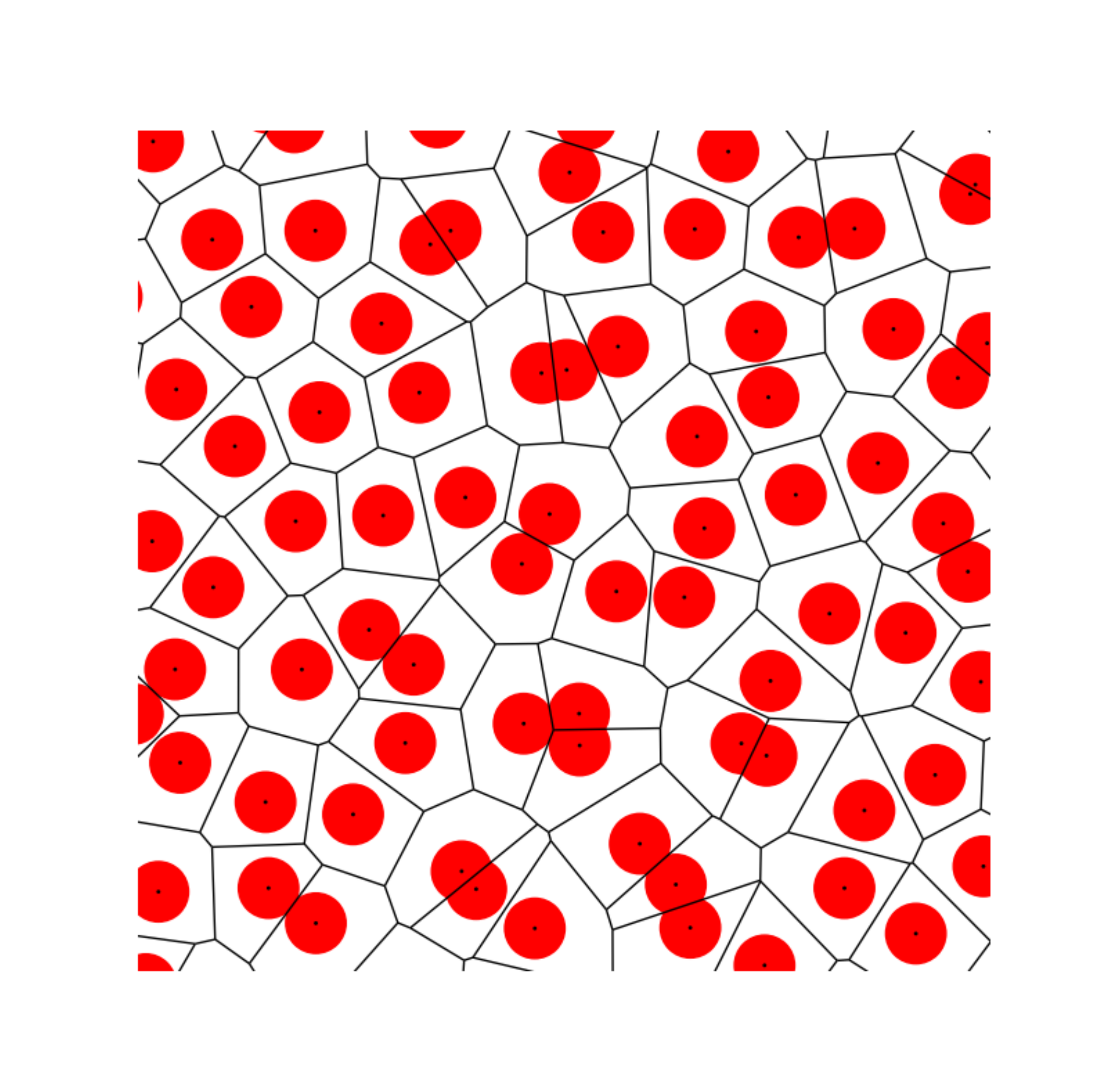}
    \caption{A disordered stealthy system decorated with spheres of radius $r$
    and its Voronoi diagram. The average over the Voronoi cells of the ratio of
    the area outside the spheres to the total area of the cell is $E_V(r)$. This
    picture of $E_V(r)$ also demonstrates its relation to the covering problem, where
    the critical-hole radius $r_c$ needed to cover all space is known as the covering radius
    \cite{torquato_reformulation_2010}.}\label{small_voronoi}
\end{figure}

This series has a fundamental geometric interpretation, which can be seen by
considering the diagram given in Fig. \ref{small_voronoi}. If one has a single
point configuration, this figure shows
that one can compute $E_V(r)$ by computing the ratio of the
volume outside a set of covering spheres of radius $r$ to the total volume in the
process, normalized appropriately by either the fundamental cell or by averaging
over the Voronoi cells \cite{hoover_exact_1979, speedy_computer_1991,
rintoul_algorithm_1995, sastry_statistical_1997}. The series
(\ref{evexpansion}) is then just the computation of this volume
fraction through the principle of inclusion-exclusion applied to the spheres. More precisely,
for a single point configuration, the above series (\ref{evexpansion}) becomes \cite{torquato_reformulation_2010}:
\begin{equation}
        E_V(r) = 1 - \rho v_1(r) + \frac{1}{v_F} \sum_{i<j} v_2^{\rm int} (x_{ij}; r)
                 - \frac{1}{v_F} \sum_{i<j<k} v_3^{\rm int} (x_{ij}, x_{ik}, x_{jk}; r) + \cdots,\label{finiteseries}
\end{equation}
where $v_F$ is the volume of the fundamental cell. This series is expected to
truncate exactly for any periodic system with a finite basis
\cite{torquato_reformulation_2010}, such as the face-centered-cubic lattice
and the hexagonal-close-packed crystal. For example, in the case of the
square and triangular lattices, this series terminates at the two-body term
\cite{torquato_reformulation_2010}. However, as we will discuss in Section \ref{tail},
it may also truncate for special disordered systems. We apply this geometric formulation of the
void nearest-neighbor functions in the numerically sampling
of computer-generated 1D stealthy configurations later in the article; see the
Appendix for details. In addition, this view of the void functions demonstrates
their close relation to important problems in discrete geometry
\cite{torquato_reformulation_2010}. For example, in the covering problem, one
defines the {\it covering radius} as the minimum radius of the spheres in Fig.
\ref{small_voronoi} required to cover all space
\cite{torquato_reformulation_2010}. Then, one can define the problem as a
search for the point configuration which minimizes the covering radius
\cite{torquato_reformulation_2010} at unit density. While the covering radius
is finite for any single periodic point configuration with a finite basis, it
is not necessarily finite for an arbitrary disordered point process. However,
in the case of stealthy point processes, it corresponds to the critical-hole
radius $r_c$. It is worthwhile to note that the optimal configurations
for the covering problem are the triangular lattice in two dimensions and the body-centered-cubic
lattice in three dimensions \cite{torquato_reformulation_2010,
conway_sphere_1999}. These lattices are also the entropically favored states
for stealthy systems in the ordered $\chi > 1/2$ regime
\cite{zhang_ground_2015}. While we focus on the disordered regime in this
paper, these optimal configurations still play an important role, since for
$\chi$ close to $1/2$, the disordered stealthy configurations will show
precursor characterstics of these lattices.

Interestingly, this representation also forms a series of successive upper and
lower bounds, which is described by a powerful general formalism developed in
Ref. \cite{torquato_microstructure_1986}. For example, for a homogeneous and
isotropic point process, one has \cite{torquato_microstructure_1986,
torquato_point_2008}
\begin{eqnarray}
    E_V(r) &\leq& 1,\label{upper1}\\
    E_V(r) &\geq& 1 - \rho v_1(r)\label{epbound1},
\end{eqnarray}
and
\begin{equation}
        E_V(r) \leq 1 - \rho v_1(r) + \frac{\rho^2}{2} s_1(1) \int_0^{2r} x^{d-1} v_2^{\rm int}(x; r) g_2(x)\,\rmd x,\label{epbound2}
\end{equation}
where $v_2^{\rm int}(x; r)$ is the intersection volume of two spheres of radius $r$
a distance of $x$ apart, which is known analytically in any
dimension \cite{torquato_new_2006}. In the first three space dimensions, these intersection volumes can be expressed, respectively, as
\cite{torquato_random_2002}
\begin{eqnarray}
    \frac{v_2^{\rm int}(x; r)}{v_1(r)} = \Theta(2r - x)\left(1-\frac{x}{2r}\right)  &\qquad d = 1,\\
    \frac{v_2^{\rm int}(x; r)}{v_1(r)} = \frac{2}{\pi}\Theta(2r - x)\left[\cos^{-1}\left(\frac{x}{2r}\right) - \frac{x}{2r}\sqrt{1- \frac{x^2}{4r^2}}\right] &\qquad d = 2,\\
    \frac{v_2^{\rm int}(x; r)}{v_1(r)} = \Theta(2r - x)\left[1 - \frac{3}{4}\frac{x}{r} + \frac{1}{16}\left(\frac{x}{r}\right)^3\right]&\qquad d = 3,
\end{eqnarray}
where $\Theta(x)$ denotes the Heaviside step function. The last inequality
(\ref{epbound2}) exactly gives $E_V(r)$ whenever only two-body terms contribute,
such as in the case of the square and triangular lattices in
2D \cite{torquato_reformulation_2010}. In the disordered case, this
series can be used to derive low-$r$ expansions for $E_V(r)$ by expanding
$g_2(r)$ in powers of $r$ \cite{torquato_point_2008}:
\begin{equation}
    g_2(r) \sim a + b r^2\qquad (r \to 0),
\end{equation}
where, for the purposes of simplification, we have used in advance
the fact that there is good evidence that the linear term in the preceding
expansion vanishes for the types of disordered stealthy hyperuniform systems
considered in this article \cite{torquato_ensemble_2015, zhang_ground_2015}.
The order to which $E_V(r)$ can then be determined depends on the spatial
dimension. In one and two dimensions, one can obtain results of the form
\cite{torquato_point_2008, torquato_random_2002}
\begin{equation}
    E_V(r) \sim 1 - \rho v_1(r) + \frac{\rho^2 a}{2} v_1(r)^2 \qquad (r\to 0).\label{1d2dlowr}
\end{equation}
In three and higher dimensions, one can obtain a fourth term
\cite{torquato_point_2008, torquato_random_2002}:
\begin{equation}
        E_V(r) \sim 1- \rho v_1(r)
        + \frac{\rho^2}{2}\left(a v_1(r)^2 +
    b\frac{2d}{d+2} r^2 v_1(r)^2 \right) \qquad (r\to 0).\label{3dlowr}
\end{equation}
We will derive expressions for the $a$ and $b$ coefficients valid at low and
intermediate values of $\chi$ in the next section.

One can repeat this analysis for $H_V(r), E_P(r),$ and $H_P(r).$
One obtains the following series expansions \cite{torquato_nearest-neighbor_1990}:
\begin{eqnarray}
        H_V(r) = \sum_{k=1}^\infty (-1)^{k+1} \frac{\rho^k}{k!}
        \int g_k({\bi R}^k) \frac{\partial}{\partial r}\prod_{j=1}^k \Theta(r - |{\bi x} - {\bi R}_j|)\,\rmd {\bi R_j},\\
        E_P(r) = 1 + \sum_{k=1}^\infty (-1)^k \frac{\rho^k}{k!}
        \int g_{k+1}({\bi R}^{k+1}) \prod_{j=2}^{k+1} \Theta(r - |{\bi R}_j - {\bi R}_1|)\,\rmd {\bi R}_j,\\
        H_P(r) = \sum_{k=1}^\infty (-1)^{k+1} \frac{\rho^k}{k!}
        \int g_{k+1}({\bi R}^{k+1}) \frac{\partial}{\partial r}\prod_{j=2}^{k+1} \Theta(r - |{\bi R}_j - {\bi R}_1|)\,\rmd {\bi R}_j.
\end{eqnarray}
Note that the sequence of partial sums can be written in the form \cite{torquato_microstructure_1986}:
\begin{equation}
    W^n = \sum_{k=0}^n X^{(k)},
\end{equation}
where $X$ represents one of the aforementioned functions, $X^{(k)}$ represents
the $k$th term of the series for that function, and we have reindexed the
series for $H_{V/P}(r)$ to start at $k=0$. Then, one obtains
bounds of the form \cite{torquato_microstructure_1986}
\begin{eqnarray}
    X \leq W^n&\qquad n\textrm{ even,}\nonumber\\
        X \geq W^n&\qquad n\textrm{ odd.}
\end{eqnarray}
We will explicitly use the first two successive bounds on $H_V(r)$
and $E_P(r)$ \cite{torquato_point_2008}:
\begin{eqnarray}
    H_V(r) \leq \rho s_1(r),\label{hvbound1}\\
    H_V(r) \geq \rho s_1(r) - \frac{\rho^2}{2}s_1(1)\int_0^{2r} x^{d-1}
    s_2^{\rm int}(x; r) g_2(x)\,\rmd x,\label{hvbound2}\\
    E_P(r) \leq 1,\\
    E_P(r) \geq 1 - Z(r)\label{epbound},
\end{eqnarray}
where $s_2^{\rm int}(x;r) = \partial v_2^{\rm int}(x; r)/\partial r$ is the intersection
surface area of two spheres of radius $r$ a distance $x$ apart and $Z(r) =
\rho s_1(1)\int_0^r x^{d-1} g_2(x)\,\rmd x$ is the cumulative coordination number.
We will also use the first bound on $H_P(r)$ \cite{torquato_point_2008}:
\begin{equation}
    H_P(r) \leq \rho s_1(r) g_2(r).\label{hpbound}
\end{equation}

To obtain upper (lower) bounds on $G_{V/P}(r)$, one can match an upper (lower)
bound on $H_{V/P}(r)$ with a lower (upper) bound on $E_{V/P}(r)$
\cite{torquato_point_2008}. For example, in this paper, we will use the bounds
\cite{torquato_point_2008}:
\begin{eqnarray}
        G_V(r) \leq \frac{1}{1-\rho v_1(r)},\label{gvbound1}\\
        G_V(r) \geq \frac{1 - \frac{\rho s_1(1)}{2 s_1(r)} \int_0^{2r} x^{d-1}
        s_2^{\rm int}(x; r) g_2(x)\,\rmd x}{ 1 - \rho v_1(r) +
        \frac{\rho^2}{2}s_1(1)\int_0^{2r} x^{d-1}v_2^{\rm int}(x; r) g_2(x)\,\rmd x},\label{gvbound2}\\
        G_P(r) \leq \frac{g_2(r)}{1 - Z(r)}.\label{gpbound}
\end{eqnarray}

As in the case of $E_V(r)$, one can use an expansion for $g_2(r)$ to derive
low-$r$ expansions for all of the nearest-neighbor functions. In one and two
dimensions, one finds \cite{torquato_point_2008}
\begin{eqnarray}
    H_V(r) \sim \rho s_1(r) - \rho^2 a v_1(r) s_1(r)&\qquad (r\to 0),\\
    G_V(r) \sim 1 + (1-a)\rho v_1(r)&\qquad (r \to 0),\label{gvSeries}\\
    E_P(r) \sim 1 - \rho a v_1(r)&\qquad (r \to 0),\\ 
    H_P(r) \sim \rho a s_1(r)&\qquad (r \to 0),\\
    G_P(r) \sim a&\qquad (r \to 0).
\end{eqnarray}
In three dimensions, one finds \cite{torquato_point_2008}
\begin{eqnarray}
    H_V(r) \sim \rho s_1(r)
        - \rho^2 \left[a v_1(r) s_1(r) + \frac{24b}{5} rv_1(r)^2\right] &\qquad (r\to 0),\\
    G_V(r) \sim 1 + (1-a)\rho v_1(r)
        - \frac{8 \rho b}{5}r^2 v_1(r)&\qquad (r\to 0),\\
        E_P(r) \sim 1 - \rho\left(av_1(r) + \frac{3b}{5}r^2 v_1(r)\right)&\qquad (r\to 0),\\
        H_P(r) \sim \rho\left(as_1(r) + \frac{12b}{5}r v_1(r)\right)&\qquad (r\to 0),\\
    G_P(r) \sim a + br^2&\qquad (r\to 0).
\end{eqnarray}

\subsection{Stealthy Hyperuniform Point Processes}

The stealthy constraint given by Eq. (\ref{sfdef}) only involves the two-point
information contained in the point process. However, we will see it has
implications for the form of the nearest-neighbor functions, which incorporate
higher-order information \cite{torquato_microstructure_1986}. The
configurational space of all stealthy systems [defined by
(\ref{sfdef})] is infinitely large in the thermodynamic limit
and extremely complex, so we make a practical restiction of our focus to a
specific distribution over this space: the canonical ensemble as $T\to 0$
\cite{torquato_ensemble_2015, zhang_ground_2015}. We will see that
the study of this well-defined ensemble provides powerful generic insights
about stealthy systems.

\subsubsection{Basic Definitions for Point Processes}

We introduce the general concepts applicable to all point processes
we will encounter throughout this article, using definitions that,
while not completely mathematically rigorous, will be
sufficient for our purposes. One can think of a $d$-dimensional
\textit{point process} as a configuration consisting of a countably infinite
number of points in $\mathbb{R}^d$ such that the density is well-defined
\cite{torquato_ensemble_2015}. If one has an ergodic process, one can compute
statistics of the point process such as the pair correlation function $g_2(r)$
through either a volume average over a single point configuration or through an
ensemble average over many such configurations \cite{torquato_random_2002}. One
important class of ordered point processes are known as \textit{lattices}, which
are point processes described by a set of linearly independent lattice vectors
$\{{\bi v}_i\}$ in $\mathbb{R}^d$. The points are placed at the integer
combinations of these lattice vectors, so that the location of an arbitrary
point is described by the expression:
\begin{equation}
    {\bi r} = \sum_{i=1}^d m_i {\bi v}_i\qquad m_i \in \mathbb{Z}.
\end{equation}
One can generalize this notion to describe a \textit{periodic} point process, which
is an arbitrary \textit{crystal}, by including a finite set of basis vectors
${\bi b}_n$, which describe the position of
the particles in the fundamental cell given by the lattice vectors.
Thus, the points of the crystal are given by the union of the sets $\{{\bi
r}_n\}$, where the members of the set for each $n$ are given by
\begin{equation}
    {\bi r}_n = \sum_{i=1}^d m_i {\bi r}_i + {\bi b}_n\qquad m_i \in \mathbb{Z}.
\end{equation}
One can then think of a disordered point
process as one in which both the size of the basis set and the volume of the
fundamental cell grows to infinity, leaving the density fixed.

\subsubsection{Computer Simulation}

The above definition and ensemble lends itself easily to computer simulation. We
will use the collective-coordinate procedure pioneered in Refs. \cite{torquato_ensemble_2015,
zhang_ground_2015} for producing ground states in the canonical ensemble. Consider a finite
system with $N$ particles under periodic boundary conditions. Then, the structure factor can be
evaluated at every ${\bi k}$ in the reciprocal lattice of the fundamental cell with
the equation
\begin{equation}
    S({\bi k}) = \frac{1}{N} \left|\sum_i e^{-i{\bi k}\cdot{\bi r}_i}\right|^2,
\end{equation}
where the sum ranges over all the particles in the fundamental
cell and ${\bi r}_i$ is the position of the $i$th particle. Note that ${\bi k}
= 0$ in the above sum corresponds to the forward scattering, and is
correspondingly omitted from the definition of stealthy hyperuniformity.
In addition, observe that the structure factor has an intrinsic
inversion symmetry:
\begin{equation}
S({\bi k}) = S(-{\bi k}).
\end{equation}
One can then define a
many-particle system in which the particles interact with energy
function \cite{zhang_ground_2015}
\begin{equation}
    \Phi = \frac{1}{v_F}\sum_{k <|{\bi k}| \leq K} N S({\bi k}) - \Phi_0,
\end{equation}
where the sum ranges over the $M$ independently constrained wavevectors and $v_F$ is
the volume of the fundamental cell. The constant $\Phi_0$ is determined by
Parseval's theorem, and can be written \cite{torquato_ensemble_2015, zhang_ground_2015}:
\begin{equation}
    \Phi_0 = \left(N(N-1) - 2NM\right).
\end{equation}
It is clear that all states of minimal energy $\Phi = -\Phi_0$ are stealthy. One can
then sample the canonical ensemble by running a molecular dynamics simulation
at a low temperature (usually around $2\times 10^{-4}$, $2\times 10^{-6}$, and
$1\times 10^{-6}$ in 1, 2 and 3 dimensions, respectively, see the Appendix). To
obtain a ground state configuration, one minimizes the energy of the molecular
dynamics configuration using the L-BFGS algorithm \cite{zhang_ground_2015}. For
more details about the algorithms used to generate configurations in this
article; see the Appendix.

The degree of short, intermediate, and long-range order depends on the
control parameter $\chi$ defined in Eq. (\ref{chieq}). For finite systems, we define
$\chi$ as \cite{torquato_ensemble_2015}
\begin{equation}
    \chi = \frac{M}{d(N-1)},\label{finitesizechi}
\end{equation}
where $M$ is the number of constrained degrees of freedom, $N$ is the number of
particles, and $d$ is the spatial dimension. We can recover
Eq. (\ref{chieq}) by going to the thermodynamic limit
\cite{torquato_ensemble_2015}. It can be shown that the system undergoes a order-disorder transition at
$\chi = 1/3$ in one dimension \cite{fan_constraints_1991} and at
$\chi = 1/2$ in two and three dimensions \cite{uche_constraints_2004,
uche_collective_2006, batten_classical_2008, batten_novel_2009,
batten_interactions_2009, suto_crystalline_2005, torquato_ensemble_2015,
zhang_ground_2015, zhang_ground_2015-1}. We will focus on the disordered low-$\chi$ regime.

\section{Pseudo-hard-sphere Approximations to Nearest-neighbor
Functions}\label{pseudohs}

We begin by deriving expressions useful at small-$r$ for the
nearest-neighbor functions of our disordered stealthy point processes. These
expressions are fundamentally based on the pseudo-hard-sphere ansatz described
below, and are valid for small enough $\chi$. We
also make heavy use of the bounding series given in Section \ref{prelim}.
Throughout, we will compare to simulation data either taken from Ref.
\cite{zhang_transport_2016} or produced by the procedure described in the
Appendix.

\subsection{Basic Theory}\label{pseudohsintro}

To use the upper and lower bounds on the nearest-neighbor
functions given in Section \ref{prelim}, we must first determine an accurate
expression for the pair correlation function $g_2(r)$. Torquato,
Zhang, and Stillinger \cite{torquato_ensemble_2015} developed an analytical
theory valid at sufficiently small $\chi$ in the limit of large systems, justifying
their work through direct simulations of stealthy sytems. They make
the ansatz that the structure factor follows the behavior of the pair
correlation function of a hard-sphere system at a density related to $\chi$
[defined by (\ref{chieq})], namely,
\begin{equation}
    S(k) = g_2^{HS}(r = k),
\end{equation}
where $g_2^{HS}(r)$ is the pair correlation function for a hard-sphere system of
diameter $K$ and packing fraction
\begin{equation}
    \eta = \frac{\chi}{\alpha(K;K)2^d},
\end{equation}
where $\alpha(r;R) = v_2^{\rm int}(r; R)/v_1(R)$ is the scaled intersection volume
of two spheres of radius $R$ separated by $r$. This approximation closely
follows the simulated $S(k)$ and $g_2(r)$ for
$\chi \leq 0.15$ \cite{torquato_ensemble_2015, zhang_ground_2015}. We will use
this approximation as a starting point to derive theories valid at small enough
values of $r$. In particular, we will make use of the following low-$\chi$
expansion \cite{torquato_ensemble_2015}:
\begin{equation}
    S(k) \approx \Theta(k-K) \left( 1 + \chi \frac{\alpha(k;K)}{\alpha(K;K)}
    \right),\label{lowchi}
\end{equation}
valid in any dimension. We will also use the
generalized Orstein-Zernike relation \cite{torquato_ensemble_2015}
\begin{equation}
    \tilde{H}(k) = \tilde{C}(k) + \frac{\eta}{v_1(K/2)} \tilde{H}(k) \otimes
    \tilde{C}(k),\label{OSeq}
\end{equation}
where $\tilde{H}(k) = S(k) - 1$ and $\tilde{C}(k) = c^{HS} (r = k)$, where
$c^{HS}(r)$ is the standard direct correlation function
\cite{jean_pierre_hansen_theory_1986} for the aforementioned hard-sphere
system. For a more detailed discussion of the pair statistics of disordered
stealthy systems; see the Supplementary Material \cite{t_middlemas_supplemental_nodate}.

\subsection{1D Results}

\begin{figure}
    \begin{center}
    \begin{subfigure}{0.3\textwidth}
        \includegraphics[width=\linewidth]{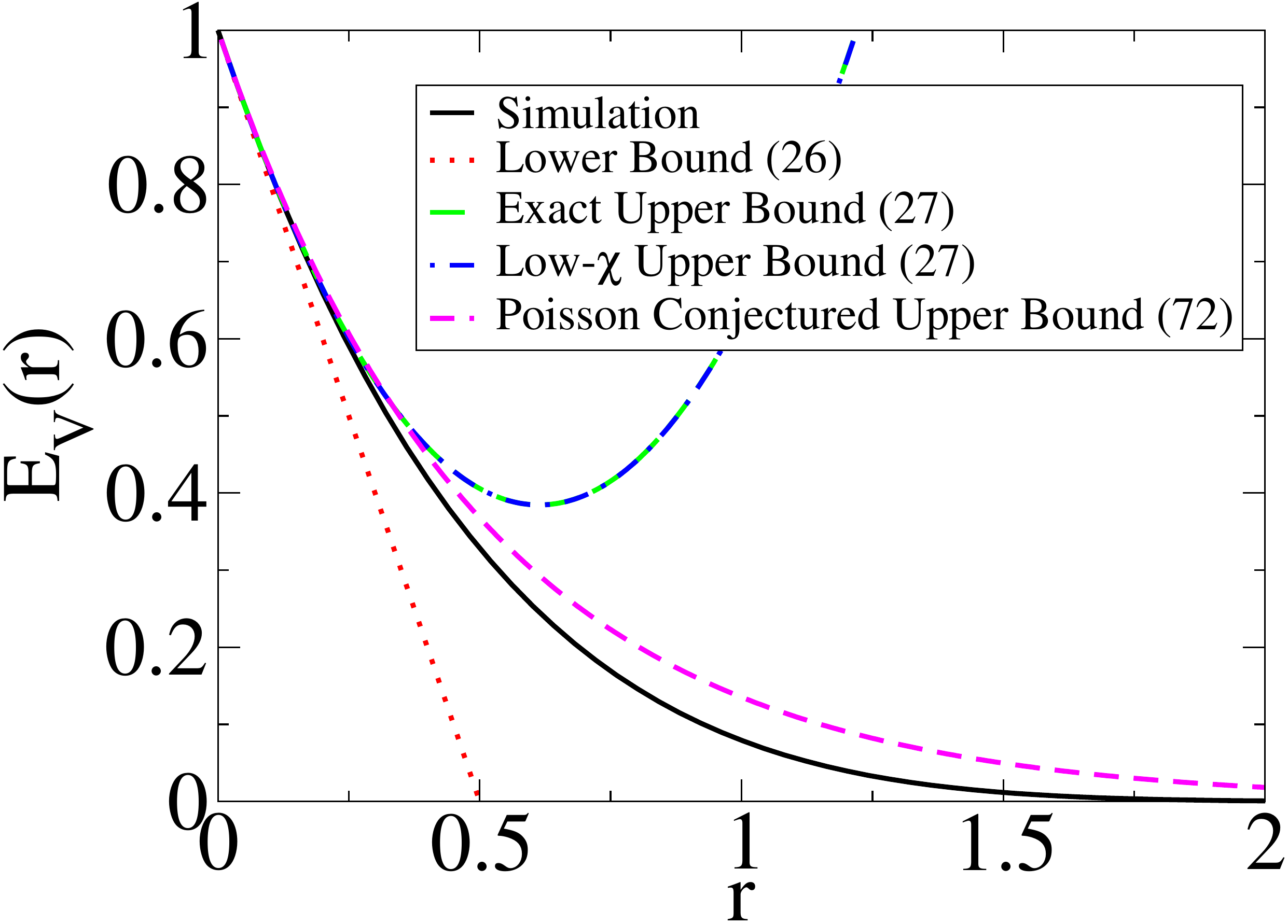}
        \caption{}
    \end{subfigure}
    \begin{subfigure}{0.3\textwidth}
        \includegraphics[width=\linewidth]{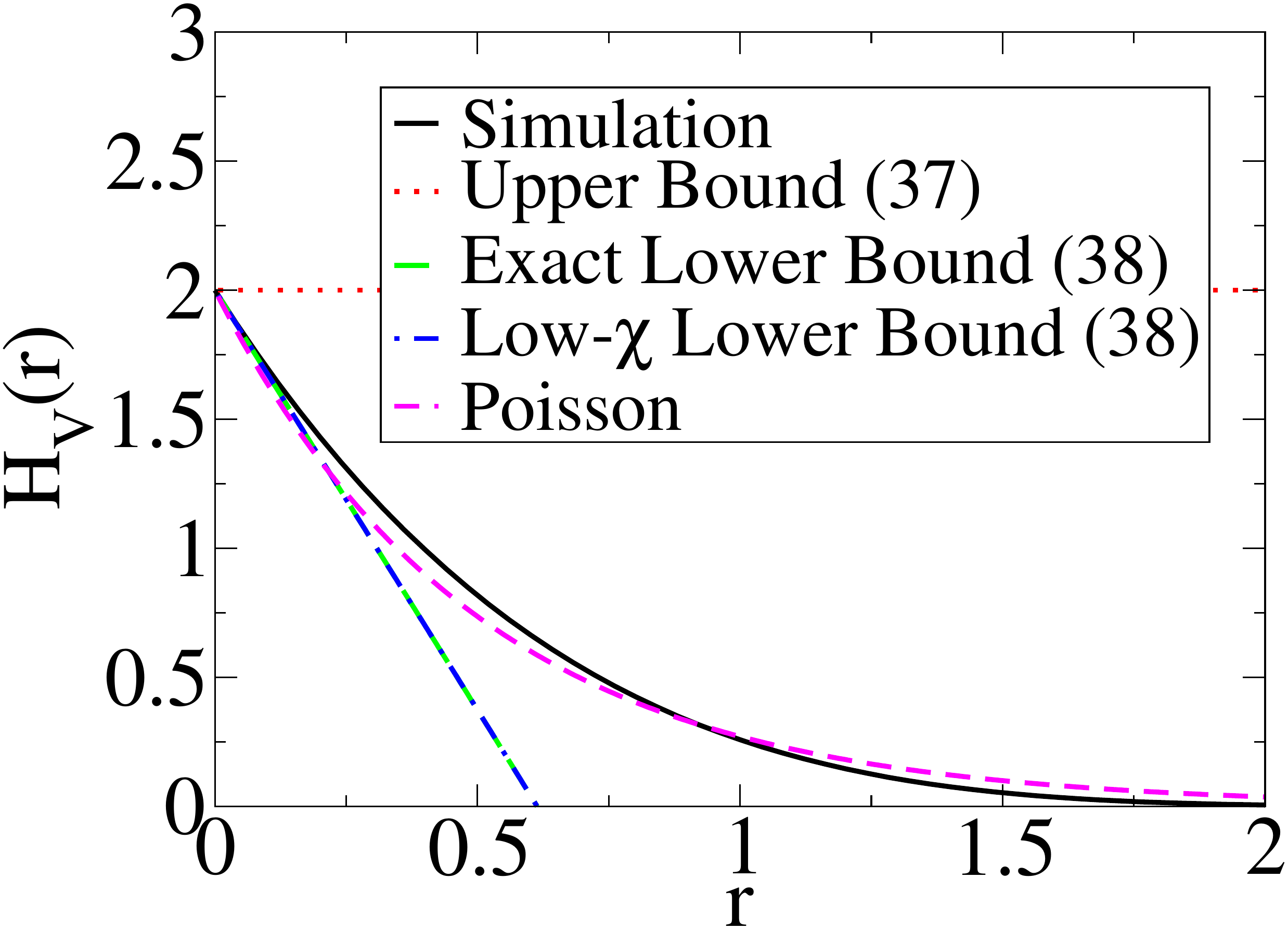}
        \caption{}
    \end{subfigure}
    \begin{subfigure}{0.3\textwidth}
        \includegraphics[width=\linewidth]{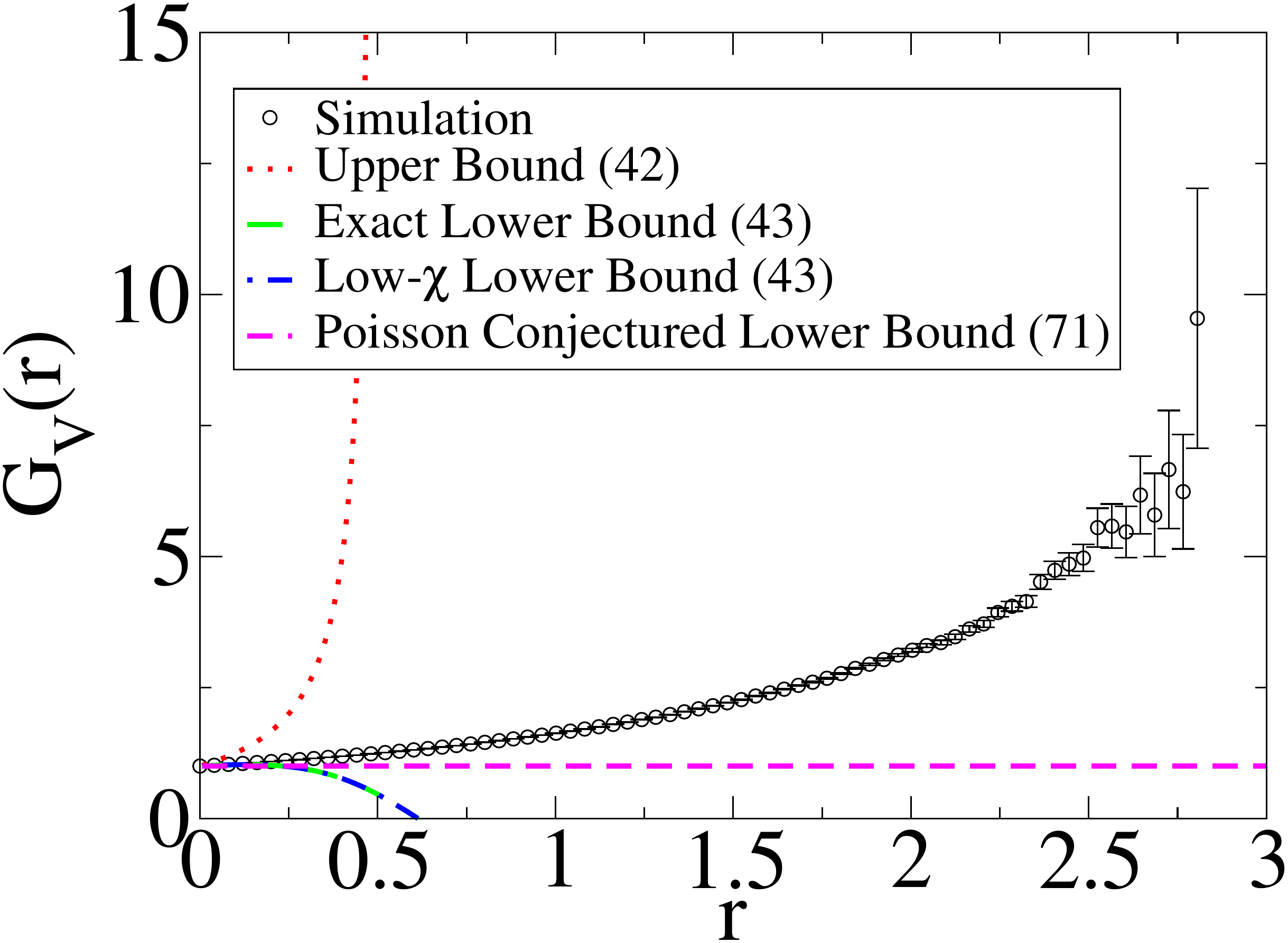}
        \caption{}
    \end{subfigure}\\
    \begin{subfigure}{0.3\textwidth}
        \includegraphics[width=\linewidth]{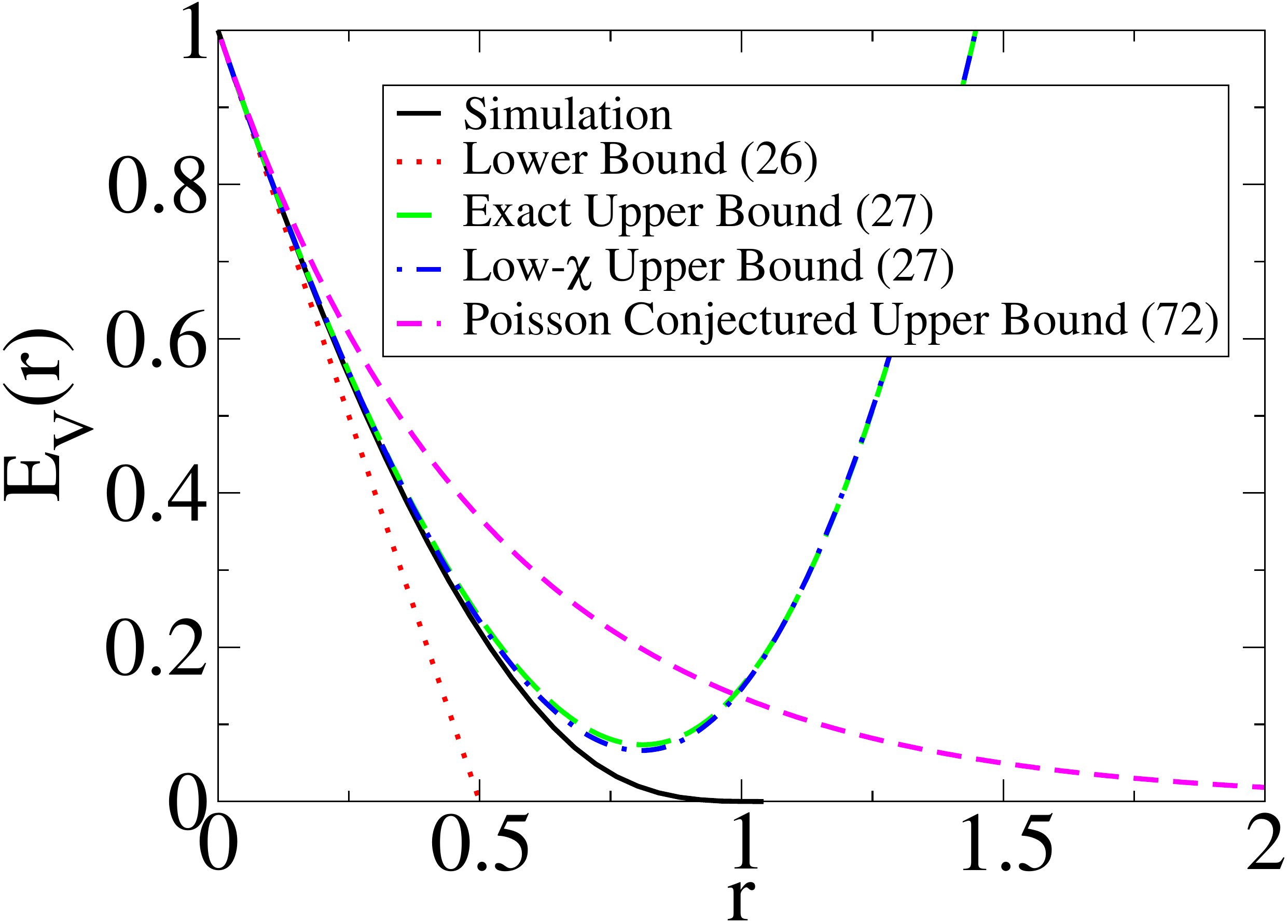}
        \caption{}
    \end{subfigure}
    \begin{subfigure}{0.3\textwidth}
        \includegraphics[width=\linewidth]{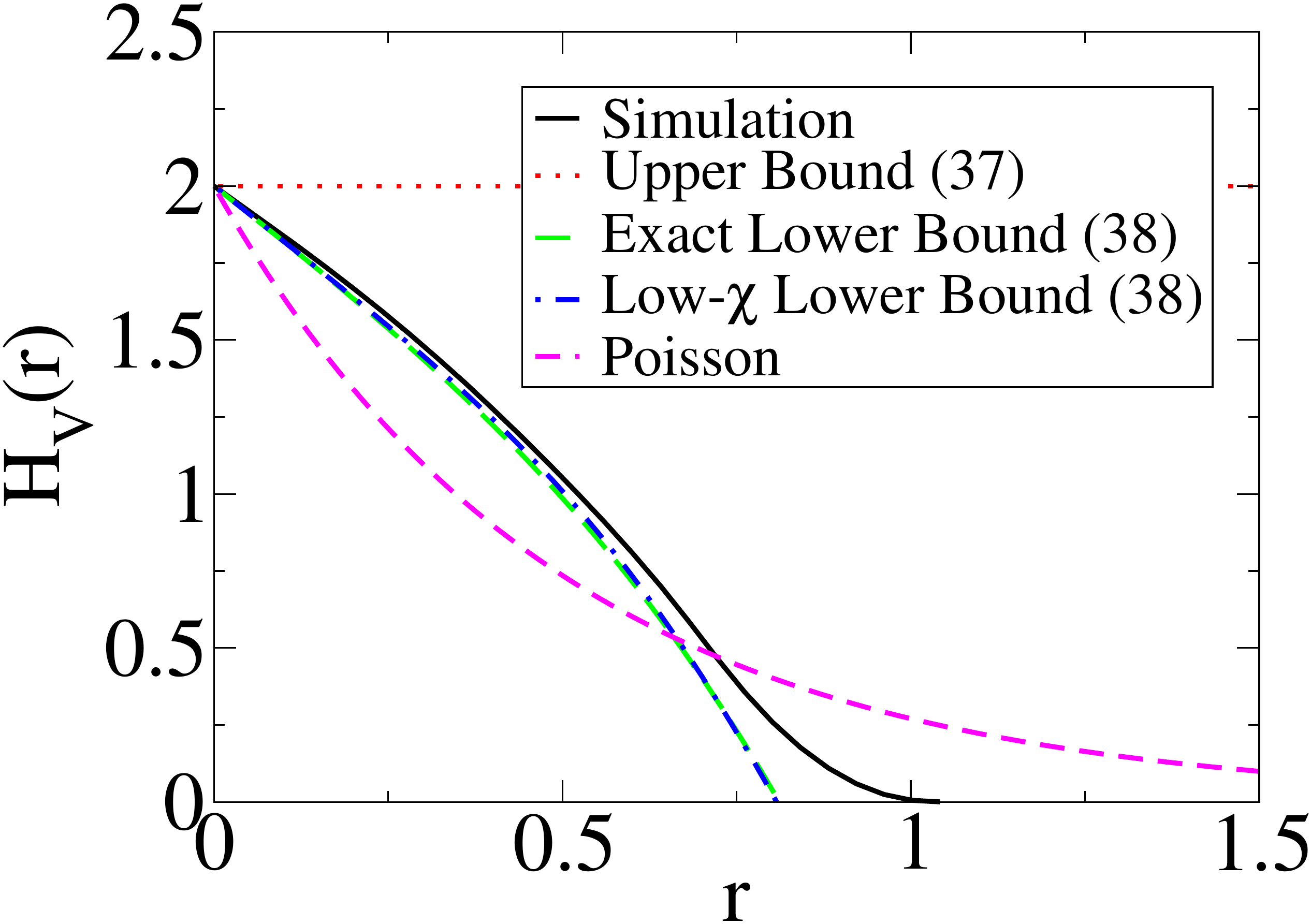}
        \caption{}
    \end{subfigure}
    \begin{subfigure}{0.3\textwidth}
        \includegraphics[width=\linewidth]{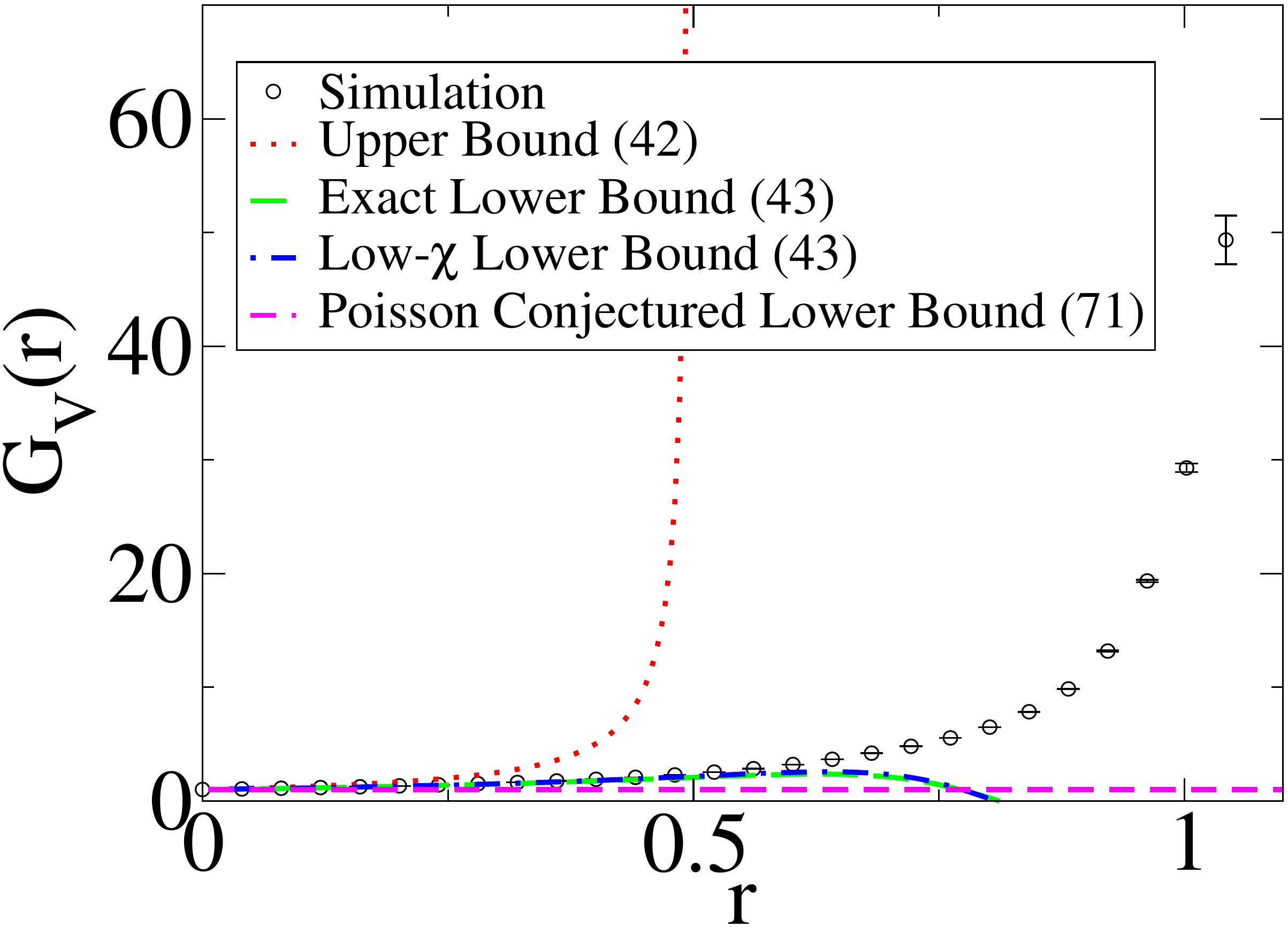}
        \caption{}
    \end{subfigure}\\
    \begin{subfigure}{0.3\textwidth}
        \includegraphics[width=\linewidth]{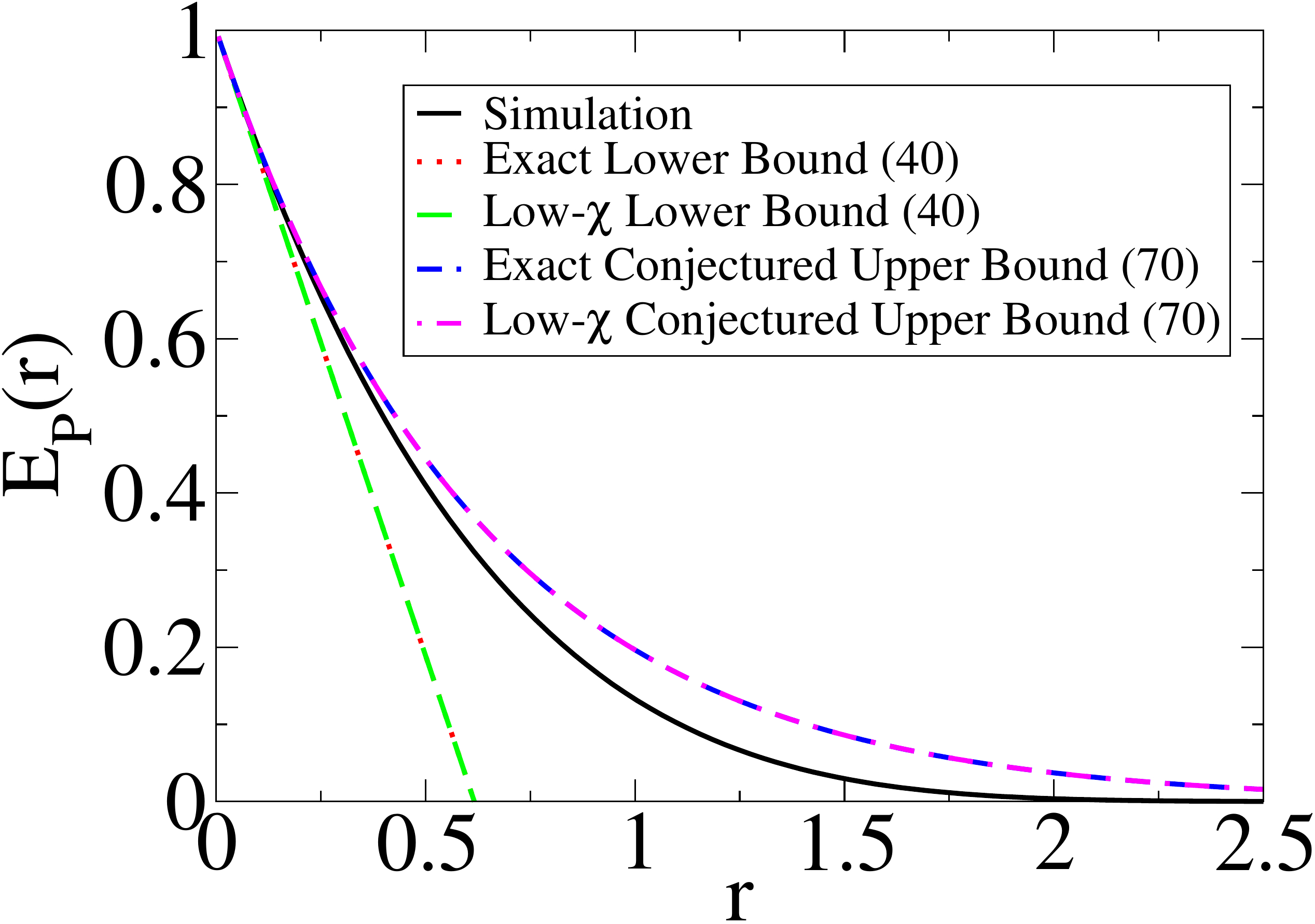}
        \caption{}
    \end{subfigure}
    \begin{subfigure}{0.3\textwidth}
        \includegraphics[width=\linewidth]{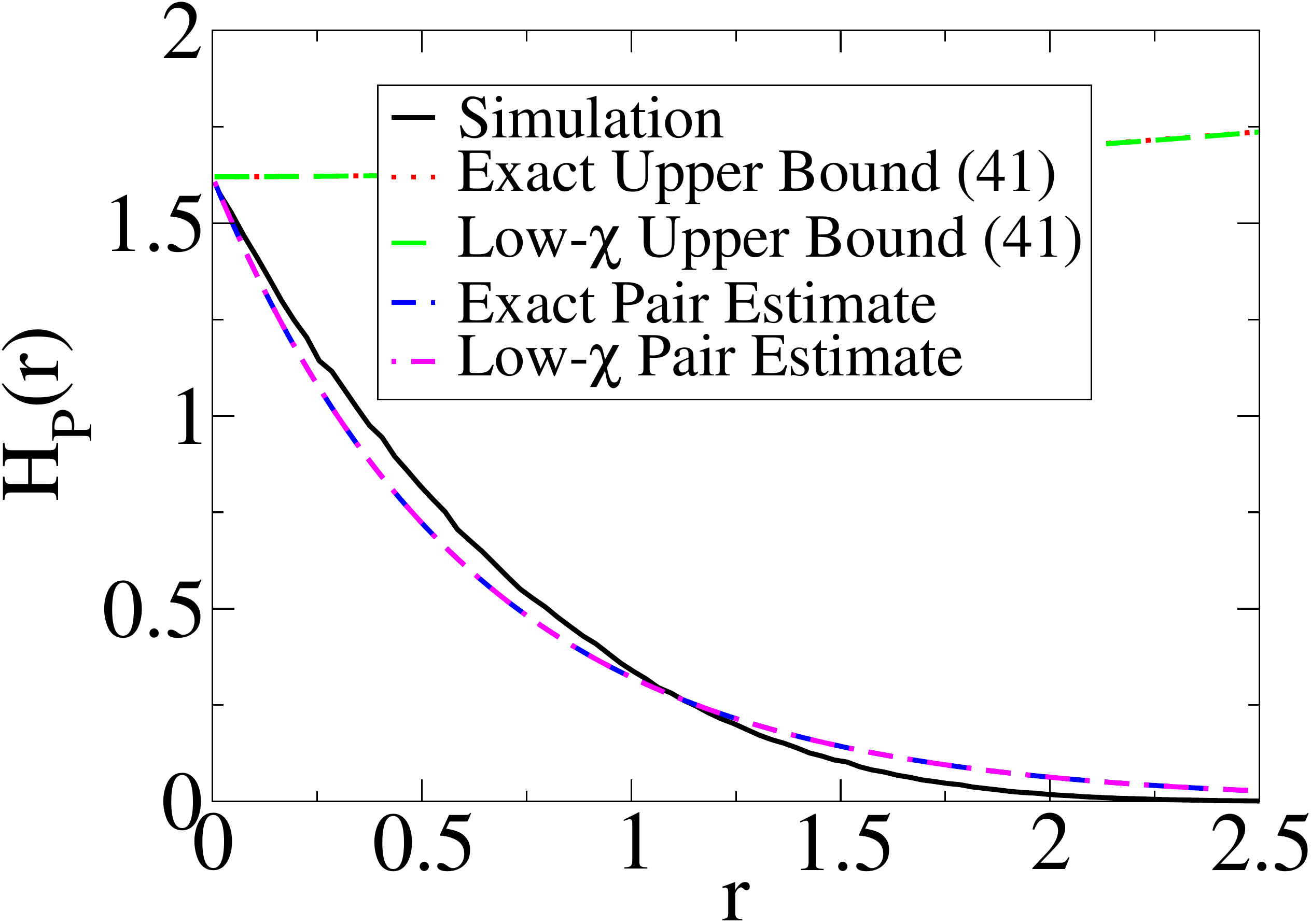}
        \caption{}
    \end{subfigure}
    \begin{subfigure}{0.3\textwidth}
        \includegraphics[width=\linewidth]{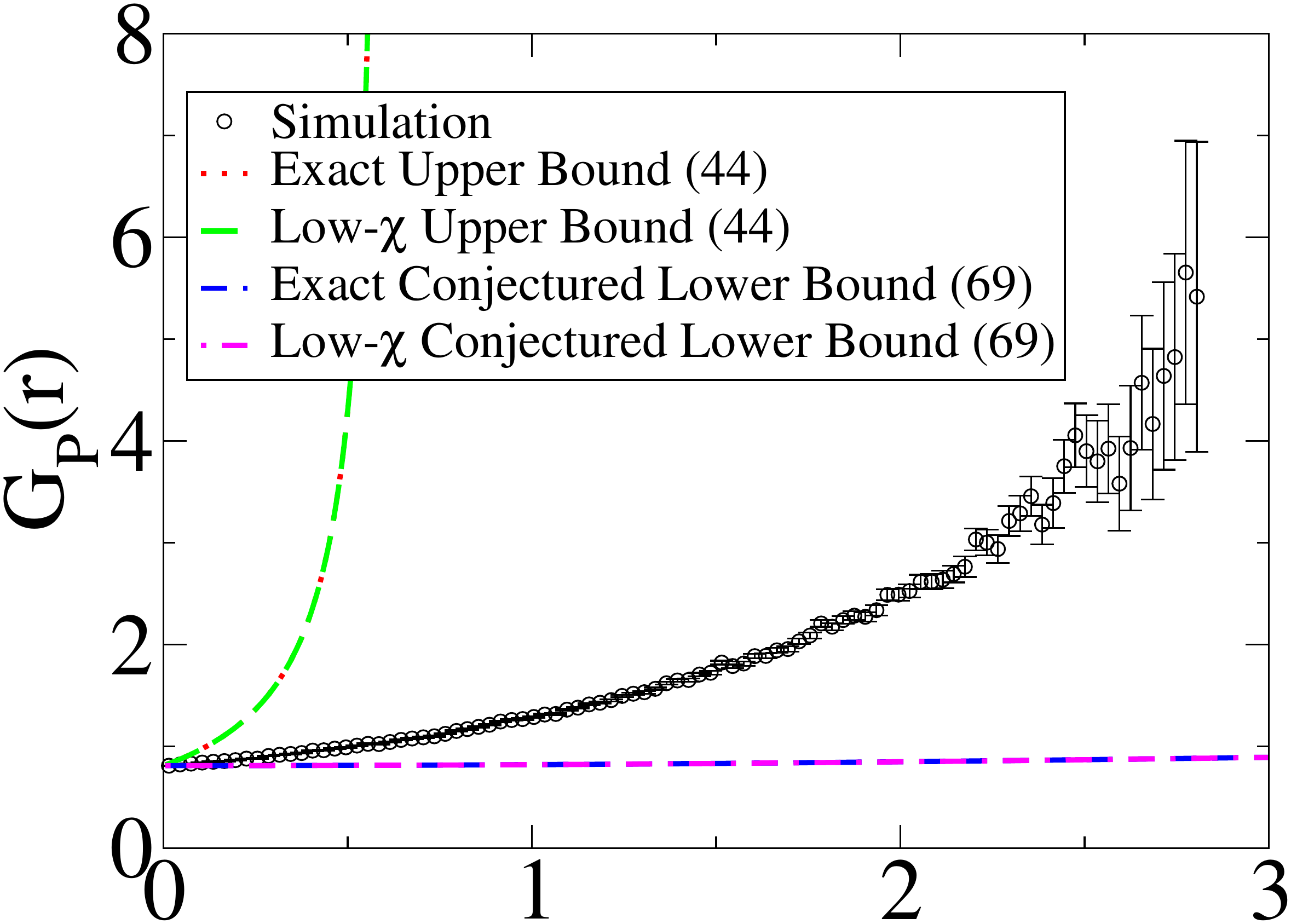}
        \caption{}
    \end{subfigure}\\
    \begin{subfigure}{0.3\textwidth}
        \includegraphics[width=\linewidth]{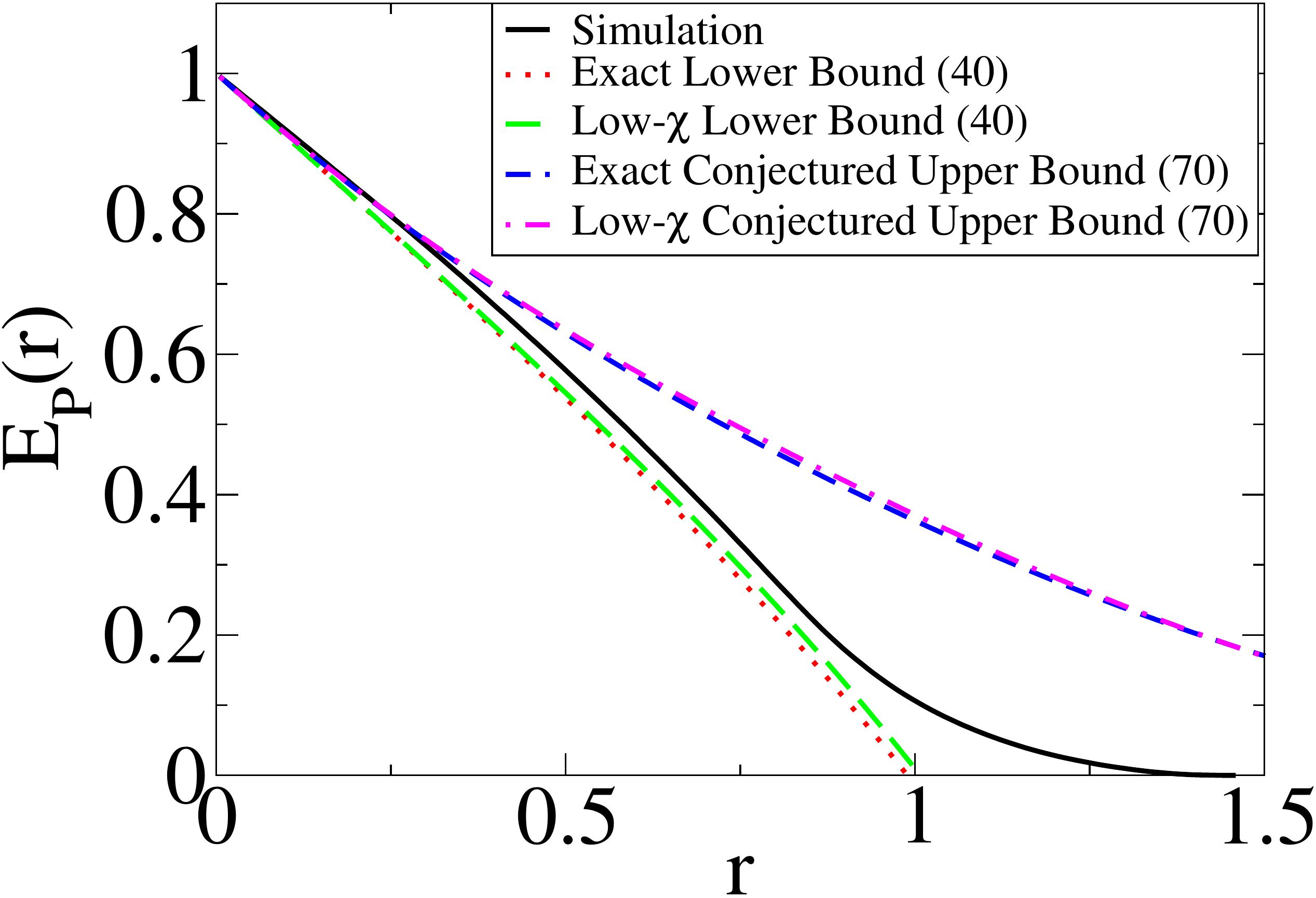}
        \caption{}
    \end{subfigure}
    \begin{subfigure}{0.3\textwidth}
        \includegraphics[width=\linewidth]{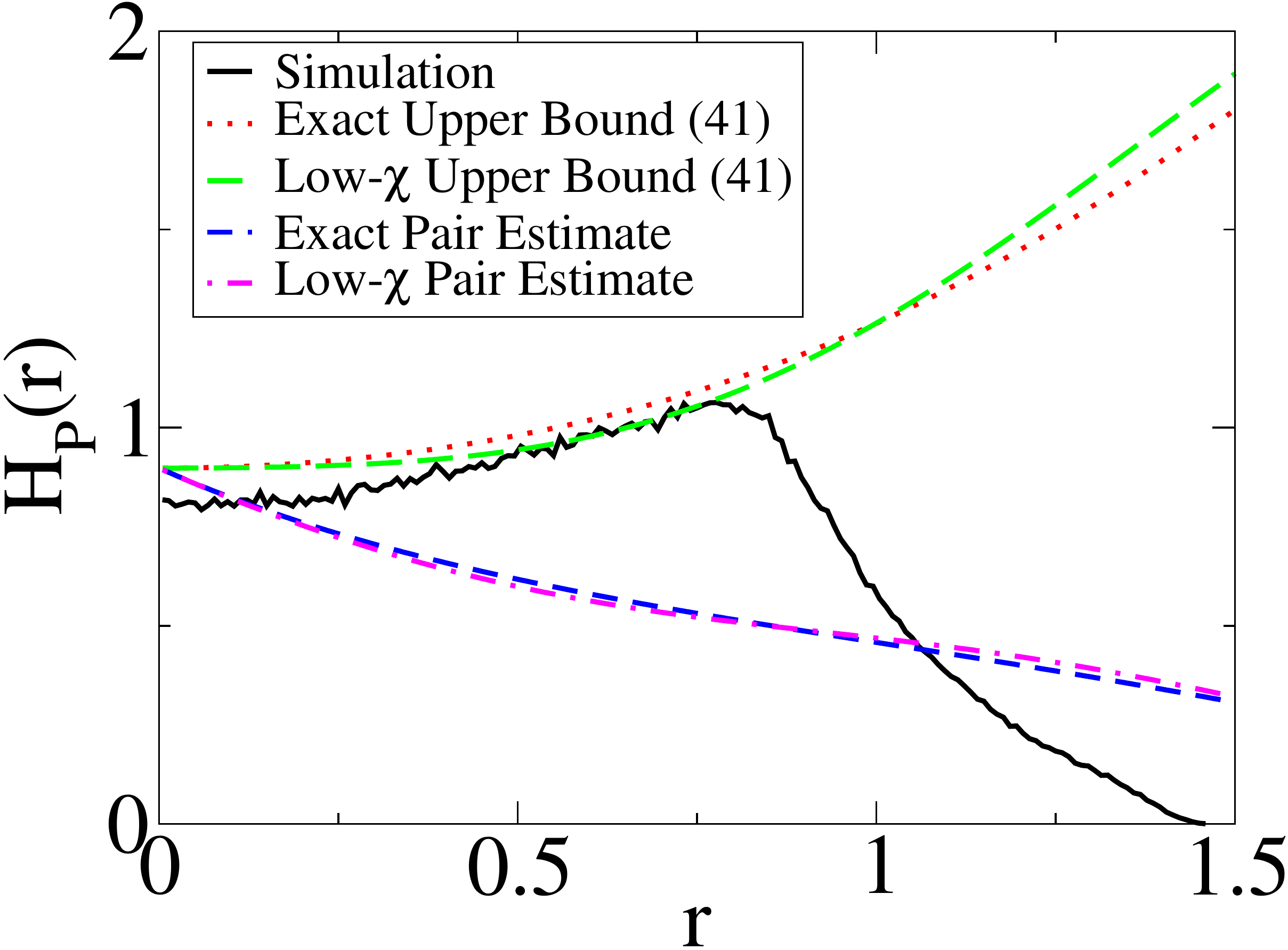}
        \caption{}
    \end{subfigure}
    \begin{subfigure}{0.3\textwidth}
        \includegraphics[width=\linewidth]{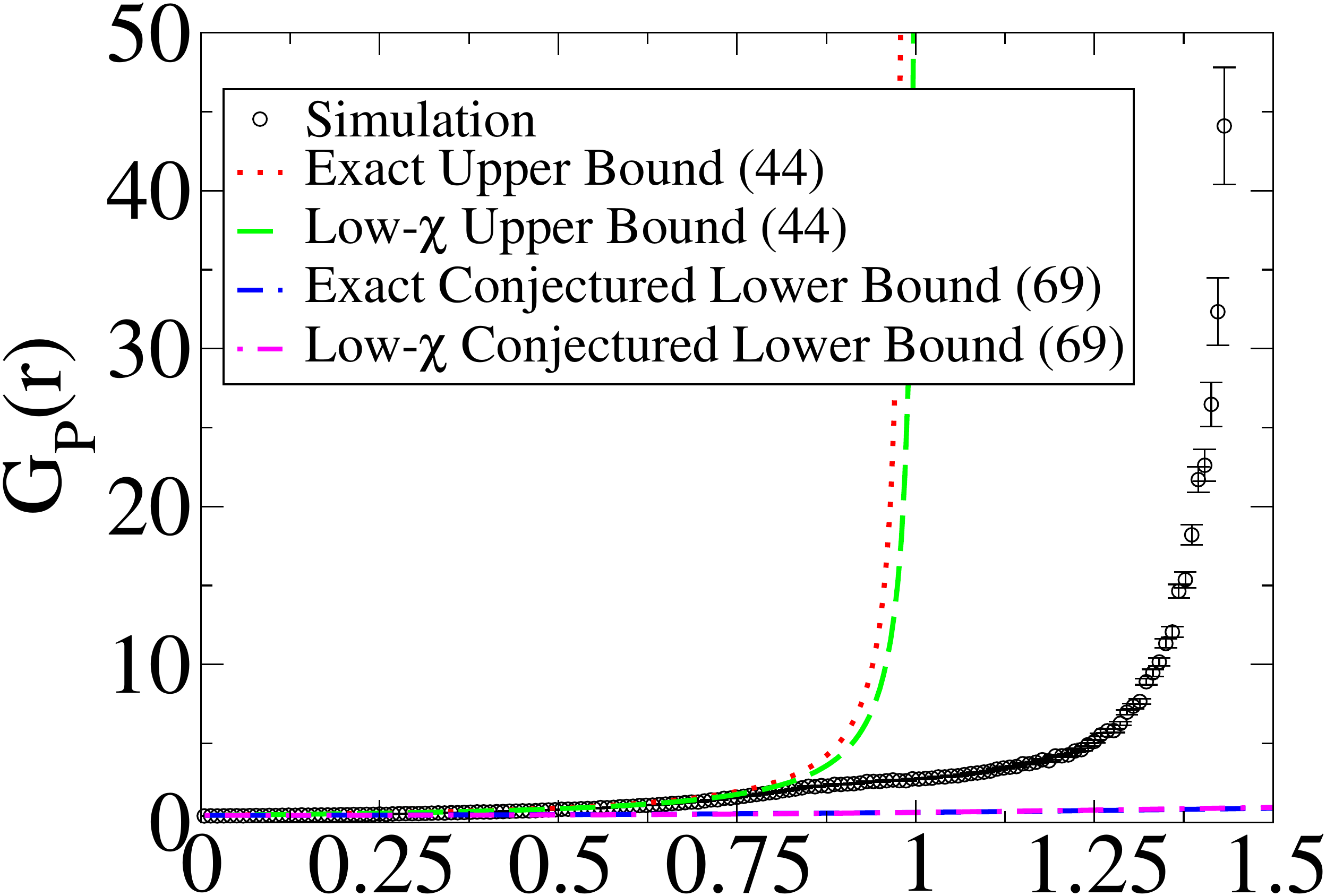}
        \caption{}
    \end{subfigure}\\
    \end{center}
    \caption{Bounds and approximations of the
    nearest-neighbor functions based on the pseudo-hard-sphere approximation
    for 1D stealthy systems. (a-c) Void
    functions for a system at $\chi=0.10$, which is within the applicability
    of the pseudo-hard-sphere approximation. (d-f) Void functions for a system at $\chi=0.33$,
    outside the applicability of the pseudo-hard-sphere approximation.
    (g-i) Particle functions for a system at $\chi=0.10$.
    (j-l) Particle functions for a system at $\chi=0.33$.}\label{1dresultsfig}
\end{figure}

In one dimension, the expression (\ref{lowchi}) can be inverted analytically
to obtain $g_2(r)$. We plot the results of numerically integrating this $g_2(r)$
with the bounds given in Section \ref{prelim} in Fig.
\ref{1dresultsfig}. We see that this approximation does quite
well at low $\chi$.

Furthermore, we can derive an analytical expression for the $g_2(r)$ of a
stealthy system using exact results for hard spheres. We take the well-known
exact solution for the direct correlation function of a hard-rod
system and interpret it as the Fourier transform of the direct correlation
function for the stealthy system \cite{torquato_ensemble_2015}:
\begin{equation}
    \tilde{C}(k) = -\Theta(K-k) \frac{1 - \chi k /K}{(1-\eta)^2}.
\end{equation}.

We used this expression in Eq. (\ref{OSeq}), and analytically took the Fourier
inversion. The resulting expression for $g_2(r)$ was used to evaluate the
bounds in Section \ref{prelim} through numerical quadrature. In
Fig. \ref{1dresultsfig}, we verify that these expressions form upper bounds at
low $\chi$ as expected. In addition, they remain useful approximations for the
void quantities even when the pseudo-hard-sphere approximation for $g_2(r)$
breaks down at intermediate $\chi$. However, in the case of the 
particle quantities, the break down of the pseudo-hard-sphere
approximation creates significant inaccuracies at intermediate $\chi$.

We can use the same approximation for $g_2(r)$ to obtain the low-$r$
series for the nearest-neighbor functions. We
obtain the coefficient $a$:
\begin{equation}
    a =1- 2\chi + \chi^2.
\end{equation}

For reference, we also show two tentative upper bounds, yet to be proven
rigorously, even in the pseudo-hard-sphere approximation. Note that combining
Eqs. (\ref{evhv}) and (\ref{evhvgv}) gives
\cite{torquato_nearest-neighbor_1990}
\begin{equation}
    E_{V/P}(r) = \exp\left( -\rho s_1(1) \int_0^r x^{d-1} G_{V/P}(x)\,\rmd x\right).
\end{equation}
Now, we make a conjecture based on observations from simulations (see the
Supplementary Material \cite{t_middlemas_supplemental_nodate} for details), that
\begin{equation}
    G_{P}(r) \geq g_2(r),\label{gconjecturedbound}
\end{equation}
yielding the putative upper bound
\begin{equation}
    E_{P}(r) \leq e^{-Z(r)}.\label{upper3}
\end{equation}
While this bound was previously presented in the context of stealthy systems
in Ref. \cite{torquato_ensemble_2015}, it was not actually proved there.
It is noteworthy that the bound is not generally obeyed by any isotropic,
homogeneous and ergodic point processes. Thus, its proof must involve some
nontrivial feature of the stealthy process, such as its propensity
to cluster to a lesser degree than a Poisson process (this can be through the
observation that $g_2(0) < 1$ for $\chi > 0$). Comparison of the
relation (\ref{upper3}) to data in Fig.
\ref{1dresultsfig} reveals that it indeed appears to form an upper
bound, as long as the pseudo-hard-sphere approximation
is applicable. We can also conjecture that bounds that apply rigorously to
fermionic point processes will also be valid for stealthy systems \cite{torquato_point_2008}:
\begin{eqnarray}
    G_V(r) \geq 1,\label{poissongv}\\
    E_V(r) \leq e^{-\rho v_1(r)}\label{poissonev}
\end{eqnarray}
Once again, comparison to simulations suggests that this is indeed the case (Fig.
\ref{1dresultsfig}). Note that formulas for $H_{V/P}(r)$ derived
from these bounds do not bound $H_{V/P}(r)$, which can be seen in Fig. \ref{1dresultsfig}.

It would be of great interest to be able to prove these
bounds. The right-hand side of relation (\ref{poissonev}) has a simple physical
interpretation, which is the void exclusion probability of a Poisson point
process \cite{hertz_uber_1909}. Thus, our conjecture is that the void exclusion
probability of a stealthy point process is bounded above by that of a Poisson
point process, aligning with physical intuition that these processes do not
tolerate large holes despite their disorder.

\subsection{2D Results}

\begin{figure}
    \begin{center}
    \begin{subfigure}{0.4\textwidth}
        \includegraphics[width=\linewidth]{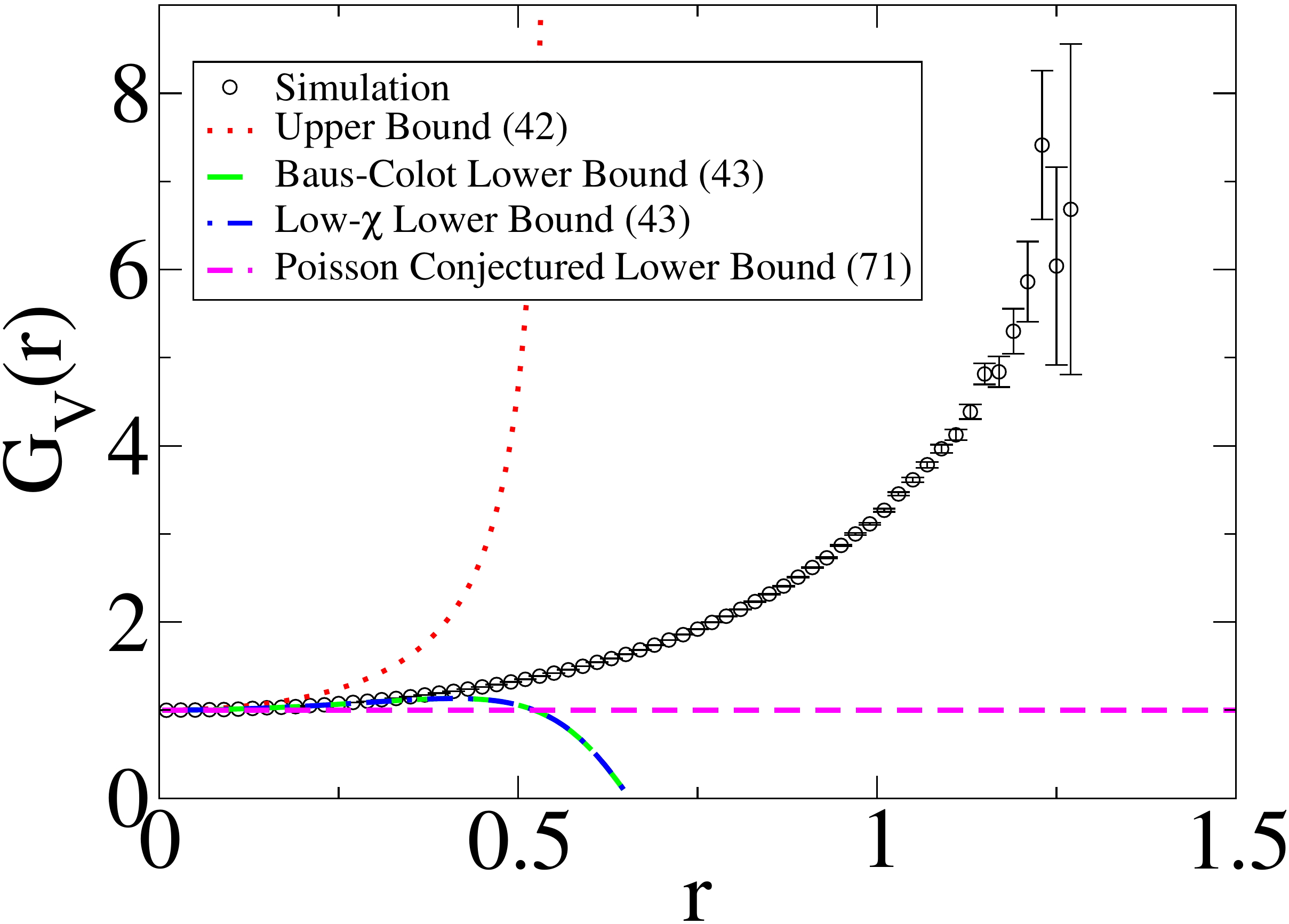}
        \caption{}
    \end{subfigure}
    \begin{subfigure}{0.4\textwidth}
        \includegraphics[width=\linewidth]{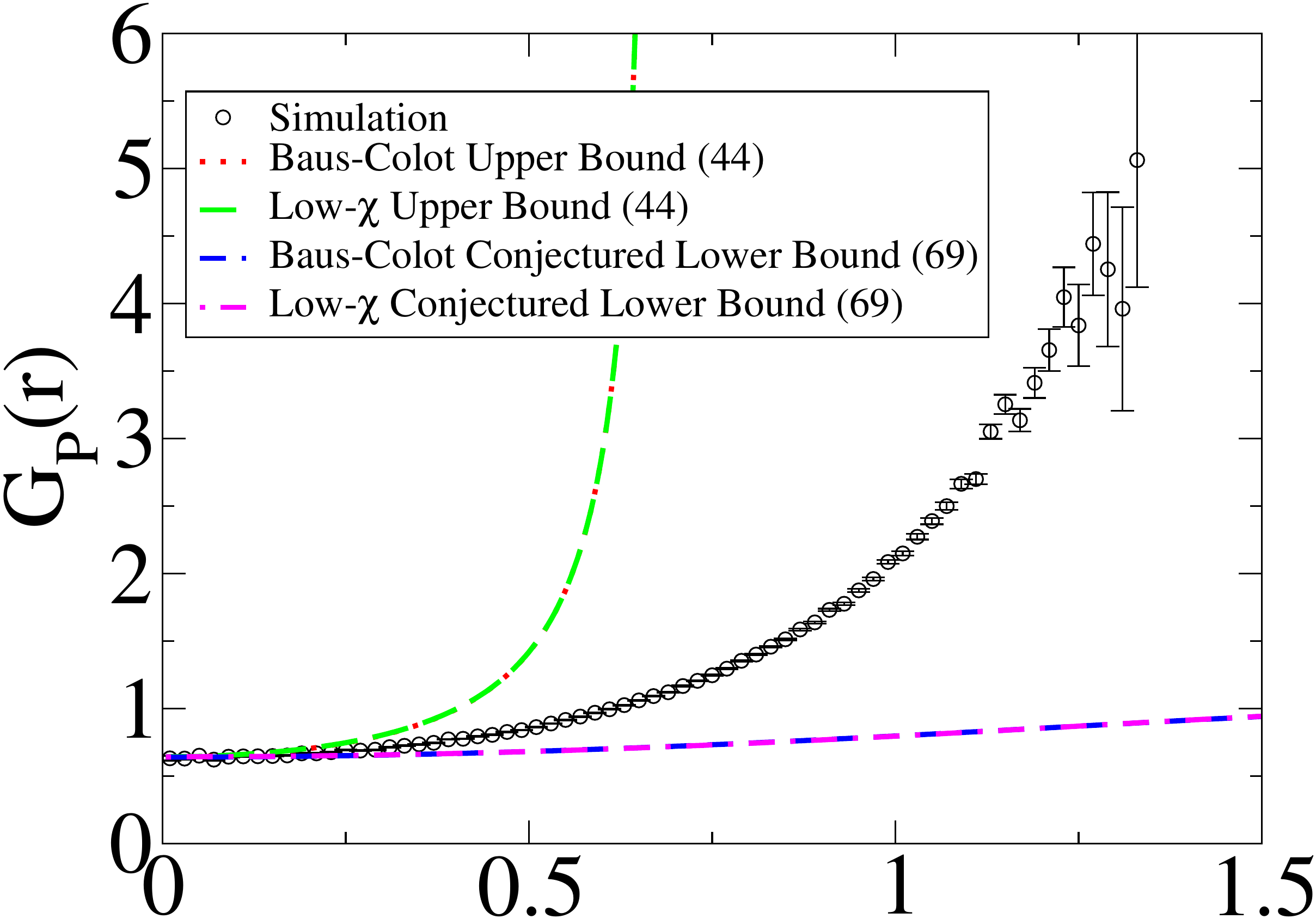}
        \caption{}
    \end{subfigure}\\
    \begin{subfigure}{0.4\textwidth}
        \includegraphics[width=\linewidth]{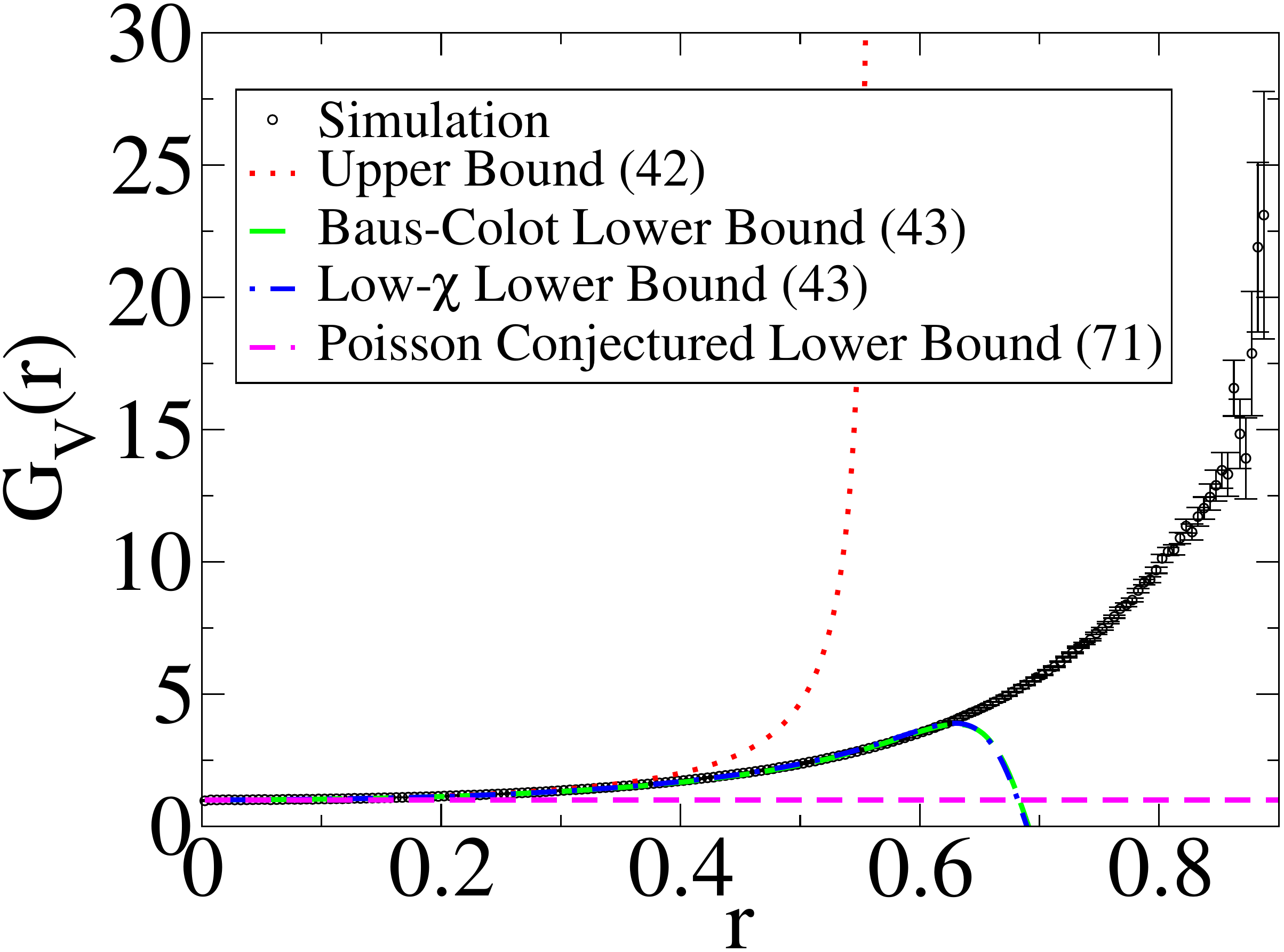}
        \caption{}
    \end{subfigure}
    \begin{subfigure}{0.4\textwidth}
        \includegraphics[width=\linewidth]{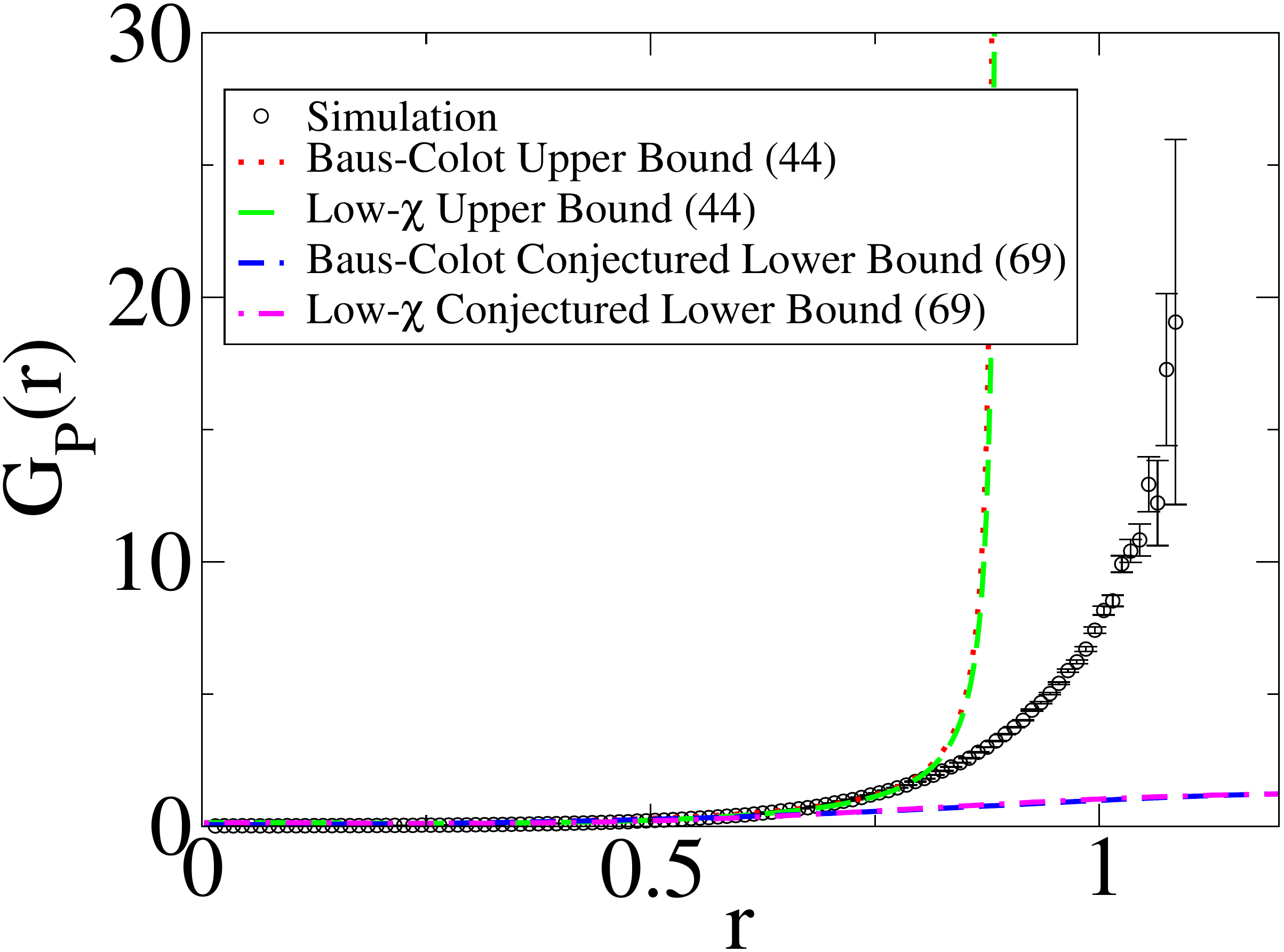}
        \caption{}
    \end{subfigure}
    \end{center}
    \caption{Bounds and approximations for the nearest-neighbor functions based on the
    pseudo-hard sphere approximation for 2D stealthy systems. (a-b) $G_V(r)$
    for a system at $\chi=0.10$ and 0.33, respectively. (c-d)
    $G_P(r)$ for a system at $\chi=0.10$ and $0.33$, respectively.}
    \label{2dresults}
\end{figure}

In two dimensions, it is generally harder to obtain results in the
pseudo-hard-sphere approximation due to the lack of exact hard-disk results. We can
still use the low-$\chi$ expansion given in Eq. (\ref{lowchi}), but we must
numerically invert to obtain $g_2(r)$. We plot the
result of using this numerical $g_2(r)$ to evaluate the bounds in Fig. \ref{2dresults}.

An accurate expression for the direct correlation function of 2D circular hard
disks is given by Baus and Colot \cite{baus_thermodynamics_1987}. They begin
with the low-density expansion for the direct correlation function, and make the
ansatz that it describes the direct correlation function for all fluid densities
so long that one uses the appropriate scaling factor. The relevant result is
that we can take the Fourier transform of the direct correlation function of the
stealthy system as
\begin{equation}
    \tilde{C}(k) = -\frac{\partial}{\partial \eta} \left[ \eta z(\eta) \right]
    \left( 1 - u^2\eta + u^2 \eta \alpha\left(\frac{k}{u},
    \frac{K}{2}\right)\right),\label{BCeq}
\end{equation}
where $z(\eta) = p v_1(K/2) /\eta k_B T$ is the compressibility factor
of the corresponding hard-disk system and $u$ is determined by the
transcendental equation
\begin{equation}
    \fl
        \frac{2}{\pi}\left( u^2(u^2-4) \sin^{-1}\left(\frac{1}{u}\right) -
        (u^2+2)\sqrt{u^2 - 1}\right) = \frac{1}{\eta^2} \left( 1- 4\eta -
        \left[\frac{\partial}{\partial \eta} \eta z(\eta)\right]^{-1} \right).
\end{equation}
To complete this description, one must specify the compressibility factor
$z(\eta)$. We use the following second-order expression
from Refs.
\cite{baus_thermodynamics_1987, torquato_random_2002}:
\begin{equation}
    z(\eta) = \frac{1+ \frac{7\pi - 12 \sqrt{3}}{3\pi} \eta^2}{(1-\eta)^2},
\end{equation}
which is accurate over the relevant hard disk packing fractions.
Then, we combine Eqs. (\ref{OSeq}) and (\ref{BCeq}) and take the Fourier
inversion numerically to determine $g_2(r)$ for our stealthy systems. We plot the result of
using this $g_2(r)$ in the approximations given in Section \ref{prelim} in Fig. \ref{2dresults}.
We see that the qualitative picture is similar to the 1D case.

We can use the preceding approximation for $g_2(r)$ to obtain the $a$
coefficient used in the low-$r$ series for the nearest-neighbor functions.
Taking $K=1$, we find that
\begin{equation}
    a =1-\frac{{\cal U}\left(\eta \left(\left(7 \pi -12 \sqrt{3}\right) (\eta-3)
            \eta-3 \pi \right)-3 \pi \right)}{24 \pi ^3 (\eta-1)^3 },
\end{equation}
where
\begin{equation}
    \fl
    {\cal U} = \frac{\left(u^2+2\right)
            \left(-\sqrt{u^2-1}\right) \eta+\left(u^2-4\right) u^2 \eta \csc
            ^{-1}(u)+2 \pi}{\frac{2 \eta
            \left(\eta \left(\left(7 \pi -12 \sqrt{3}\right) (\eta-3) \eta-3 \pi
            \right)-3 \pi \right) \rho \left(\left(u^2+2\right)
            \left(-\sqrt{u^2-1}\right) \eta+\left(u^2-4\right) u^2 \eta \csc
            ^{-1}(u)+2 \pi \right)}{3 \pi ^2 (\eta-1)^3}+\rho}.
\end{equation}

\subsection{3D Results}

\begin{figure}
    \begin{center}
    \begin{subfigure}{0.4\textwidth}
        \includegraphics[width=\linewidth]{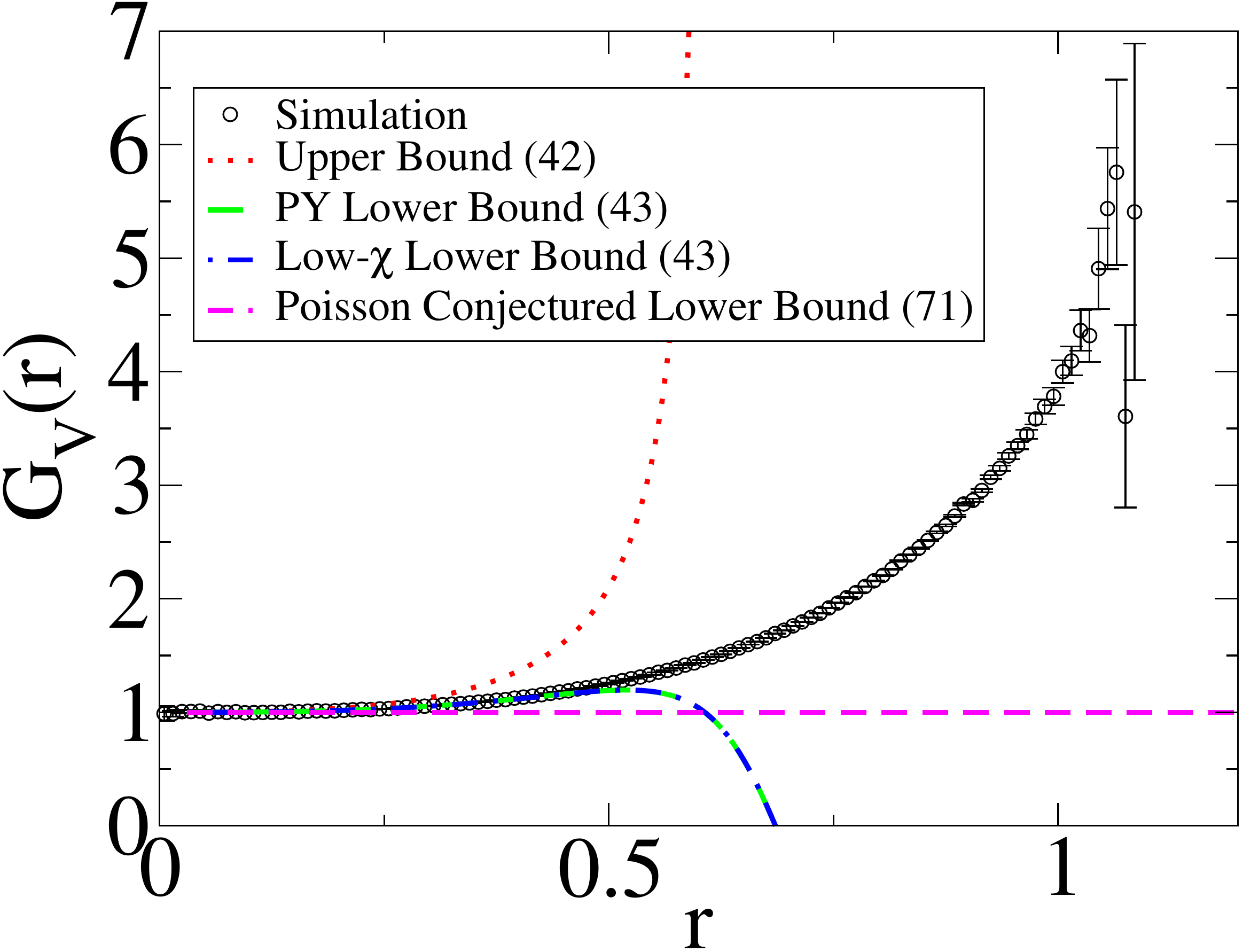}
        \caption{}
    \end{subfigure}
    \begin{subfigure}{0.4\textwidth}
        \includegraphics[width=\linewidth]{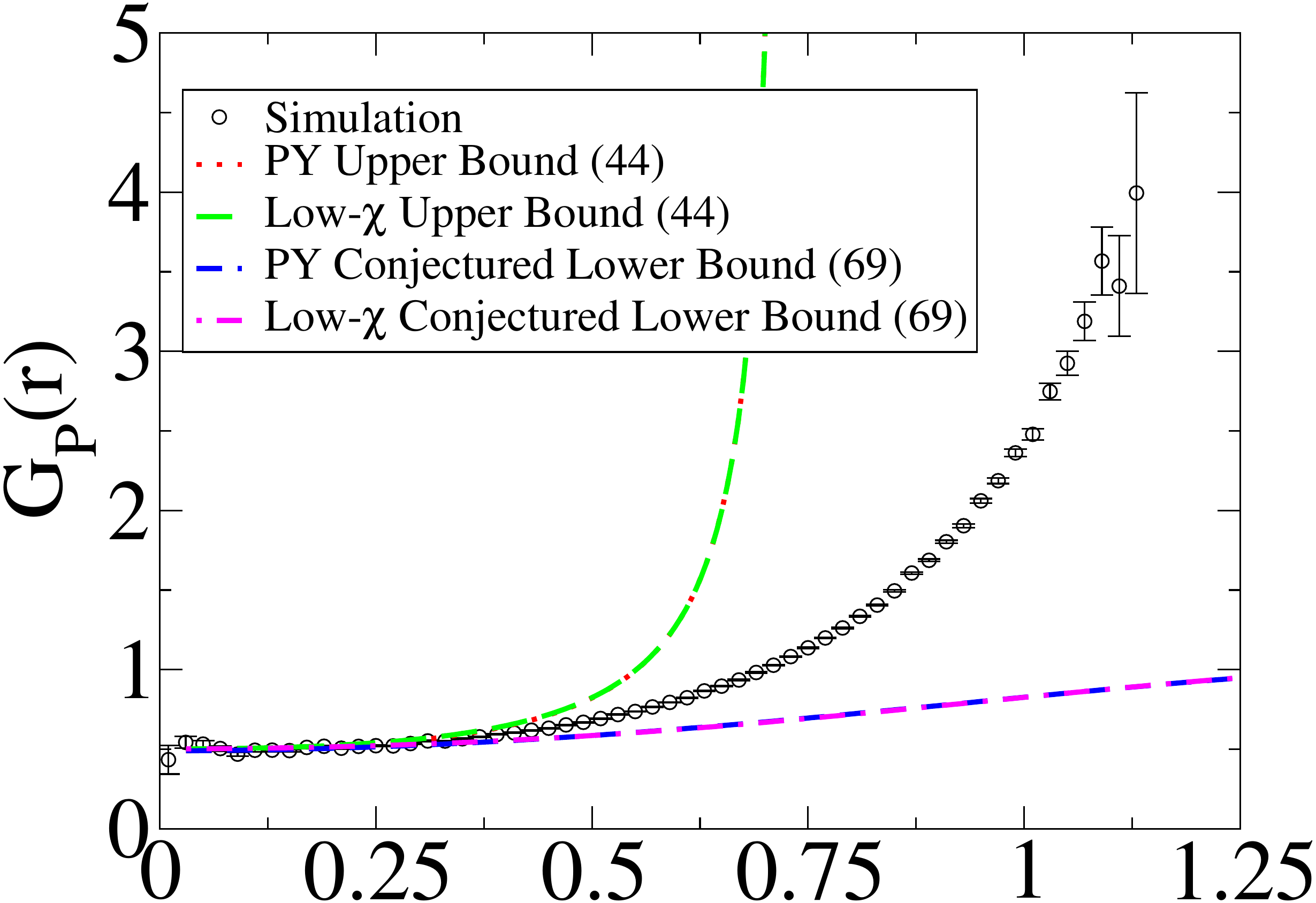}
        \caption{}
    \end{subfigure}\\
    \begin{subfigure}{0.4\textwidth}
        \includegraphics[width=\linewidth]{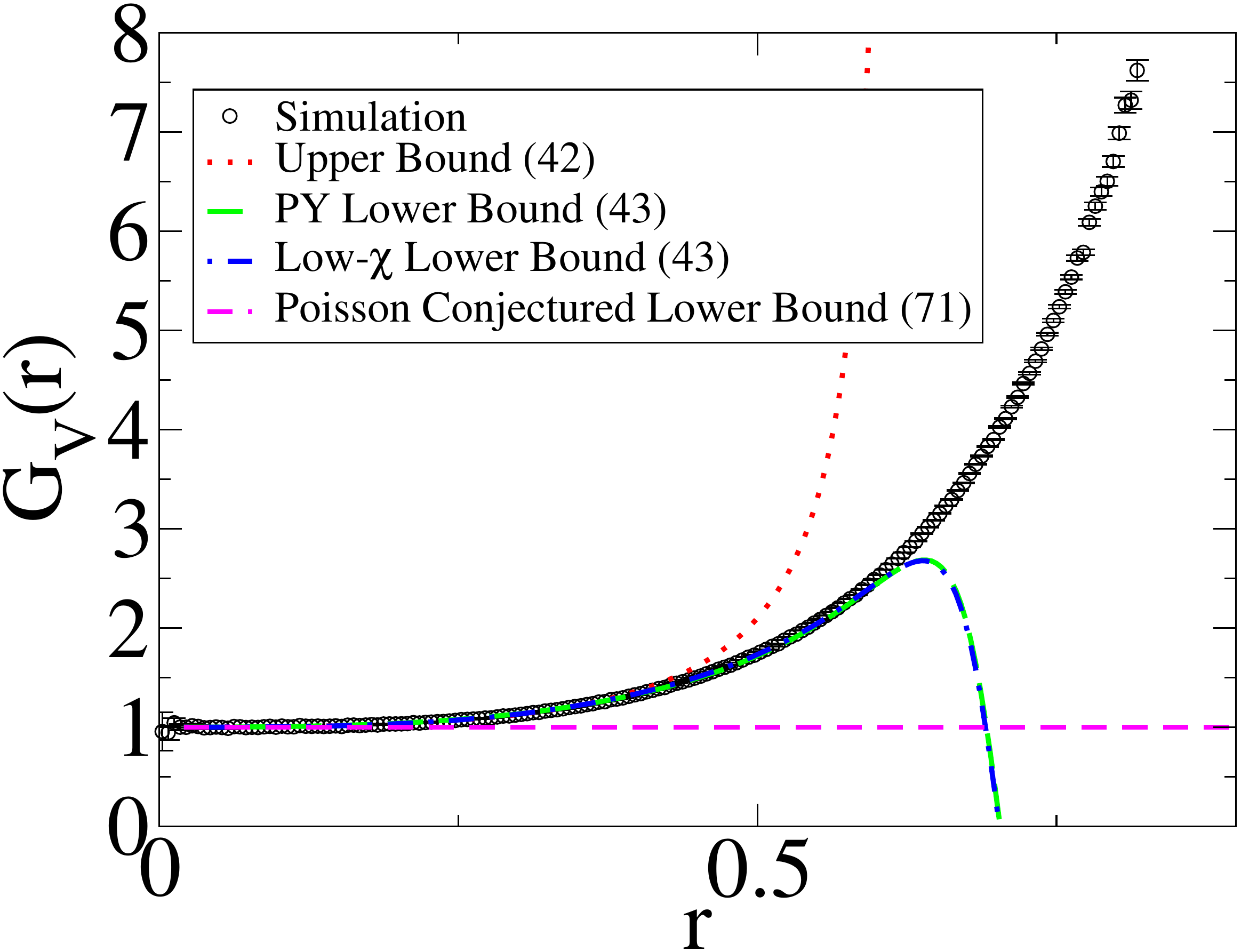}
        \caption{}
    \end{subfigure}
    \begin{subfigure}{0.4\textwidth}
        \includegraphics[width=\linewidth]{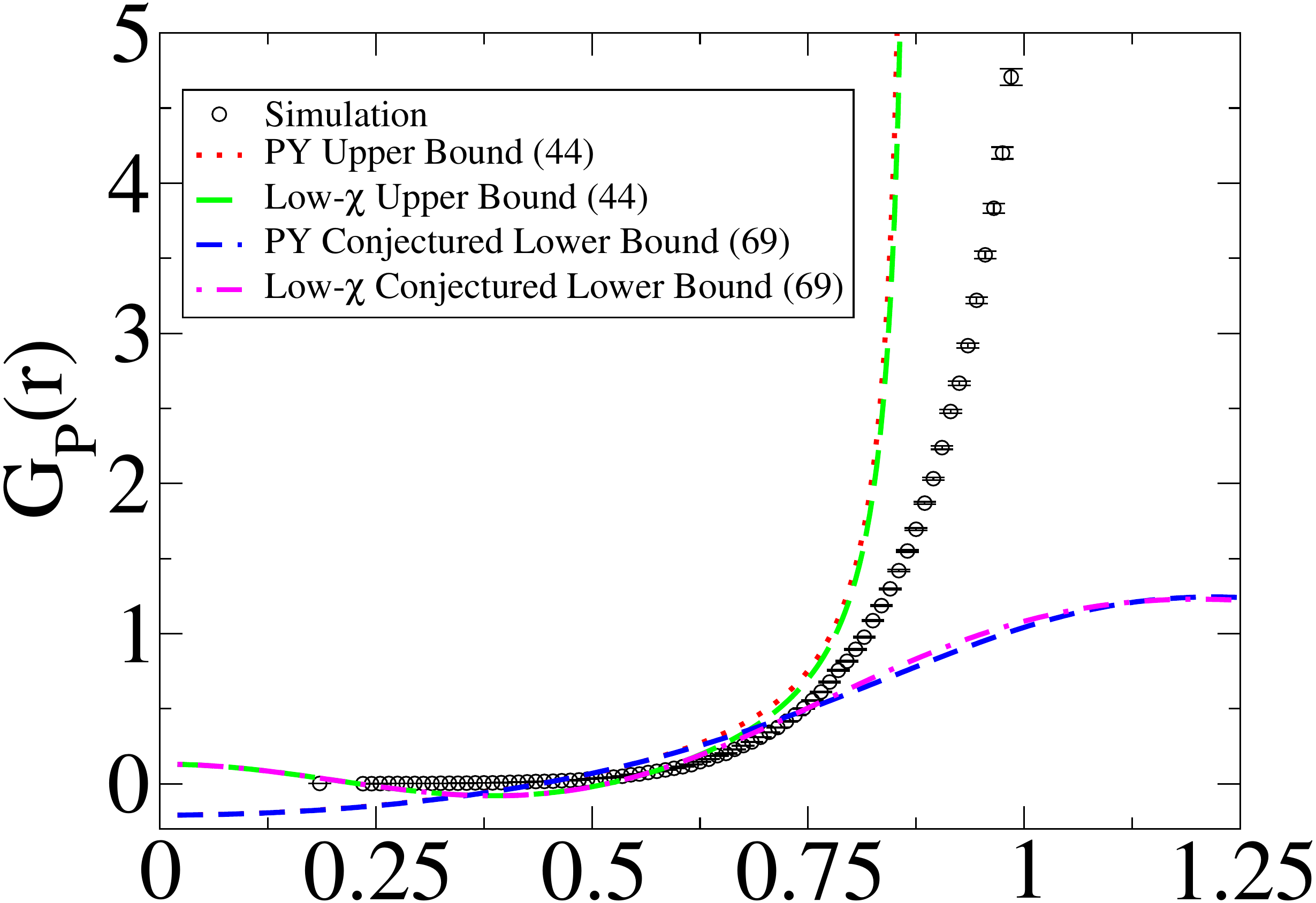}
        \caption{}
    \end{subfigure}
    \end{center}
    \caption{
    Bounds and approximations to the nearest-neigbor functions based on the pseudo-hard-sphere approximation
    for 3D stealthy ensembles. (a) Void functions
    for $\chi = 0.10$. (b) Void functions
    for $\chi = 0.33$. (c) Particle functions for
    $\chi = 0.10$. (d) Particle functions
    for $\chi = 0.33$.}\label{3dresults}
\end{figure}

Since we do not have an exact expression for $g_2(r)$ in 3D, we once again begin
with the low-$\chi$ expansion for $S(k)$ in Eq. (\ref{lowchi}).  We compute the
Fourier inverse of this equation analytically, and use it to
evaluate the bounds of Section \ref{prelim} numerically. This is compared with simulation data in
Fig. \ref{3dresults}.

A more accurate approximation is obtained using the Percus-Yevick approximation for
$g_2^{HS}(r)$. This gives \cite{jean_pierre_hansen_theory_1986}
\begin{equation}
    \tilde{C}(k) \approx \Theta(K-k) \left( -a_1 - 6 \eta a_2 \frac{k}{K} -
    \frac{\eta a_1}{2} \left(\frac{k}{K}\right)^3\right),
\end{equation}
where $a_1 = (1+ 2\eta)^2/(1-\eta)^4$ and $a_2 = -(1+\eta/2)^2/(1-\eta)^4$. We
follow the same steps as for the 1D case with an exact $g_2(r)$, and compare the
results to simulation in Fig \ref{3dresults}. We find qualitatively similar trends
to the 1D and 2D cases.

Using a similar argument to the 1D case, we find that the low-$r$ expansion
using the PY approximation to the pseudo-hard-sphere scheme \cite{jean_pierre_hansen_theory_1986},
taking $K=1$, is, given in terms of the coefficients for Eq. (\ref{3dlowr}):
\begin{equation}
    a = \frac{625 - 2750\chi + 775\chi^2 -300\chi^3 + 30\chi^4}{25(5+4\chi)^2},\label{3dA}
\end{equation}
and
\begin{equation}
        b = \frac{3(200\chi - 55\chi^2 + 8\chi^3)(625 - 1000\chi + 600\chi^2
        -160\chi^3 +16\chi^4)}{1000(5+4\chi)^4}.
\end{equation}

\subsection{Extension to Larger $\chi$}

\begin{figure}[h!]
    \centering
    \begin{subfigure}{0.33\textwidth}
        \includegraphics[width=\linewidth]{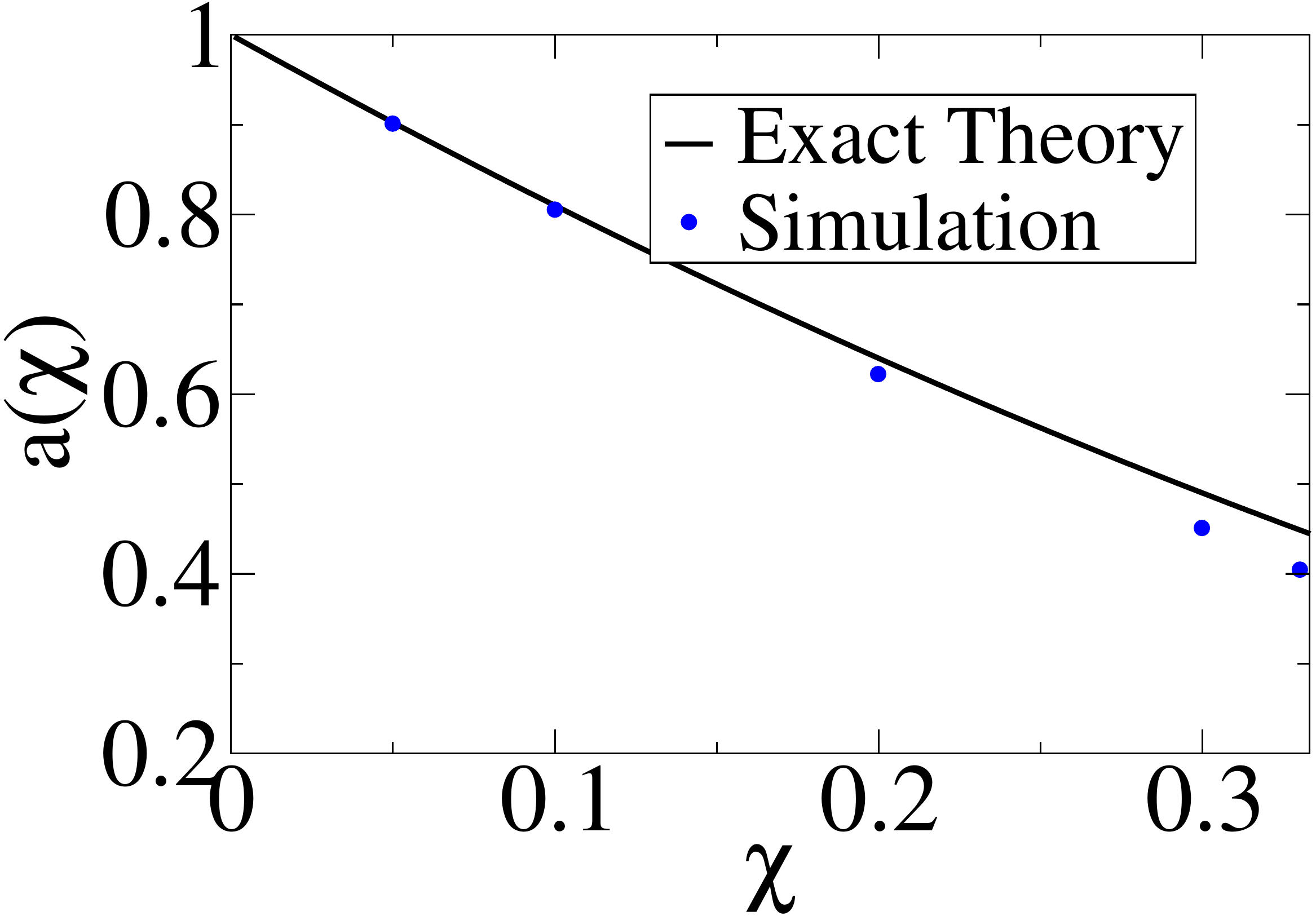}
        \caption{}
    \end{subfigure}%
    \begin{subfigure}{0.33\textwidth}
        \includegraphics[width=\linewidth]{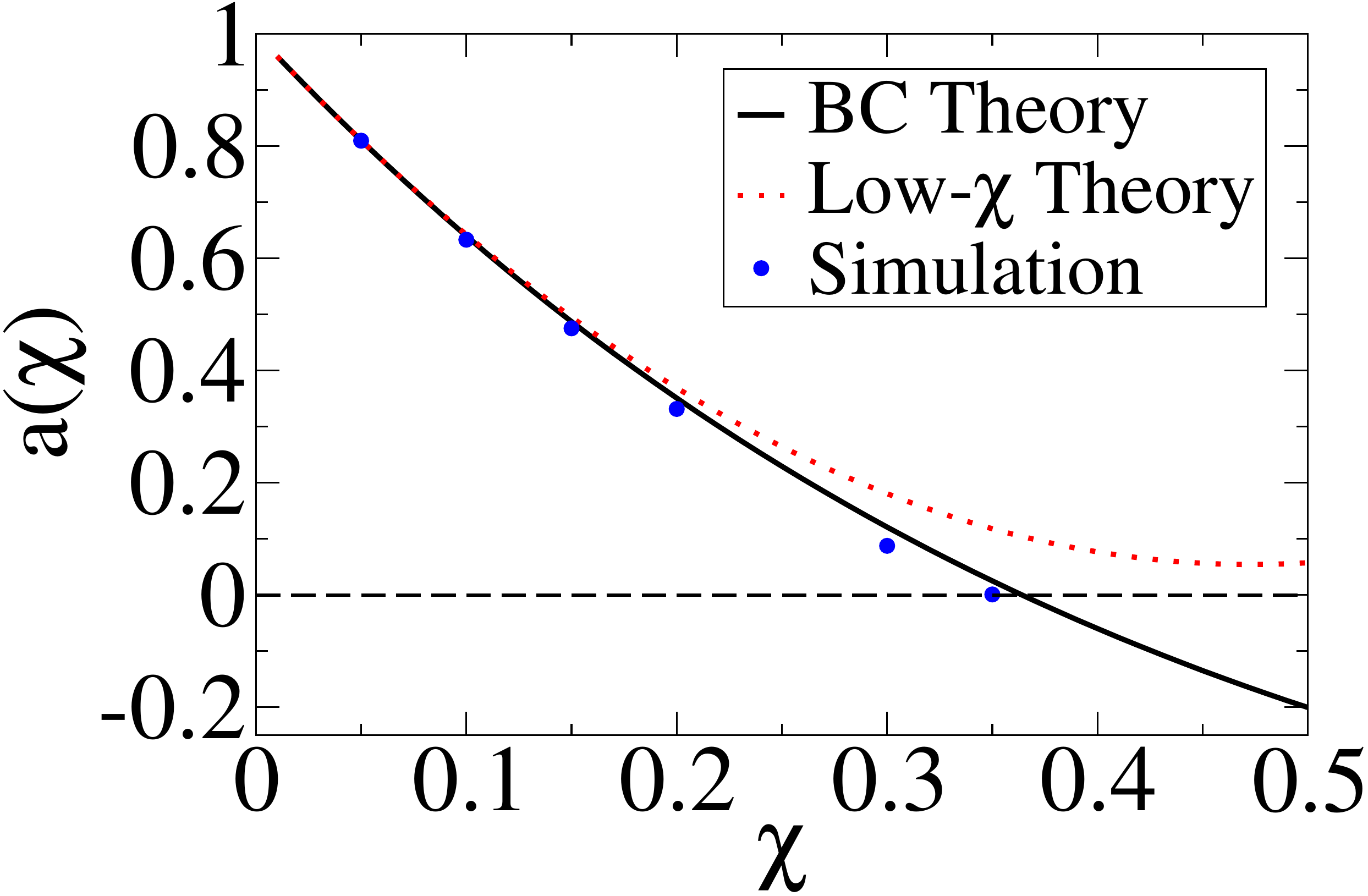}
        \caption{}
    \end{subfigure}%
    \begin{subfigure}{0.33\textwidth}
        \includegraphics[width=\linewidth]{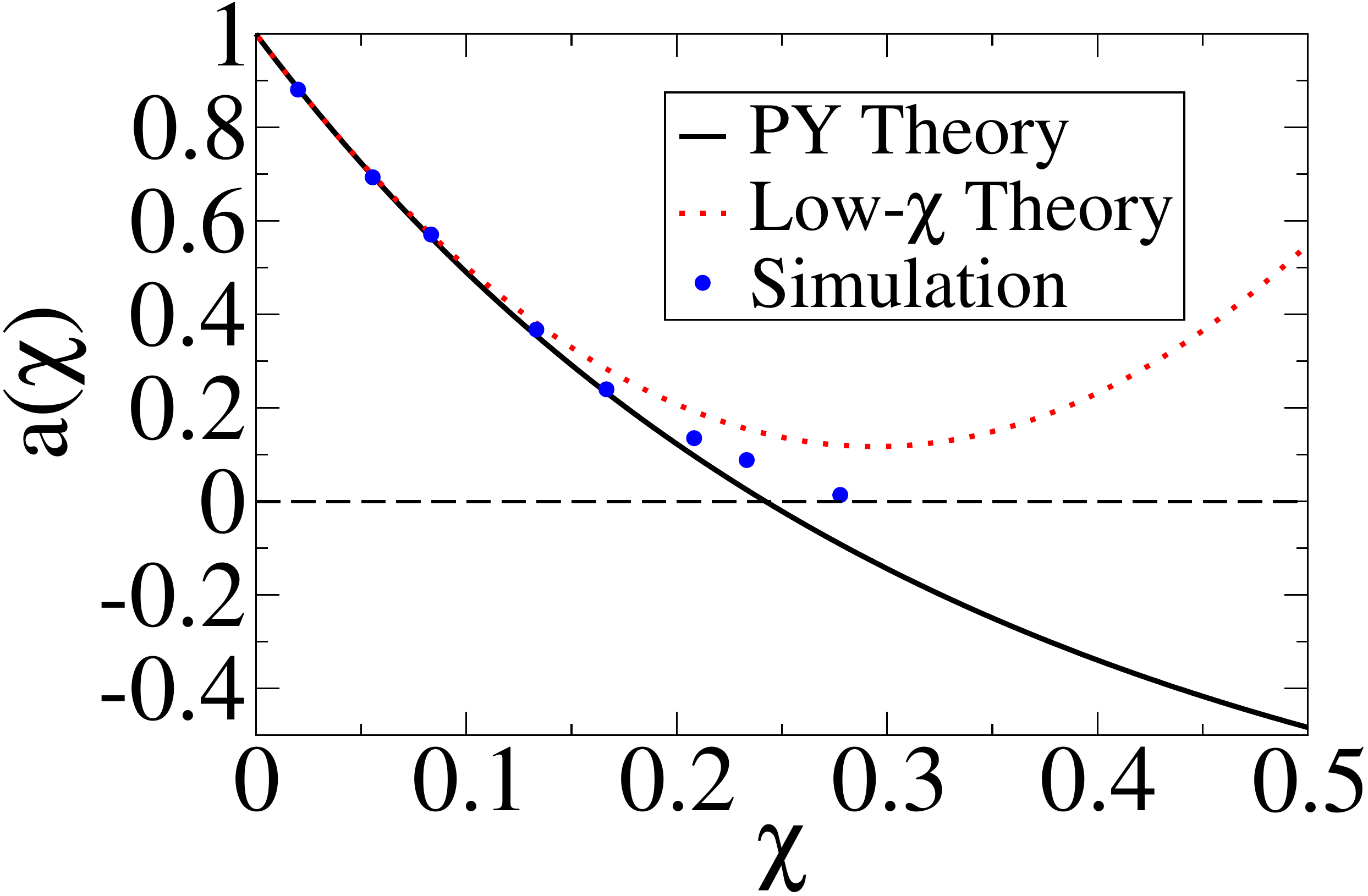}
        \caption{}
    \end{subfigure}
    \caption{A comparison of the prediction for $a(\chi)$ from
    pseudo-hard-sphere theory to values estimated from
    simulation. The value of $a$ is obtained by using a quadratic
    extrapolation on a numerically measured $g_2(r)$. Data for
    two and three dimensions were obtained from high quality pre-existing
    simulations on small systems reported in Ref. \cite{zhang_transport_2016}.
    (a) A comparison for 1D. As the low-$\chi$ approximation and exact solution
    of the pseudo-hard-sphere ansatz give the same $a$, we only show one curve.
    (b) A comparison for 2D. We include predictions based on the low-$\chi$
    approximation and the Baus-Colot approximation.  (c) A comparison for 3D.
    We include predictions based on the low-$\chi$ approximation and the
    Percus-Yevick approximation.}\label{compare_A}
\end{figure}

While the pseudo-hard sphere approximation breaks down well
before the order-disorder transition at $\chi = 0.5$, it is still possible to
derive useful results from this approximation all the way up to the transition point.
The basic observation is that while the functional form of the pair correlation function
differs from the pseudo-hard-sphere approximation above $\chi \approx 0.15$ \cite{zhang_ground_2015},
the value of $g_2(0)$, and thus the coefficient $a$ that determines the leading order
contribution to the nearest-neighbor functions, can be well modeled using a simple extension of this
theory. In Fig. \ref{compare_A}, we compare the analytical results for
$a(\chi)$ given in the preceding sections to simulation data. For the 1D case, the
pseudo-hard-sphere result becomes steadily worse as $\chi$ increases, but, as we will see
in Section \ref{domain}, this result can still be used to form a useful theory of the
void functions.
For the case of 2 and 3 dimensions, we see that the
analytical prediction for $a(\chi)$ works very well until it crosses zero
and becomes negative. In Fig. \ref{compare_A}, we only report simulation data with a non-zero value
of $a$, as our method of obtaining $a$ relies on a quadratic extrapolation
that becomes invalid when $\chi$ becomes large. However, for our particular
collection of finite configurations at large $\chi$, we observe $g_2(r) = 0$
for an entire range of $r$ near origin. Thus, one can obtain a useful
analytical approximation by setting $a(\chi)$ through the pseudo-hard-sphere
approximation up to the zero crossing, and setting it to zero thereafter.

\section{Inclusion of Higher-order Information}\label{higherorder}

\begin{figure}
    \centering
    \includegraphics[width=0.3\linewidth]{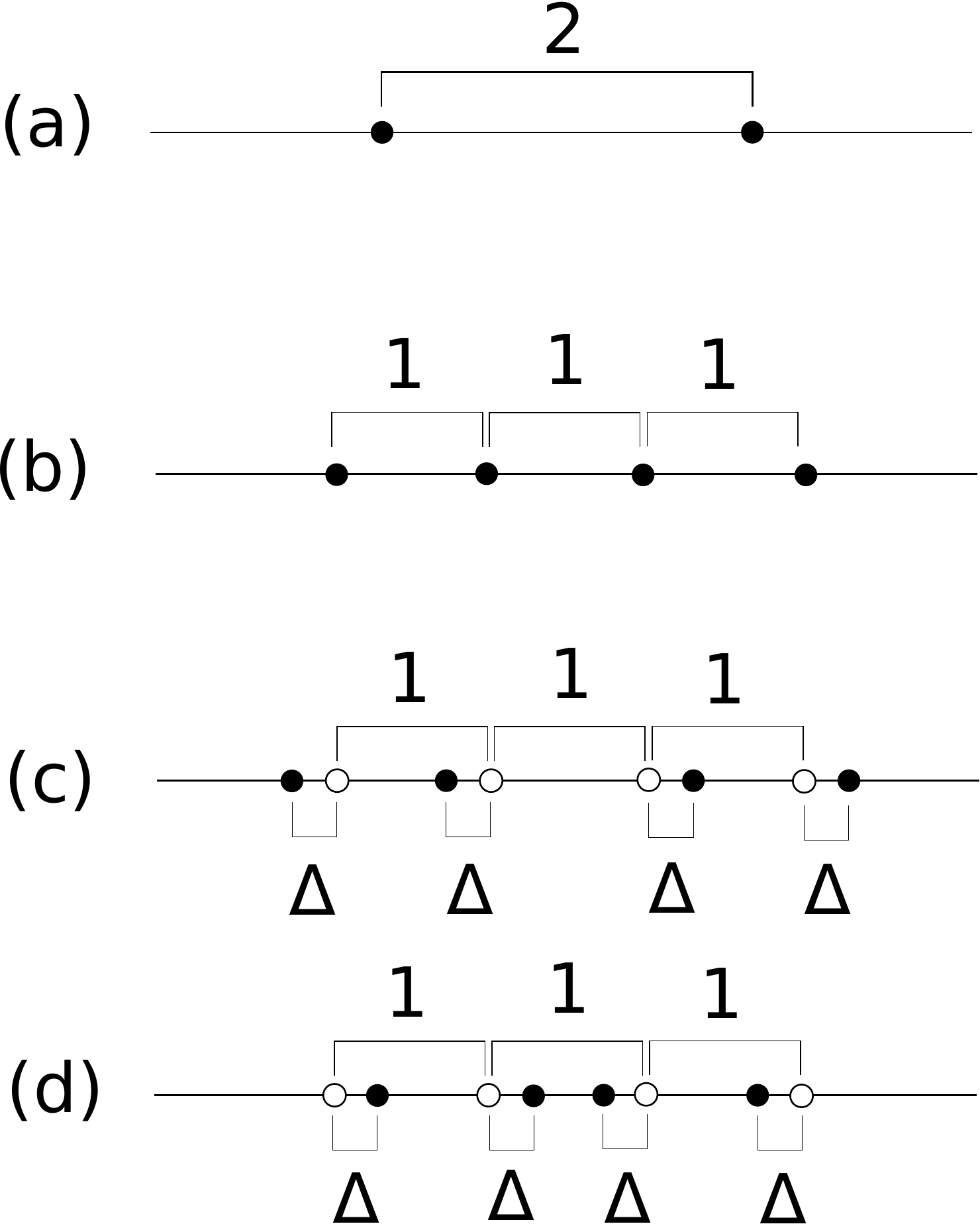}
    \caption{Clusters used to determine the highest order in the series
    (\ref{evexpansion}) needed to evaluate $E_V(r)$ for the corresponding
    point process. (a) The least dense cluster of two particles for a one-dimensional
    saturated RSA process with particles of unit diameter. (b) The densest
    cluster of four particles for the same RSA process. (c) One of the degenerate
    least dense clusters of four particles for the one-dimensional perturbed
    lattice described in the text. The open circles represent the underlying
    lattice locations, while the filled circles represent the points of the
    configuration. (d) The one of the degenerate least dense clusters of four
    particles for the same perturbed lattice.}\label{cluster_schematic}
\end{figure}

In the previous section, the results were derived using only the
one and two-point correlation functions. However, as can be seen
in series such as Eq. (\ref{evexpansion}), the nearest-neighbor
functions in principle depend on the $n$-point correlation functions
$g_n({\bi r}^n)$ up to arbitrary order in the infinite
system size limit. In this section, we will discuss
conditions under which series of this nature can be truncated using the
bounded holes property, so that $E_V(r)$ is determined by a finite number
of terms in Eq. (\ref{evexpansion}). This discussion 
also applies to single finite configurations using the series (\ref{finiteseries}),
in which case the key observation is that the series can be
truncated far before the last $v_N^{\rm int}({\bi r}^N)$ term.

The bounded holes property plays a fundamental role in the truncation of this
series. This can be seen in the following way. If one has a condition that
prevents arbitrary clustering of points, such as a requirement to be a packing
\cite{torquato_reformulation_2010}, one can show that the series in Eq. (\ref{evexpansion})
must terminate after a finite number of terms for any given value of $r$. The bounded holes
property lets us then extend this observation to show truncation of the series for $E_V(r)$
at all $r$, with an $r$-independent number of terms. Since the number of necessary terms to keep
generally grows with the value of $r$ considered, the existence of a critical-hole size $r_c$
allows us to compute the number of terms needed by finding the number of terms needed to evaluate
$E_V(r_c)$. Thus, the preceding argument shows that the series (\ref{evexpansion}) must
terminate for any packing with the bounded hole size property. This argument
has been used to show the truncation of the series
(\ref{finiteseries}) for all crystals \cite{torquato_reformulation_2010}, but
we note that it applies equally to the series (\ref{evexpansion})
for random sequential addition at saturation.

We can also show the truncation property for the case of stealthy point
processes. To do this, we use the anti-concentration property proved in Ref.
\cite{ghosh_generalized_2018}. This states that for a box of side length
$C/K$, the number of particles in the box is bounded above by $C'\rho/K^d$,
where $C$ and $C'$ are generic constants \cite{ghosh_generalized_2018}. Since
stealthy systems have bounded hole sizes, it is once again sufficient to
consider $E_V(r_c).$ Since we have the strict upper bound on the number of
particles in a large enough box, we can also bound the maximum number of
particles contained in the decorated sphere surrounding each particle in the
geometric interpretation given in Fig. \ref{small_voronoi}. Thus, the series
must terminate after a finite number of terms. Note, however, that the number of terms
that we may need to consider in the series expansion (\ref{evexpansion}) increases with
decreasing $\chi$.

This last observation raises an interesting fundamental question concerning
the number of terms of Eq. (\ref{evexpansion}) needed to describe a stealthy
system. While we are not aware of a method to solve this problem analytically
for disordered stealthy systems, we present analytical solutions for two
interesting systems with the bounded holes property: the case of one-dimensional
random sequential addition at saturation and a specific one-dimensional perturbed lattice.

For 1D random sequential addition at saturation, the
truncation property is established by the previous general principle concerning
packings. One can find that the series truncates at the $g_2(r)$ term by considering
the two local configurations of four particles (or clusters)
given in Fig. \ref{cluster_schematic}. We will take the diameter of the spheres
to be unity. The first cluster shows that $r_c = 1$ in this system, since
starting at a separation of two, one can insert another particle between the
neighbors, contradicting the saturation assertion. The second cluster
represents the densest cluster possible while respecting the packing
condition. Overlaying the covering spheres as in Fig. \ref{small_voronoi}
readily shows that one only needs to consider up to intersections of two
covering spheres, which corresponds to the $g_2(r)$ term.

The perturbed lattice we will consider is a one-dimensional lattice of unit
spacing where the points (indexed by $i$) are displaced by independent
random variables $\delta_i$ are independently drawn from an
arbitrary distribution with compact support over $[-\Delta, \Delta]$.
We further restrict $0 <\Delta < 1/2$, to prevent transposition of particles.
It is interesting to note that this system is hyperuniform
\cite{kim_effect_2018}, but not stealthy hyperuniform. One should also
be aware that this is very specific case of a perturbed lattice; in general,
one can have correlations between the $\delta_i$ or unbounded displacement
distributions \cite{kim_effect_2018}. The series (\ref{evexpansion}) truncates by virtue of this
system being a packing with a bounded hole size. One can see this by
considering Fig. \ref{cluster_schematic}. The first cluster shows that $r_c =
(1 + 2\Delta)/2$, while the second cluster shows that the system can be
considered a packing, since there is always a gap of $1- 2\Delta$
between the particles. These clusters also show that the number of terms needed
is dependent on $\Delta$. For $\Delta \in (0, 1/4]$, one only needs up through
the $g_2(r)$ term, however, for $\Delta \in (1/4, 1/2)$, one requires the
addition of the third order term.

\begin{table}
    \centering
        \begin{tabular}{|c|c|}
            \hline
            $\chi$ & Highest Order in Series (\ref{finiteseries})\\
            \hline\hline
            0.050 & 16\\
            \hline
            0.10 & 10\\
            \hline
            0.20 & 6\\
            \hline
            0.30 & 5\\
            \hline
            0.33 & 4\\
            \hline
        \end{tabular}
        \caption{A table containing the highest order necessary to evaluate
        $E_V(r)$ through the series (\ref{finiteseries}) for 1D stealthy
        systems at various $\chi$.}\label{ordertable}
    \end{table}

The number of terms needed for stealthy systems in the series (\ref{finiteseries}) can be
determined numerically.  While this is computationally expensive in two and
higher dimensions, it can be determined in an efficient manner in one dimension
by using the fact the intersection volume of $n$ 1D spheres can be written
as the intersection volume of the two spheres farthest apart in the collection.
The interested reader can refer to the Appendix for details.  We have reported
the highest order needed for our 1D stealthy systems in Table \ref{ordertable}.
We see that the number of terms needed increases with decreasing
$\chi$, as predicted from the general argument above. We expect this trend to
continue in higher dimensions.

\section{Behavior on Approach to Critical-Hole Size}\label{tail}

One of the surprising yet fundamental properties of any stealthy system is a
bounded hole size \cite{zhang_can_2017, ghosh_generalized_2018}.
This in turn implies that $E_V(r)$ and $H_V(r)$ have compact
support and that $G_V(r)$ diverges as it approaches the critical-hole size.
We investigate the asymptotic behavior of the nearest-neighbor
functions as they approach this maximum hole size, using simple theoretical
arguments and computer simulations as our primary tools.

\subsection{Fundamental Considerations}

\begin{figure*}
    \centering
    \begin{subfigure}{0.4\textwidth}
        \includegraphics[width=\linewidth]{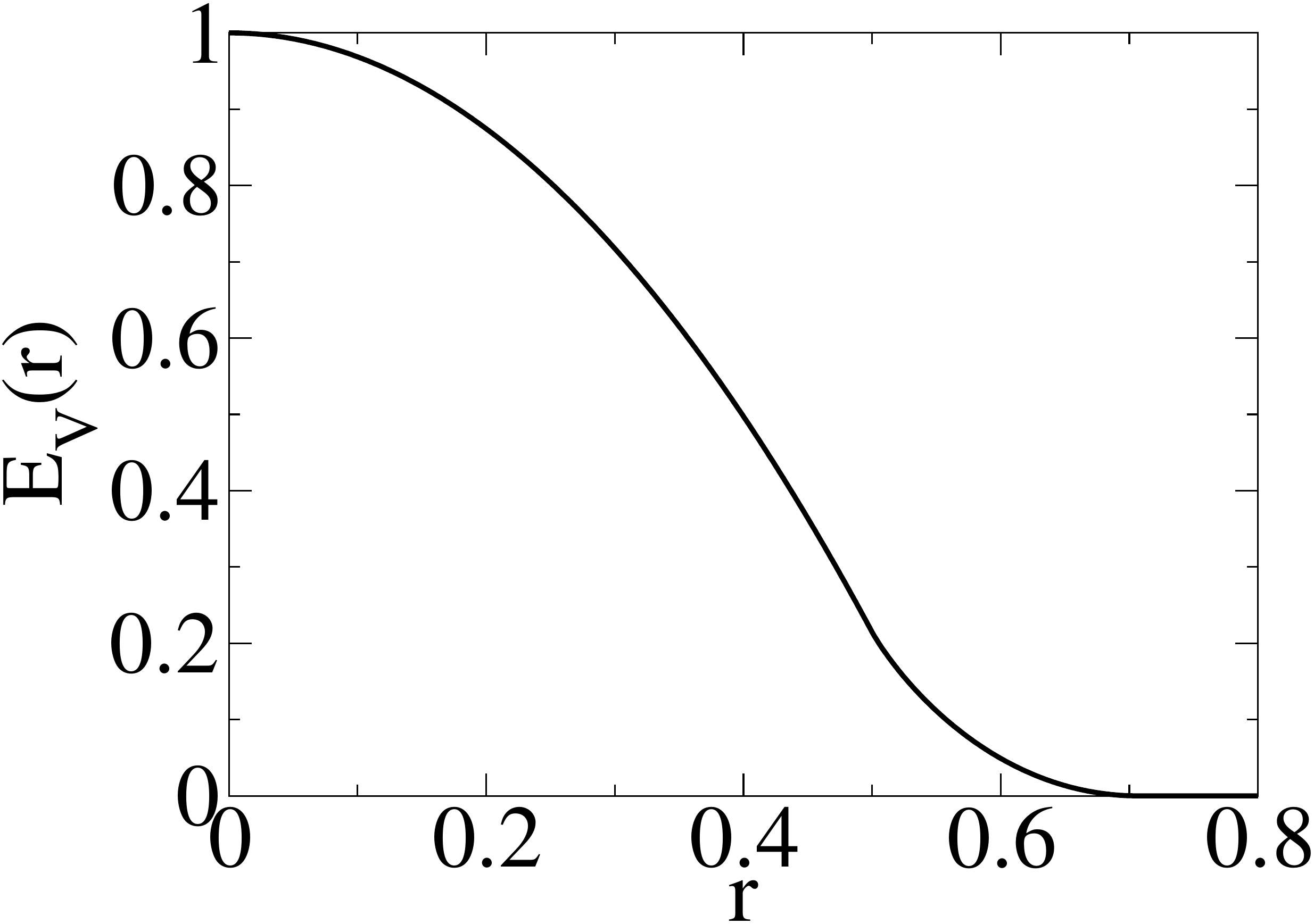}
        \caption{}
    \end{subfigure}
    \begin{subfigure}{0.4\textwidth}
        \includegraphics[width=\linewidth]{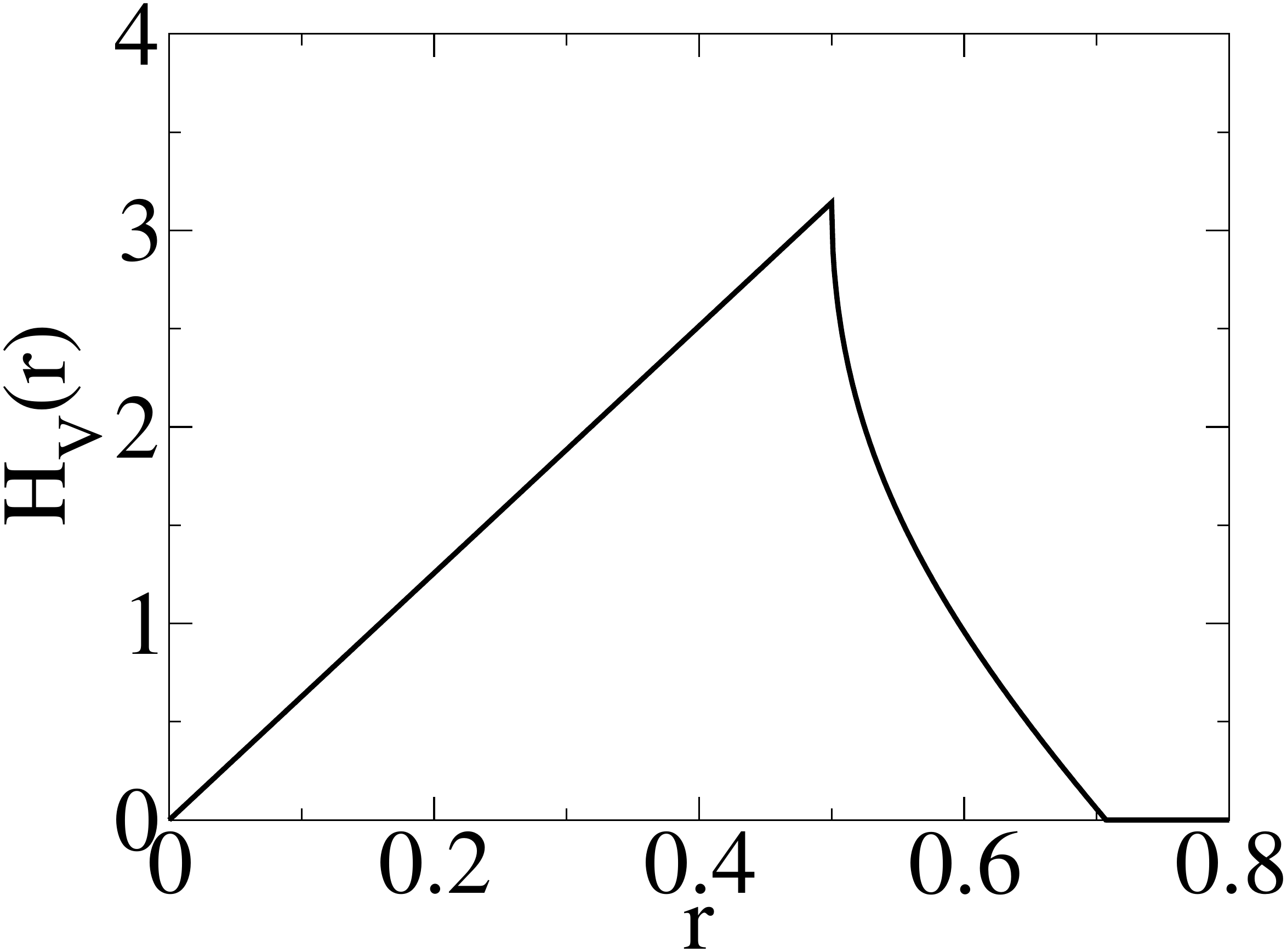}
        \caption{}
    \end{subfigure}\\
    \begin{subfigure}{0.4\textwidth}
        \includegraphics[width=\linewidth]{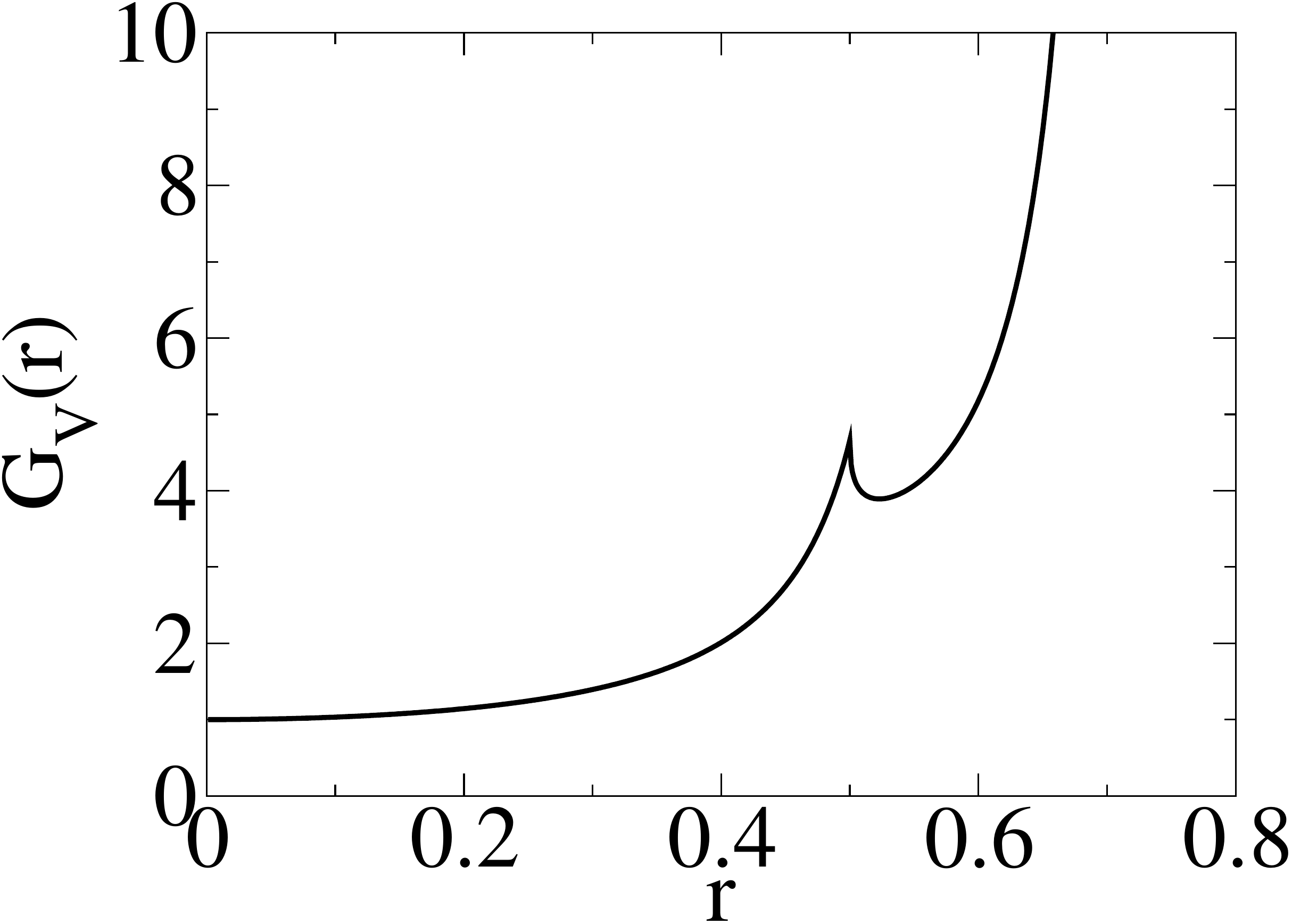}
        \caption{}
    \end{subfigure}
    \begin{subfigure}{0.4\textwidth}
        \includegraphics[width=\linewidth]{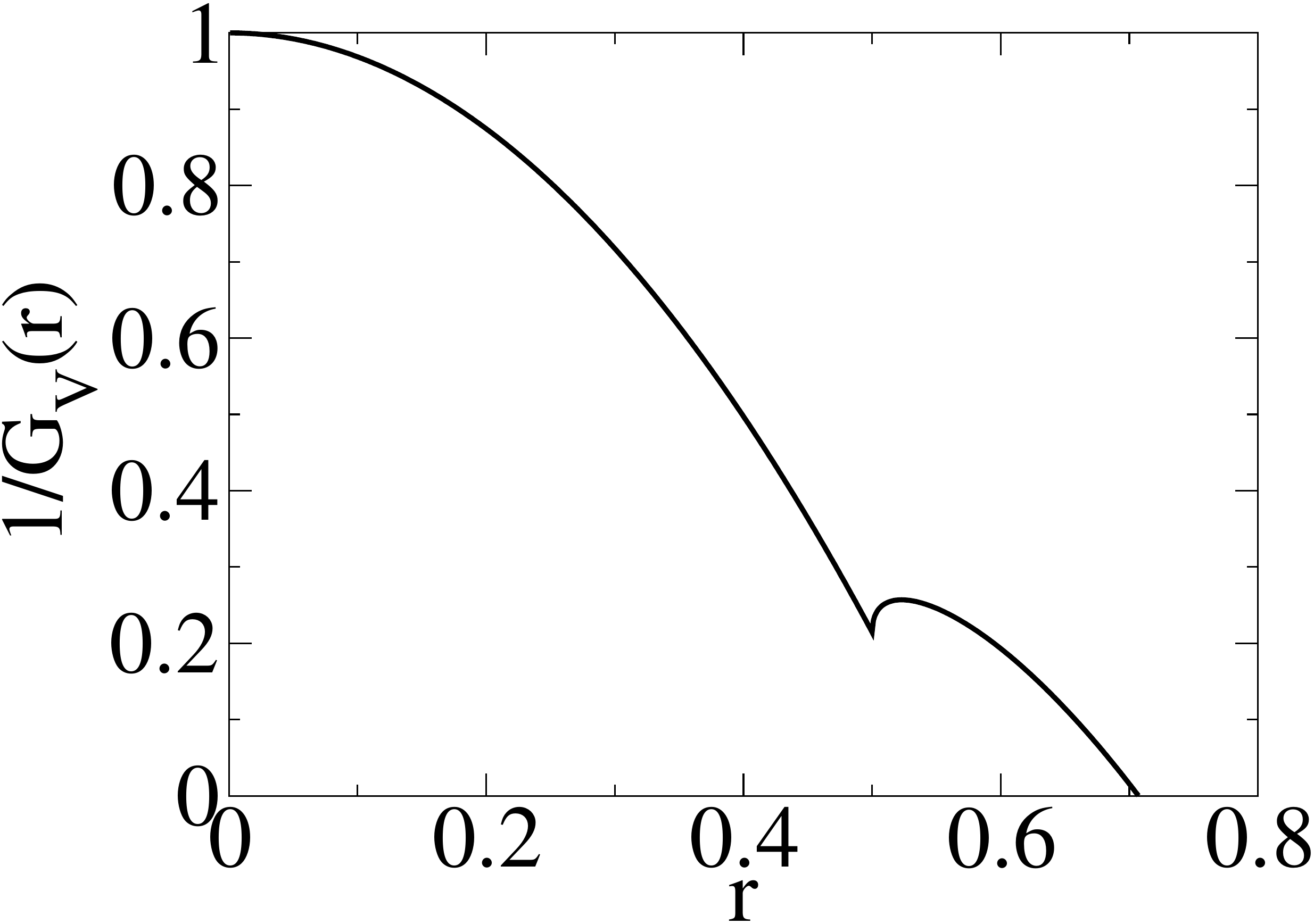}
        \caption{}
    \end{subfigure}
    \caption{The nearest-neighbor functions for a square lattice. We can see
    that $E_V(r)$ and $H_V(r)$ have compact support, which we know is guaranteed by
    stealthiness. (a) The void exclusion probability decays with a power law
    with exponent $\gamma = 2$. (b) The previous behavior of $E_V(r)$ implies a linear decay of $H_V(r)$.
    (c) We see that $G_V(r)$ diverges with a pole of order one. (d) By plotting
    $1/G_V(r)$, we can ascertain the asymptotic behavior of the other functions by
    the slope of the linear zero-crossing.}\label{squarelatfig}
\end{figure*}

\begin{figure}[h!]
    \centering
    \includegraphics[width=0.8\linewidth]{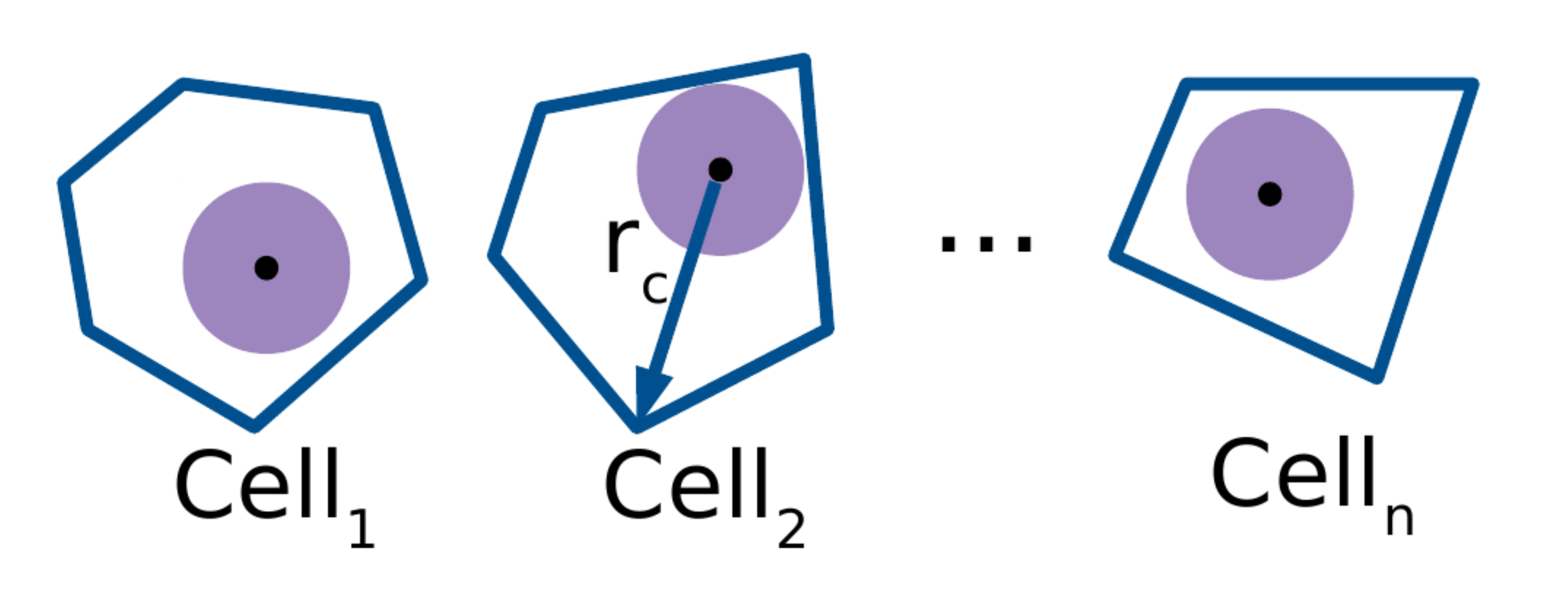}
    \caption{A schematic of Voronoi cells and spherical covering areas
    that demonstrates that the $E_V(r)$ of a crystalline system
    decays as a power-law with exponent $\gamma=d$ close to the critical-hole
    size. A crystal has a finite number of distinct Voronoi cells, and thus
    there exists a gap between the set of vertices at a distance $r_c$ and the
    next farthest set of vertices. Thus, we can conclude the proposed behavior
    through considering the decrease in volume of a small uncovered
    corner.}\label{schematic}
\end{figure}

It is useful to generally classify the asymptotic
behavior of the nearest-neighbor functions as they approach the critical-hole
size. One begins with the study of crystals, which are both
trivially stealthy due to the presence of Bragg peaks and have a trivially
bounded hole size which is found by identifying the location of the ``deep
holes'' in the crystal \cite{conway_sphere_1999}. In this case, the
hole probability function decays to zero as a power law
\cite{torquato_reformulation_2010}
\begin{equation}
    E_V(r) \sim C (r_c - r)^\gamma\qquad(r\to r_c^-),\label{latev}
\end{equation}
where $\gamma$ is a positive
exponent. It is possible to compute the exponent of this power law
analytically in the case of a crystal, and it takes the value
$\gamma = d$ for spatial dimension $d$ \cite{torquato_reformulation_2010}. To see this,
note that there are a finite number of distinct Voronoi
cells (Fig. \ref{schematic}), so we can always find the set of deepest holes in the
system, and no other hole will be infinitesimally close to being as deep. In
the intepretation of $E_V(r)$ as the ratio of the uncovered volume to the total
volume (Fig. \ref{small_voronoi}), the uncovered volume around these holes will
vanish according to a power law consistent with the dimension of the system as
the covered radius grows larger. In practice, this characteristic $\gamma = d$
power-law decay is most easily observed in systems with high degrees of
crystallographic symmetry. Examples include lattices and crystals with only a
few particles in the basis such as the hexagonal close-packed crystal. As the
number of particles in the smallest basis increases, the domain in which this
power law is guaranteed to be found shrinks, and disappears as the basis grows
to infinity.

The asympotic form (\ref{latev}) then implies
\begin{eqnarray}
    H_V(r) \sim \gamma C (r_c - r)^{\gamma - 1}&\qquad (r\to r_c^-)\\
    G_V(r) \sim \frac{\gamma}{\rho s_1(r) (r_c - r)}&\qquad (r\to r_c^-).\label{gvlin}
\end{eqnarray}
The behavior of $G_V(r)$ is particularly interesting. It diverges
with a pole of order one, which we will see is a generic feature
of the $G_V(r)$ of stealthy systems. For the case of the disordered stealthy
systems studied here, it is a reflection of the fact that we are taking the
limit $T\to 0$ while the pressure remains positive
\cite{torquato_ensemble_2015} (Eq. (\ref{thermo})). We can visualize the
near-$r_c$ behavior of these functions easily by plotting $1/G_V(r)$ (Fig.
\ref{squarelatfig}). We see that a linear decay of $1/G_V(r)$ with a specified
relation between the slope to the zero crossing is associated with a
crystal-like power law decay.

However, in the disordered case, we find that the exponent is typically not
given by $\gamma = d$. One in general expects to find a larger value, implying
that the holes vanish more quickly as one approaches $r_c$, and we
first give an intuitive argument for this fact. Since the number of
distinct Voronoi cells is infinite in the general disordered case, it is
possible to have a continuum of vertices with distances close to $r_c$. While
the deepest hole in each cell still closes with the characteristic $\gamma = d$
power-law decay, the fraction of uncovered holes is also decreasing as one gets
closer to $r_c.$ This is in contrast with the crystalline case, where the gap
between the farthest and next-to-farthest vertex ensures that this fraction is
constant. Thus, we expect the hole probability to reach zero asymptotically
faster in the case of a disordered system, since one is both covering up volume
and decreasing the fraction of cells in which there are uncovered holes near
$r_c.$

While we are not aware of a method to compute the exponent $\gamma$ analytically in the case
of a disordered stealthy system, we will work through the
two examples of non-stealthy systems with bounded holes introduced
in Section \ref{higherorder}, and verify that $\gamma > d$. In the case of a
one-dimensional random sequential addition process at saturation, one has that
the void exclusion probability assuming spheres of unit diameter is
given by \cite{rintoul_nearest-neighbor_1996}
\begin{equation}
        E_V(r) = 1 - 2 (1-r) \int_0^\infty \frac{H(t)}{t^2}\,\rmd t
                -2\int_0^\infty \frac{H(t)}{t^3}\left[1-e^{-(2r -1) t}\right]\,\rmd t,
\end{equation}
where
\begin{equation}
    H(t) = e^{-2[\gamma_e - {\rm Ei}(-t)]}, 
\end{equation}
where $\gamma_e$ is Euler's constant and ${\rm Ei}(t)$ is the exponential
integral. We then expand the exponential in the second integrand around $r=1$ and find
\begin{equation}
    E_V(r) \sim 4(1-r)^2 \int_0^\infty \frac{H(t)e^{-t}}{t}\,\rmd t + \cdots\qquad (r\to 1^-),
\end{equation}
where we have crucially used the fact that
\begin{equation}
    \int_0^\infty \frac{H(t)}{t^2}\,\rmd t = 2\int_0^\infty \frac{H(t)e^{-t}}{t^2}\,\rmd t,
\end{equation}
which can be shown by integration by parts. Thus, for this one-dimensional disordered process,
the exponent $\gamma$ has increased to $\gamma = d+1$. One interesting but
currently unresolved question is whether the formula $\gamma = d+1$ holds for saturated
RSA processes in all dimensions.

\begin{figure}[h!]
    \centering
    \includegraphics[width=0.6\linewidth]{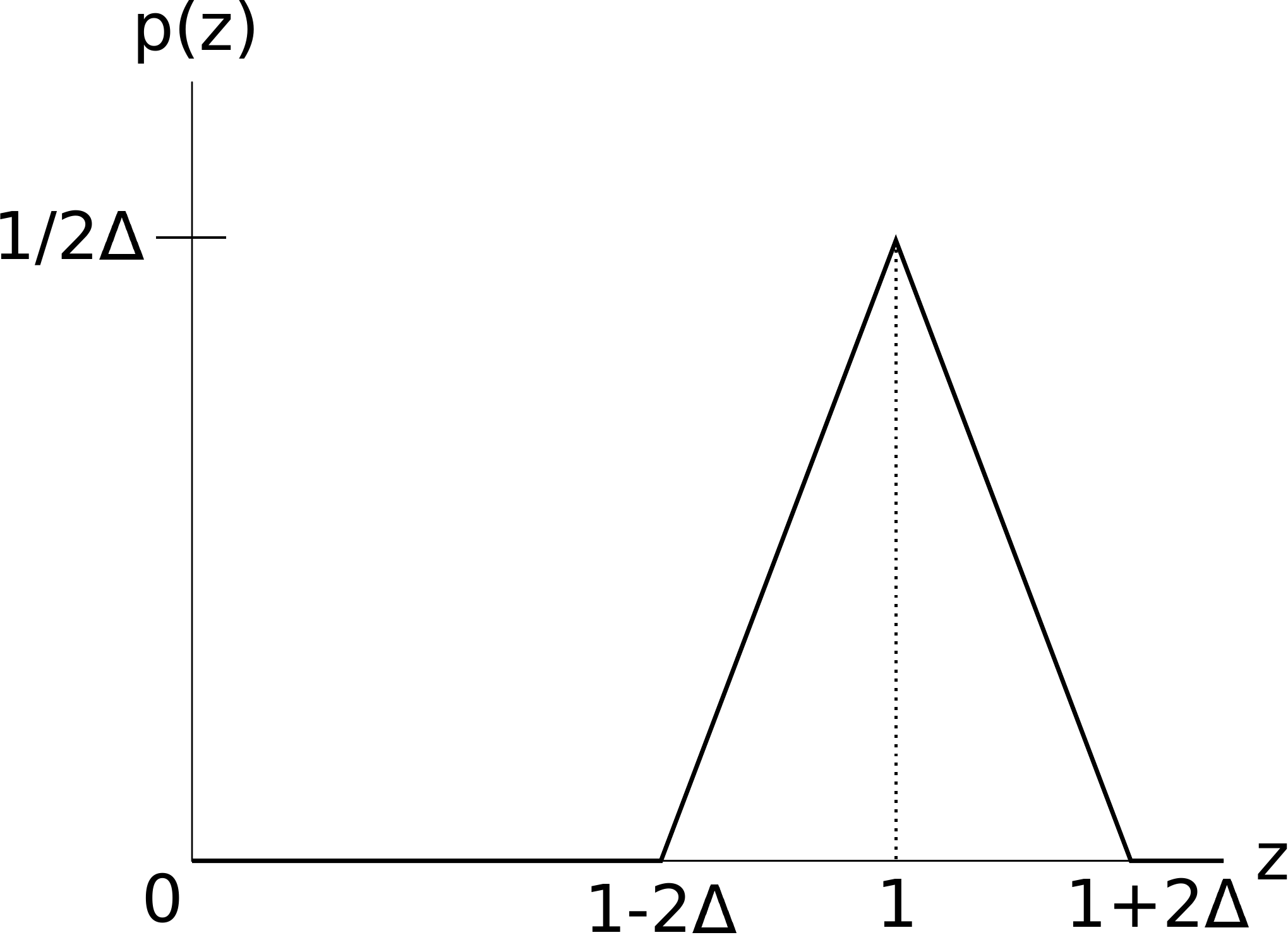}
    \caption{The gap distribution function $p(z)$ for the
    one-dimensional perturbed lattice described in the
    text.}\label{gap_distribution}
\end{figure}

The approach implicitly used above by taking results from Ref.
\cite{rintoul_nearest-neighbor_1996} is also of fundamental theoretical interest.
Thus, we will now describe it in some detail, and derive new general results
concerning the near-$r_c$ behavior of the functions involved.
In one dimension, we analyze systems by introducing the {\it gap distribution
function} $p(z)$, which gives the probability density to observe a gap with length
between $z$ and $z + dz$ between neighboring particles \cite{torquato_point_2008}.
Since one can relate \cite{rintoul_nearest-neighbor_1996, torquato_point_2008}
\begin{equation}
    E_V(r) = \rho \int_{2r}^\infty p(z) (z-2r)\,\rmd z,\label{efromp}
\end{equation}
one can derive the near-$r_c$ behavior of $p(z)$ for systems with bounded
holes by expanding around $r = r_c$:
\begin{equation}
        E_V(r) = 2\rho p(2r_c)(r_c -r )^2
        + \frac{4\rho}{3}\frac{d}{dz}p(z)\bigg|_{z=2r_c} (r_c - r)^3 + \cdots.
    \label{pzseries}
\end{equation}
From this expansion, it is seen that if $E_V(r)$ decays with a power law
with exponent $\gamma=n$ as $r\to r_c$, then $p(z)$ decays as a power law with exponent
$\gamma=n - 2$ as $z\to 2r_c$.

We apply this observation to determine the asymptotic behavior
of the 1D perturbed lattice considered in Section \ref{higherorder}, given a
specific form of the displacement distribution. As a concrete example, we
consider a uniform displacement distribution. By using the fact that the gap
between particles $i$ and $i+1$ can be written in terms of the displacement
variables of Section \ref{higherorder} as $z = 1 - \delta_i + \delta_{i+1}$,
and that the distribution of the sum of independent random variables is the convolution
of their individual distributions \cite{grinstead_introduction_1997},
one can verify that the gap distribution of this system
is that given in Fig. \ref{gap_distribution}. Upon inserting this form of $p(z)$ into Eq.
(\ref{efromp}) and expanding around $r = r_c = (1 + 2 \Delta)/2$, we find that
\begin{equation}
    E_V(r) \sim \frac{(r_c -r)^3}{3\Delta^2} \qquad (r\to r_c^-).
\end{equation}
Thus, this 1D system has a power-law decay of
$E_V(r)$ with exponent $\gamma = d+2.$ However, we emphasize that this
system is a special type of perturbed lattice, with a
specific bounded displacement distribution and uncorrelated displacements. It is
clear from Eq. (\ref{pzseries}) that one can obtain any $\gamma \geq 2$
by specifying the asymptotic behavior of $p(z)$, but it would be interesting
to also determine whether adding correlations between displacements or going
to higher spatial dimensions would change the result.

The intuitive argument for the increase of the exponent $\gamma$ also suggests
another intriguing possibility, which we will use in Section \ref{domain}. In principle,
one can have that the hole probability function decays faster than any
power law. One simple functional form that exhibits this asymptotic behavior
is \cite{hunter_applied_2004}:
\begin{equation}
E_V(r) \sim C \exp \left(-\frac{\zeta}{r_c - r} + \cdots\right)\qquad (r\to r_c),\\
\end{equation}
giving rise to the following asymptotic forms for $H_V(r)$ and $G_V(r)$:
\begin{eqnarray}
        H_V(r) \sim \left(\frac{\zeta C}{(r_c - r)^2}+\cdots\right) \exp
        \left(-\frac{\zeta}{r_c - r} + \cdots\right)&\qquad (r\to r_c)\\
        G_V(r) \sim \frac{\zeta}{\rho s_1(r) (r_c - r)^2} + \cdots&\qquad (r\to r_c).\label{gvquad}
\end{eqnarray}
This behavior gives a divergent $G_V(r)$ with a pole of order two. In general, a pole of
any order would be permissible, but we only explicitly consider the order two case.
If we consider instead the reciprocal function $1/G_V(r)$, we see that a quadratic (or any
higher order) decay is associated with an $E_V(r)$ that decays faster than any crystal
on approach to the critical-hole size. It is possible to specify a specific distribution
for the 1D perturbed lattice considered previously that can be shown to possess
an $E_V(r)$ that decays faster than any power law, although we have not been
able to compute the exact form of Eq. (\ref{gvquad}). One takes the
displacement distribution $u(\delta)$ as
\begin{equation} u(\delta) =
    \frac{1}{I}\exp\left(\frac{1}{\delta^2-\Delta^2}\right) \qquad -\Delta< \delta <
    \Delta,
\end{equation}
and zero elsewhere, where $I$ is the normalization constant needed to create a
well-defined probability density function. One can then verify through
judicious replacements of pieces of the convolution integrand for $p(z)$ by
constants that form upper bounds that $p(z)$ decays faster than any power law
as $z \to 2r_c$. This implies through Eq. (\ref{pzseries}) that $E_V(r)$ also
decays faster than any power law, since the coefficient for each term in
the series will be zero \cite{hunter_applied_2004, feldman_supplementary_2008}.
However, the use of these coarse upper bounds in the argument prevents us from
extracting the exact asymptotic behavior. It would be interesting to identify a
system for which a form of $G_V(r)$ with a higher-order pole could be exactly
computed.

Given the fundamental importance of the near-$r_c$ behavior described above, it
is essential to develop intuition for the case of stealthy systems. Since we currently
lack strong enough direct theoretical tools, we provide some preliminary data
via computer simulations that contextualizes these simple arguments.

\subsection{Simulation Results}
\begin{figure}
    \centering
    \begin{subfigure}{0.33\textwidth}
        \includegraphics[width=\linewidth]{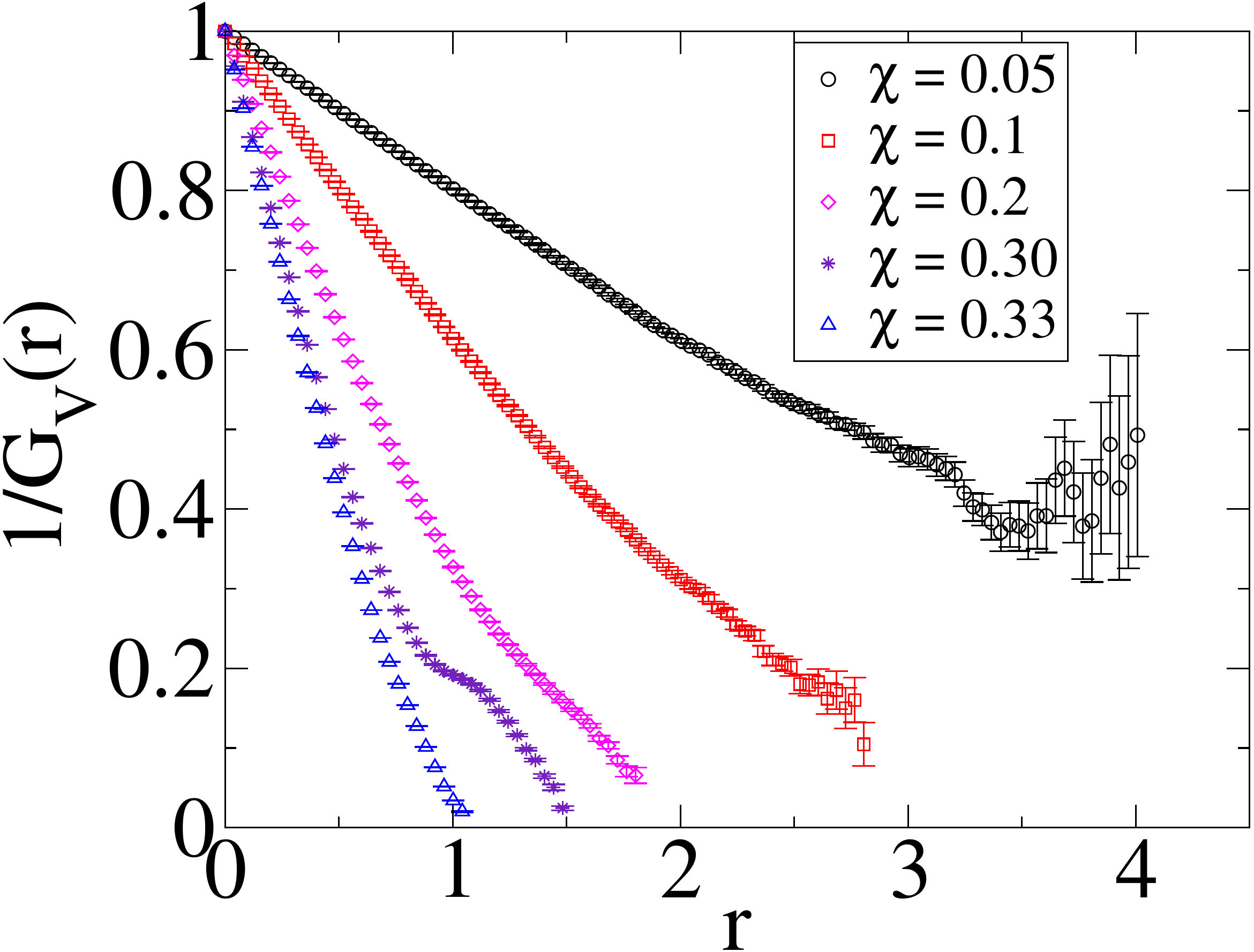}
        \caption{}
    \end{subfigure}%
    \begin{subfigure}{0.33\textwidth}
        \includegraphics[width=\linewidth]{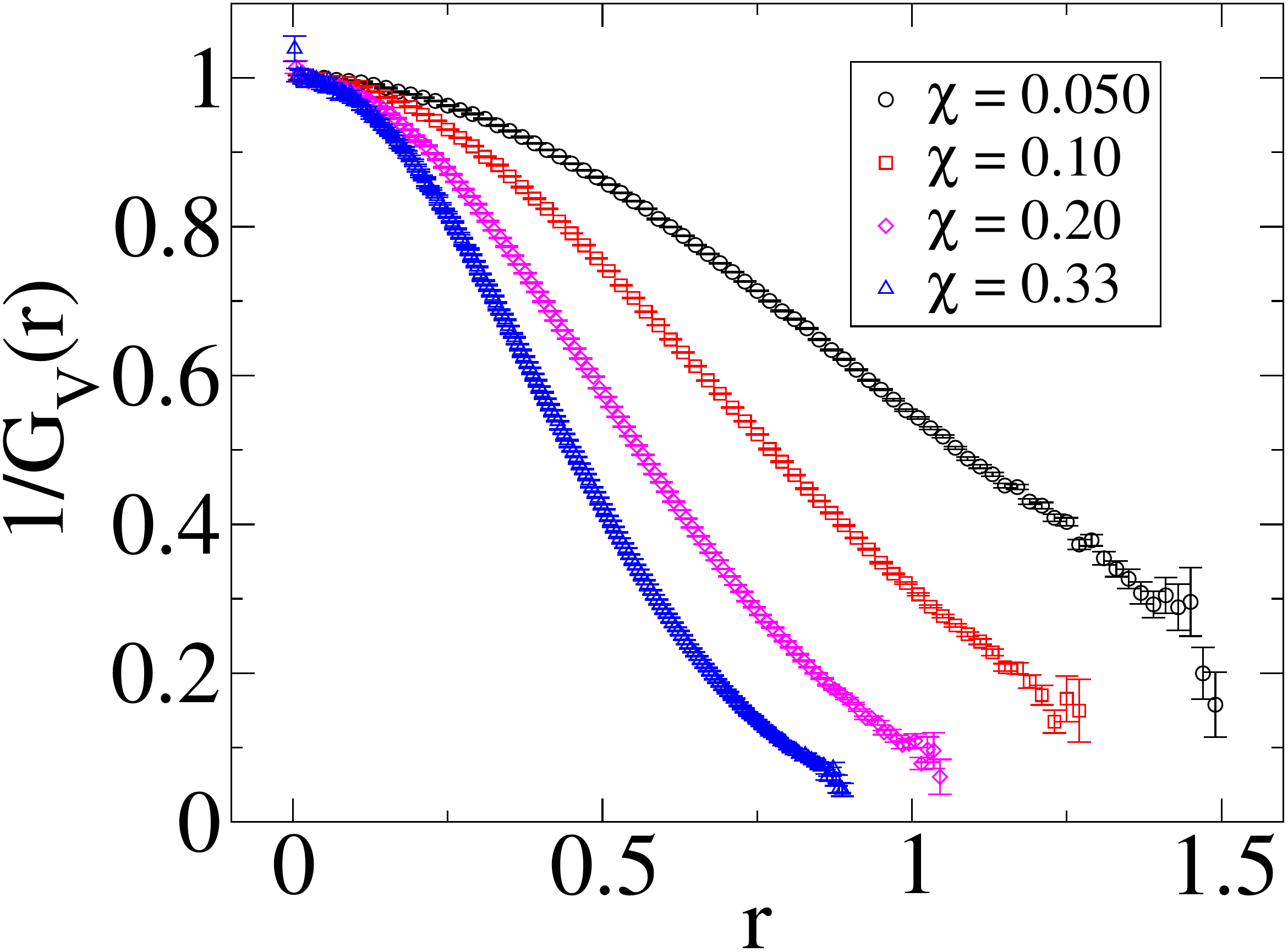}
        \caption{}
    \end{subfigure}%
    \begin{subfigure}{0.33\textwidth}
        \includegraphics[width=\linewidth]{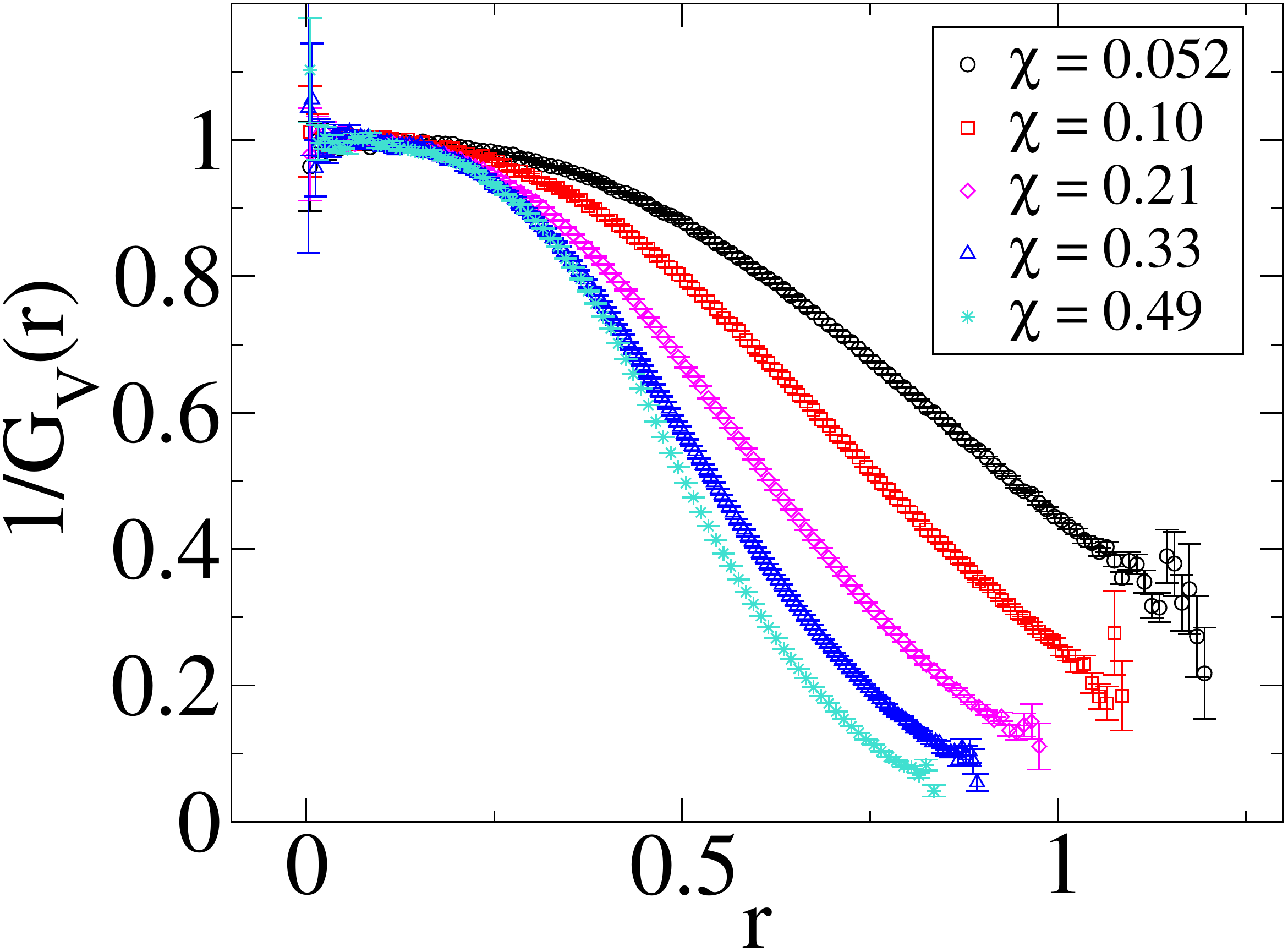}
        \caption{}
    \end{subfigure}
    \caption{Simulations of $1/G_V(r)$ for the first
    three spatial dimensions. See the Appendix for details on system and
    ensemble size. (a) Simulations for 1D. (b) Simulations for 2D. (c)
    Simulations for 3D.}\label{simulationfig}
\end{figure}

Here, we present simulation results for $G_V(r)$ in 1, 2, and 3 dimensions (Fig.
\ref{simulationfig}). We sample $G_V(r)$ through a geometric method in 1D and
by binning the nearest neighbors in 2D and 3D; see the Appendix for details.
While the obtained statistics are not good enough to draw robust conclusions, in
two and three dimensions, we can see the beginning of a cross-over in the form of a decreased slope for 
$1/G_V(r)$ on the larger-$\chi$ samples. This suggests that $E_V(r)$ for these
configurations will either have power-law tails with large values
of $\gamma$, or that they will have an $E_V(r)$ that decays
faster than any power law. In one dimension, the behavior at higher-$\chi$
is somewhat more complicated, with a plateau forming at $\chi=0.20$ and disappearing
in the $\chi=0.33$ data. This disappearance is likely due to a subtle finite
size error, where crystallization is enhanced close to the $\chi =1/3$
order-disorder transition that exists in the thermodynamic limit
\cite{fan_constraints_1991}. Indeed, we observe
that a small amount of long-order develops in the form of slowly decaying oscillations
in the pair correlation function for our $\chi= 0.33$ data. However, we still report these curves throughout
the article, since we expect this error to be much less noticeable both in
$G_V(r)$ far from the critical-hole size and in the less sensitive quantities
$E_V(r)$ and $H_V(r)$. We observe that the likely behavior close to $r_c$ for $E_V(r)$ is
a power-law decay, however, this does not necessarily imply
that this is the case for all dimensions. Whether the true asympotics for $E_V(r)$
are power-law decays or not, these results
suggest that while both crystals and disordered stealthy systems have bounded
hole sizes, the functions $H_V(r)$ and $E_V(r)$ of disordered systems approach
their asymptotic value much more quickly as one moves toward the critical-hole
size.

This observation explains why it is extremely difficult to sample the
near-$r_c$ behavior for disordered stealthy systems. Since
$E_V(r)$ vanishes as the critical-hole size is approached, the faster decay
implies that for any finite configuration, the event of observing a hole
with size sufficiently close to the critical-hole size is rarer than that of a
crystalline system. Thus, one needs to sample much larger systems, unlike in the case of a
crystal, for which a single copy of the fundamental cell
suffices. This observation is also closely linked to one made by Zhang,
Stillinger, and Torquato \cite{zhang_can_2017}, which is that it is easier to
observe large holes for $\chi$ close to $1/2$ than for smaller $\chi$, since
the relative lack of close-range order for small $\chi$ implies that the event
of finding a large hole becomes correspondingly rarer.

\section{Towards Accurate Expressions for Nearest-neighbor Functions Over the Whole
Domain}\label{domain}

\begin{figure*}
    \centering
    \begin{subfigure}{0.33\textwidth}
        \includegraphics[width=\linewidth]{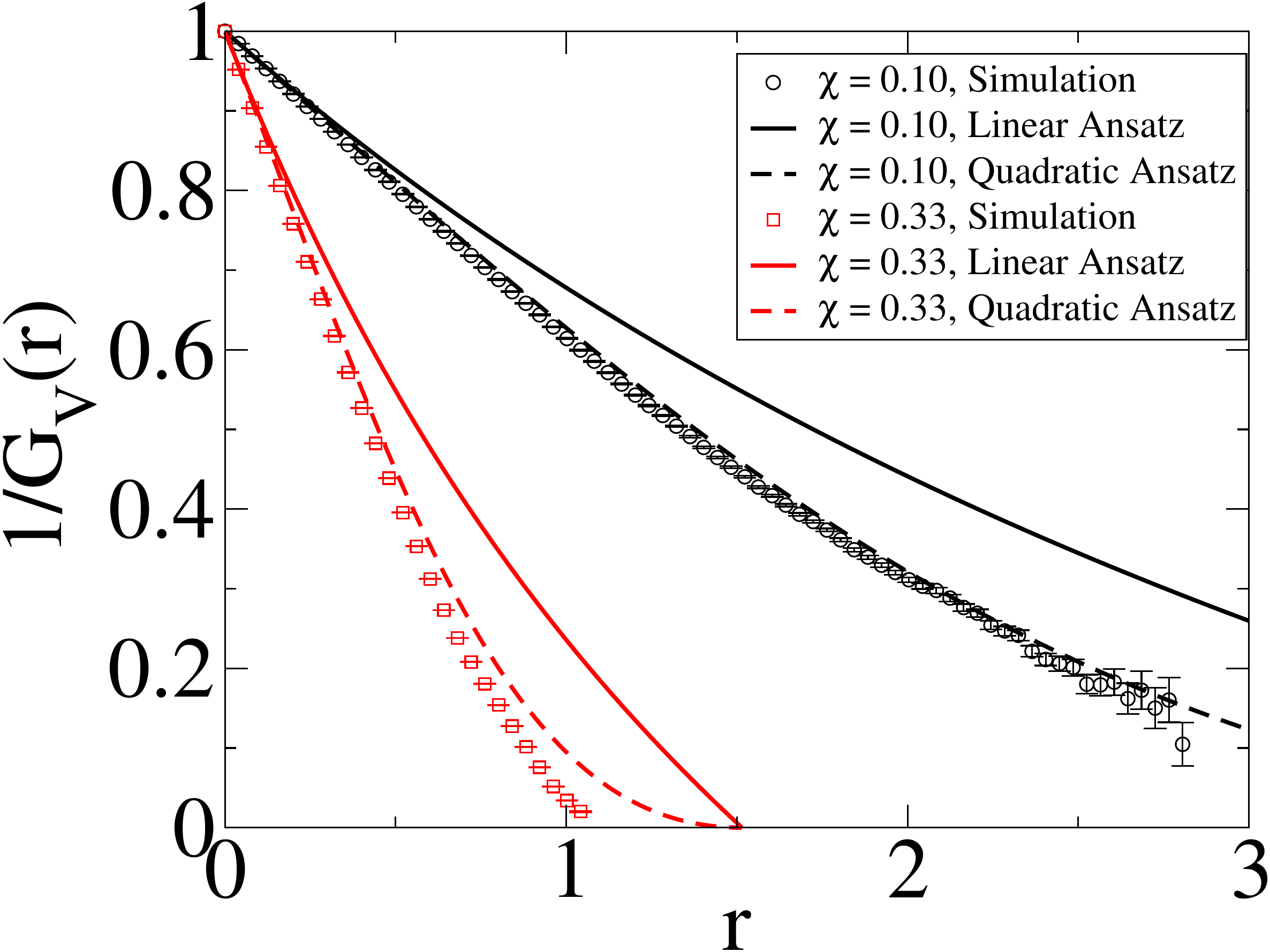}
        \caption{}
    \end{subfigure}%
    \begin{subfigure}{0.33\textwidth}
        \includegraphics[width=\linewidth]{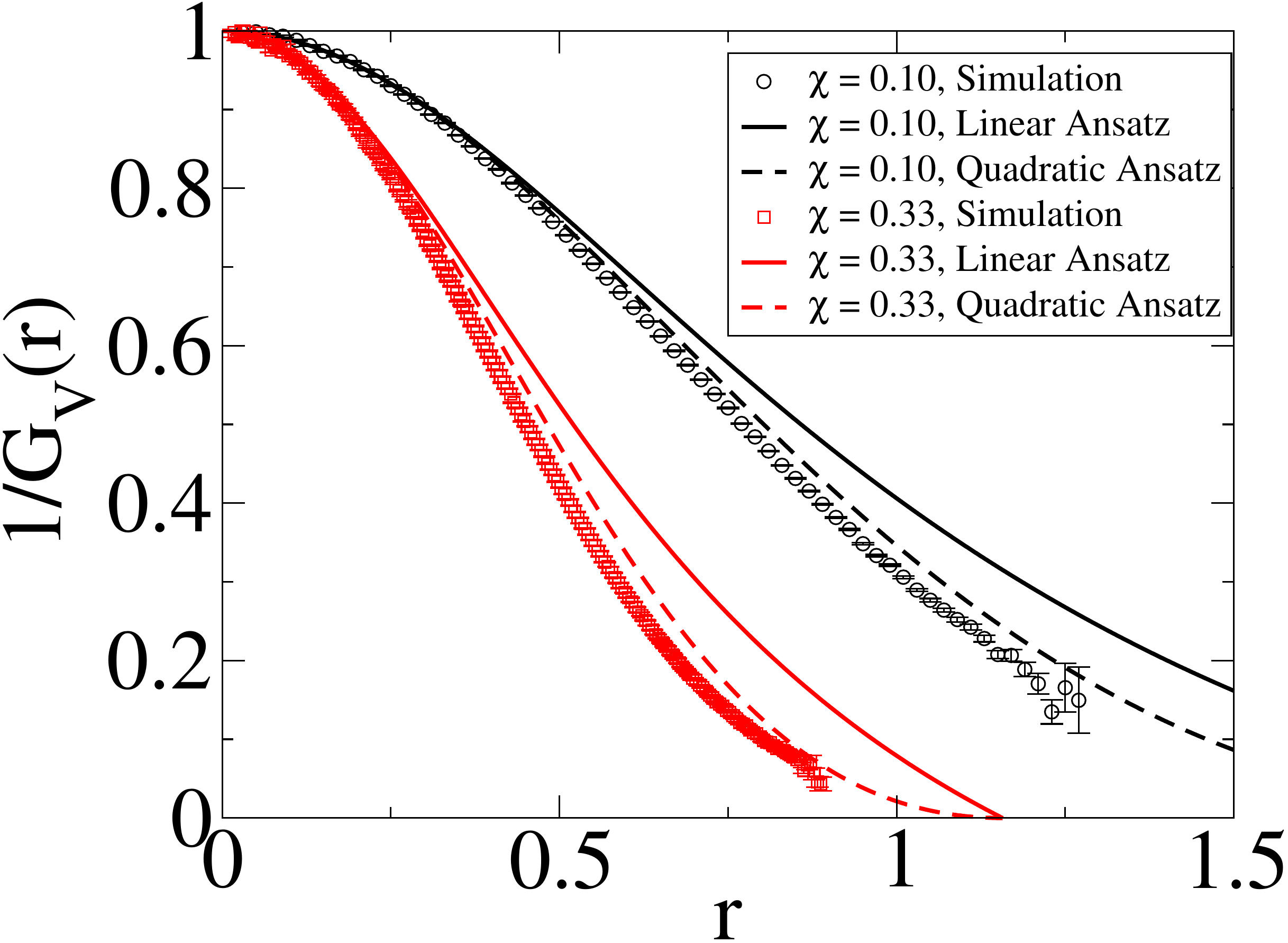}
        \caption{}
    \end{subfigure}%
    \begin{subfigure}{0.33\textwidth}
        \includegraphics[width=\linewidth]{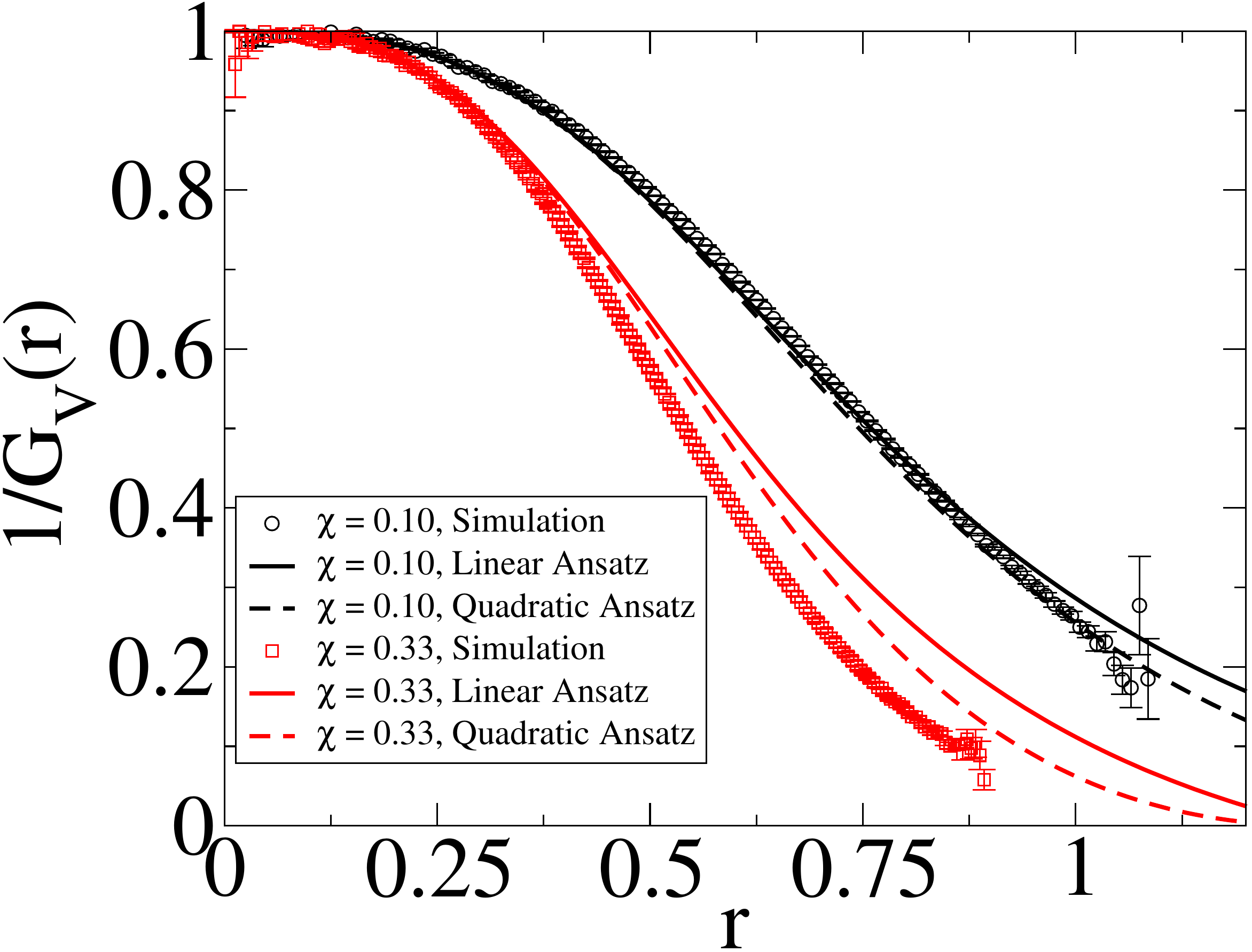}
        \caption{}
    \end{subfigure}\\
    \begin{subfigure}{0.33\textwidth}
        \includegraphics[width=\linewidth]{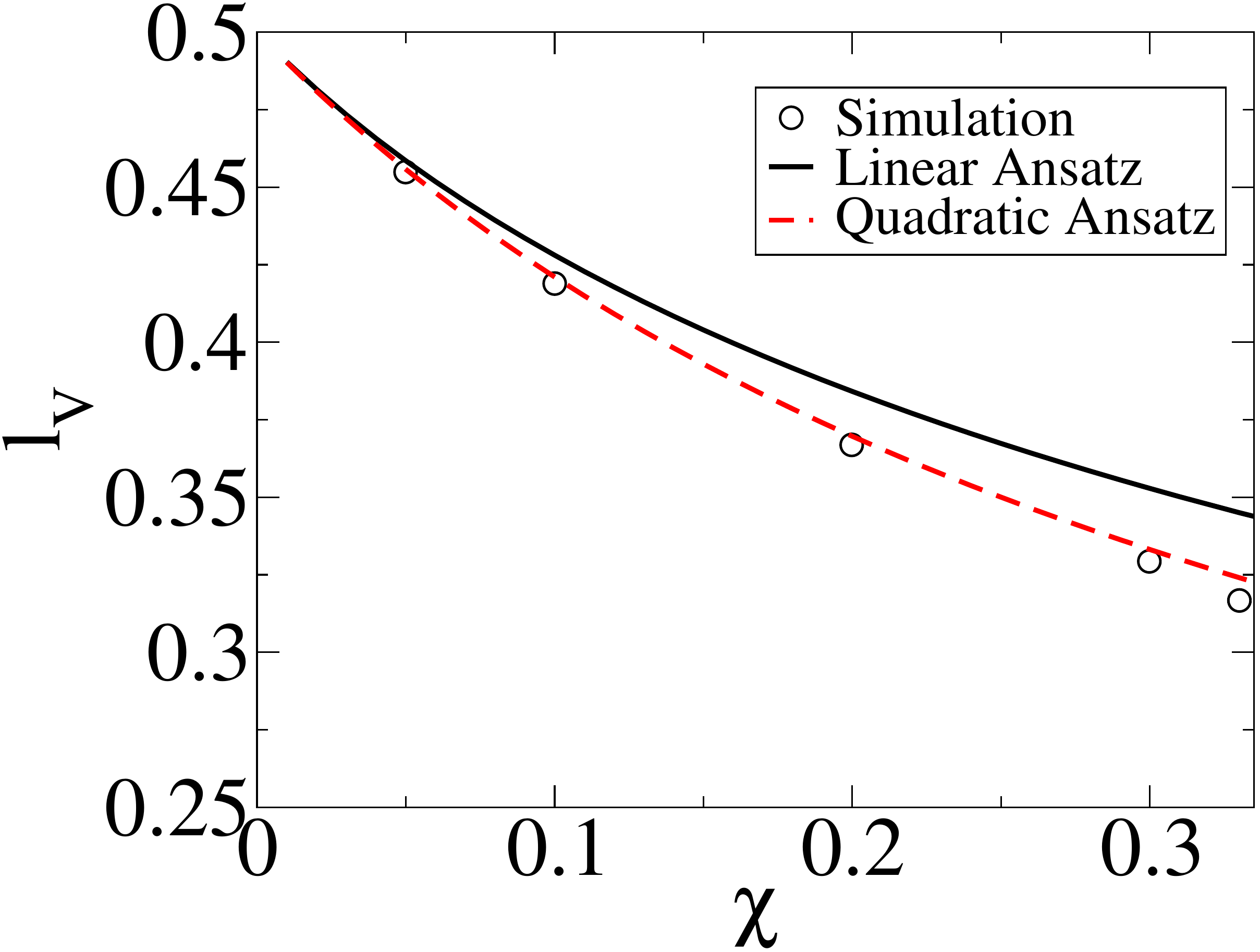}
        \caption{}
    \end{subfigure}%
    \begin{subfigure}{0.33\textwidth}
        \includegraphics[width=\linewidth]{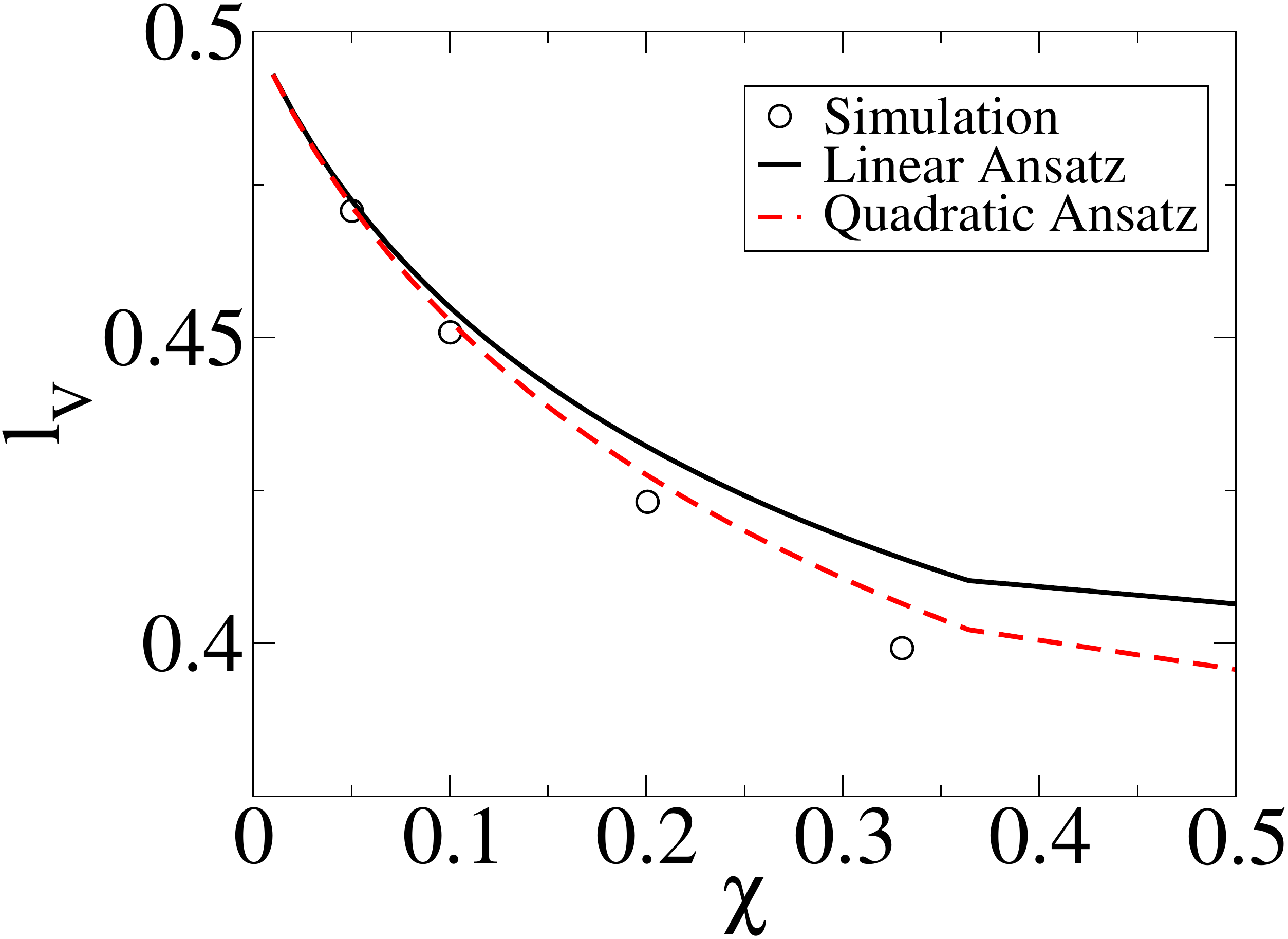}
        \caption{}
    \end{subfigure}%
    \begin{subfigure}{0.33\textwidth}
        \includegraphics[width=\linewidth]{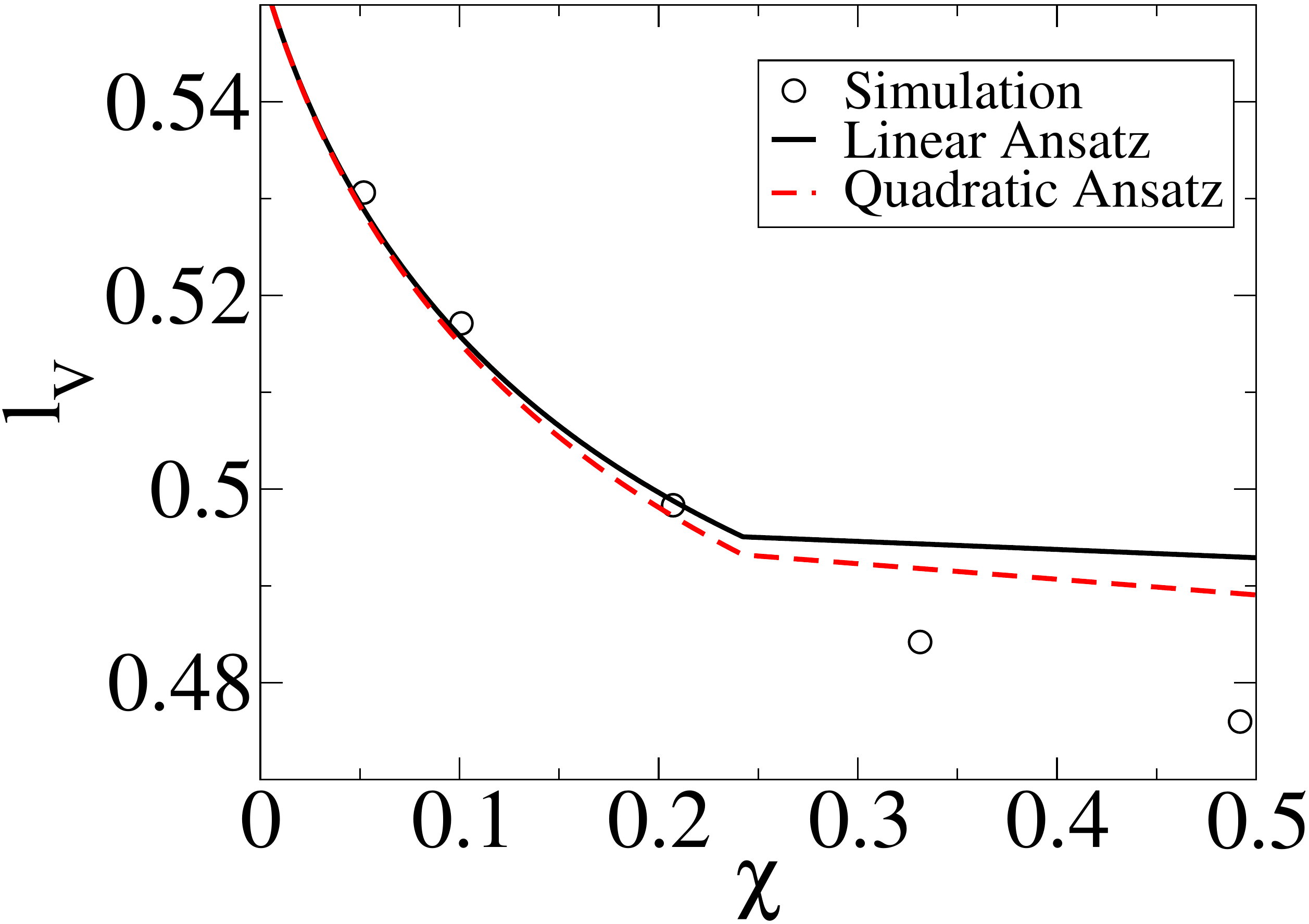}
        \caption{}
    \end{subfigure}
    \caption{Comparisons between simulations and the theories
    given in Eqs. (\ref{linear}) and (\ref{quadratic}). (a-c) $1/G_V(r)$ for two different values of $\chi$ in one, two, and three dimensions, respectively.
    (d-f) The void mean nearest-neighbor distance $l_V$ as a function of
    $\chi$ for one, two, and three dimensions, respectively.}\label{whole_domain}
\end{figure*}

We now devise an approximation that matches the contributions to the small-$r$ and near-$r_c$
expressions discussed above. The basic strategy is to make a change of asymptotic scale
on the small-$r$ asymptotic expansion for $G_V(r)$ given in Eq. (\ref{gvSeries}) so that it
matches either the pole-of-order-one asymptotics of Eq. (\ref{gvlin}) that gives
rise to a power law decay of $E_V(r)$ or the pole-of-order-two asymptotics of
Eq. (\ref{gvquad}) that gives rise to an exponential decay of $E_V(r)$ as $r$
approaches $r_c$. While there are many ways of doing this, one fruitful choice
is to take either the pole-of-order-one formula:
\begin{equation}
    G_V(r) = 1 + \frac{(1-a) v_1(r_c) \rho v_1(r)}{v_1(r_c)-v_1(r)},\label{linear}
\end{equation}
or the pole-of-order-two
\begin{equation}
    G_V(r) = 1 + \frac{(1-a) v_1(r_c)^2 \rho v_1(r)}{[v_1(r_c)-v_1(r)]^2},\label{quadratic}
\end{equation}
where the maximal hole size $r_c$ is given by the formula \cite{zhang_can_2017}:
\begin{equation}
    r_c = \frac{(d+1)\pi}{2K}.
\end{equation}
In addition to connecting behaviors consistent
with the small-$r$ expansions given in Section \ref{pseudohs} and our
observations concerning the close-to-critical-hole-size regime
presented in Section \ref{tail}, we believe they satisfy the bounds given by
the inequalities (\ref{gvbound1}) and (\ref{gvbound2}) (although we have
not constructed a rigorous proof of this proposition). They have been compared to simulation
data for $G_V(r)$ and $l_V$ in Fig. \ref{whole_domain}. We can see that in one and two dimensions,
the pole-of-order-two formula (\ref{quadratic}) is more accurate, with good
agreement at low-$\chi$ and tolerable agreement at intermediate $\chi$. While this formula gives
different asymptotics than is apparent in the data for $1/G_V(r)$ for the 1D system at $\chi =
0.30$ given in Fig. \ref{simulationfig}, this is likely balanced by lessening
the error at smaller $r$, which dominates the contribution to $l_V$. In three dimensions, the two
formulas give similar predictions, with the more accurate approximation being
determined
by the value of $\chi$. We also see that the prediction for $l_V$ qualititatively breaks down past
a certain $\chi$, where we believe higher-order coefficients such as $b$ must be included.

\section{Positive Temperature}\label{finitetemp}

\begin{figure}
    \centering
    \begin{subfigure}{0.45\textwidth}
        \includegraphics[width=\linewidth]{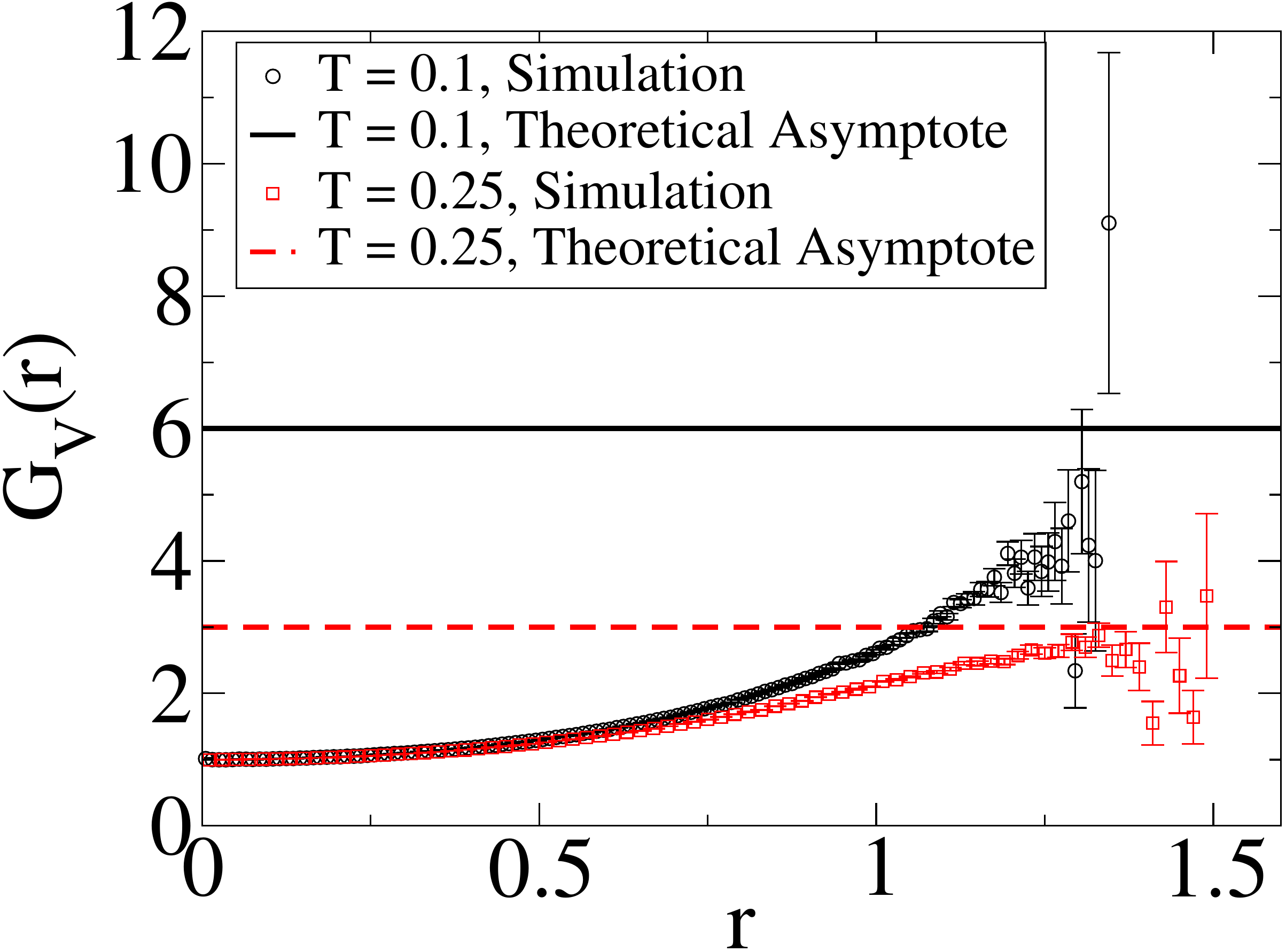}
        \caption{}
    \end{subfigure}%
    \begin{subfigure}{0.45\textwidth}
        \includegraphics[width=\linewidth]{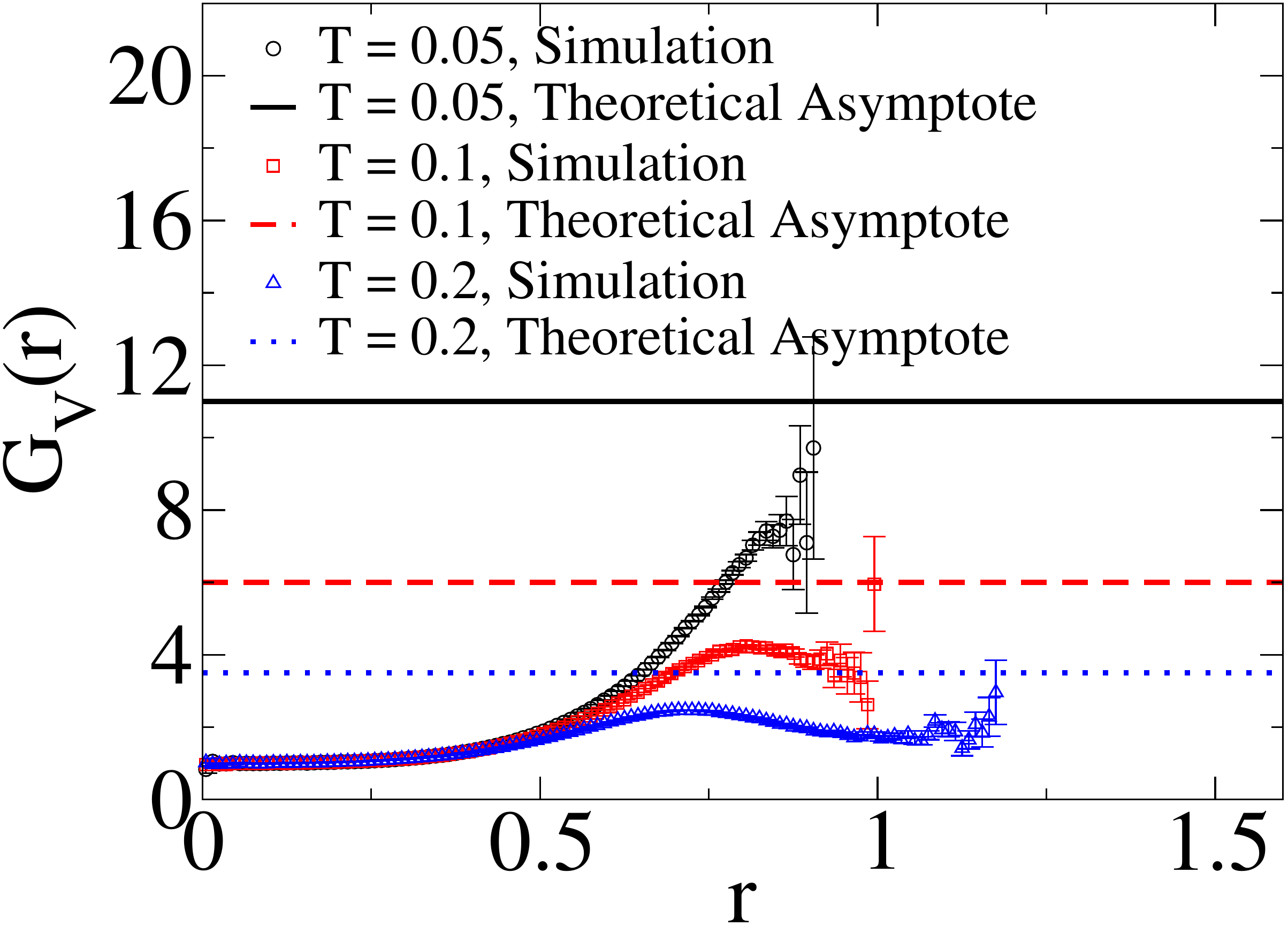}
        \caption{}
    \end{subfigure}
    \caption{The function $G_V(r)$ sampled at positive temperatures for
    (a) a 2D ensemble at $\chi = 0.10$ and (b) a 3D ensemble at
    $\chi=0.49$, as well as the saturating values predicted by Eq.
    (\ref{saturate}).}\label{finitet}
\end{figure}

For sufficiently small temperatures, we have that the pressure of a stealthy system
is expliclitly given by \cite{torquato_ensemble_2015}
\begin{equation}
    p \sim \rho T + \frac{\rho^2 v_0}{2}.\label{positiveTpressure}
\end{equation}
From Eq. (\ref{thermo}), we then have a prediction for the asympotic value of $G_V$ as (taking $\rho =
1$):
\begin{equation}
    G_V(r\to \infty) \sim 1 + \frac{1}{2T}.\label{saturate}
\end{equation}
Note that this formula implies that there will be a singular change
in behavior of holes in the system as one increases the temperature, even
infinitesimally, from $T=0$.
The presence of a finite asymptote suggests that one can in
principle expend an arbitrarily large amount of work to create an arbitrarily
large hole. Thus, since this system is in equilibrium, the maximum hole size
will be unbounded in the infinite volume limit at positive
temperature. This is in stark contrast to the ground state behavior,
where Eq. (\ref{saturate}) does not apply due to the presence of the divergence.
This divergence is ultimately derived from the fact that the relation between $G_V(r)$
and the work required to produce a hole becomes singular at $T=0$ \cite{reiss_statistical_1959}.

We have plotted
the results of simulations at positive temperatures for $G_V(r)$
in Fig. \ref{finitet}. However, none of these cases definitively asymptote to
the predicted value before the simulated data becomes very
imprecise. We are unsure whether this discrepancy is caused by the
difficulty of sampling $G_V(r)$ at large $r$ and positive $T$, or whether the
linear approximation (\ref{positiveTpressure}) simply breaks down. We can also
see that for some values of $\chi$ and $T$, the behavior of $G_V(r)$ can become
non-monotonic.

\section{Conclusions and Discussion}\label{conclusion}

In summary, we have obtained bounds and approximations to the
nearest-neighbor functions valid in the small-$r$ and $\chi$ regimes through
the use of the pseudo-hard-sphere ansatz, formally advanced a pair
of conjectured bounds, showed that the nearest-neighbor functions
of stealthy systems can be determined by a finite number of $g_n({\bi r}^n)$,
investigated the close-to-critical-hole-size regime through theoretical
arguments and simulation, and combined insights from these analyses to form an
approximation valid for small $\chi$ and all $r$. We showed that disordered
stealthy processes appear to possess different behavior from
their ordered counterparts as they approach their critical-hole size. Finally,
we have given the asymptotic behavior of $G_V(r)$ for finite temperature
systems, and concluded that we expect stealthy systems to lose their bounded
hole size property, even at arbitrarily small temperatures.

These results both answer fundamental questions about the statistical properties
of stealthy hyperuniform systems and raise new avenues of inquiry. They suggest that
the asysmptotic behavior of the nearest-neighbor functions near the critical-hole
size is different for ordered and disordered stealthy systems, but
obtaining more complete evidence in favor of this proposition will
likely require the development of new methods for the
investigation of stealthy systems. This may either take the form of numerical
methods to sample large holes or an increase in efficiency in which these unusual
potentials can be simulated, or theoretical methods to directly obtain the
asymptotic behavior. In addition, the singular disappearance of a bounded
hole size at positive temperature further incentivizes studies of the
positive temperature regime, as one may find other unusual statistical
characteristics of these systems.

In addition to the implications for these systems as point processes, those
results which apply at intermediate $\chi$ can also be used to comment on the
structure of disordered packings of intermediate density, as any finite
stealthy system can be decorated with spheres whose diameter depends on $\chi$
to obtain a packing. Considered as two-phase systems, these packings are also
stealthy hyperuniform \cite{zhang_transport_2016}.

Having outlined methods for obtaining good analytical approximations to these
functions, we can then investigate applications to the field of heterogenous
materials. Accurate expressions for the nearest neighbor functions can be used
to place bounds on or estimate the trapping constant \cite{keller_extremum_1967,
rubinstein_diffusioncontrolled_1988, torquato_diffusioncontrolled_1989,
torquato_random_2002} and fluid permeability \cite{torquato_random_2002} of
two-phase systems derived from these point processes. These bounds may find
use in identifying applications for stealthy processes in materials engineering.

\section*{Acknowledgements}
    We gratefully acknowledge the support of the National Science Foundation
    under Grant No. CBET-1701843 and the use of computer time from Princeton
    Research Computing. We also would like to thank Ge Zhang for providing the
    stealthy system simulation code, data from previous studies, and valuable
    discussions, as well as Michael Klatt and Jaeuk Kim for useful discussions.

{\appendix
\section{Simulation Details}

    In this Appendix, we give details on the numerical methods used to produce results
    in this article, with the exception of the two and three dimensional results
    given in Fig. \ref{compare_A}, which is partially based on a resampling of
    data presented in Ref. \cite{zhang_transport_2016}.

    \subsection{Collective Coordinate Procedure}
    \begin{table}
        \centering
        \begin{tabular}{|c|c|c|c|c|c|c|c|}
            \hline
            $d$ & $\chi$ & Figs. & $N$ & $N_{\rm snap}$ & $N_{\rm step}$ & $N_{\rm eq}$ & Unit Cell\\
            \hline\hline
            1 & 0.0499 & T1,6,11,12 & 9300 & 500 & 5000 & 200 & Integer Lattice\\
            \hline
            1 & 0.10002 & T1,3,6,11,12 & 6600 & 500 & 5000 & 200 & Integer Lattice\\
            \hline
            1 & 0.1998 & T1,6,11,12 & 4600 & 500 & 5000 & 200 & Integer Lattice\\
            \hline
            1 & 0.2998 & T1,6,11,12 & 3800 & 500 & 5000 & 200 & Integer Lattice\\
            \hline
            1 & 0.3301 & T1,3,6,11,12 & 3600 & 500 & 5000 & 200 & Integer Lattice\\
            \hline
            2 & 0.0502 & 6,11,12,S3 & 9300 & 500 & 5000 & 200 & Triangular Lattice\\
            \hline
            2 & 0.1002 & 4,6,11,12,S3,S4 & 6600 & 500 & 5000 & 200 & Triangular Lattice\\
            \hline
            2 & 0.201 & 6,11,12,S3 & 4600 & 500 & 5000 & 200 & Triangular Lattice\\
            \hline
            2 & 0.3301 & 4,6,11,12,S3,S4 & 3600 & 500 & 5000 & 200 & Triangular Lattice\\
            \hline
            3 & 0.519 & 11,12 & 9300 & 500 & 5000 & 200 & BCC Lattice\\
            \hline
            3 & 0.101 & 5,11,12,S1,S2 & 6600 & 500 & 5000 & 200 & BCC Lattice\\
            \hline
            3 & 0.207 & 11,12 & 4600 & 500 & 5000 & 200 & BCC Lattice\\
            \hline
            3 & 0.331 & 5,11,12 & 3600 & 500 & 5000 & 200 & BCC Lattice\\
            \hline
            3 & 0.492 & 11,12 & 3000 & 500 & 5000 & 200 & BCC Lattice\\
            \hline
            2 & 0.101 & 13 & 6600 & 300 & 5000 & 200 & Square Lattice\\
            \hline
            3 & 0.491 & 13 & 2000 & 200 & 5000 & 600 & Cubic Lattice\\
            \hline
        \end{tabular}
        \caption{A table containing simulation parameters for systems used
        throughout article. The figure numbers prefixed with a T refer to tables and
        those prefixed with an S refer to the Supplementary Material
        \cite{t_middlemas_supplemental_nodate}.}\label{simtable}
    \end{table}

    Our collective coordinate procedure is similar to the one used in Refs. \cite{torquato_ensemble_2015, zhang_ground_2015, zhang_ground_2015-1, zhang_can_2017, zhang_transport_2016},
    but with a few key differences. The first is that the the time step choice
    algorithm and general structure of the program has been modified. We still
    adjust the time step based on the log-ratio of the energy between
    snapshots, but the threshold depends on the total number of snapshots
    $N_{\rm snap}$ taken. Denote the number of steps between samples to be
    $N_{\rm step}$. During the initial time step choice and equilibration
    phase, one first evolves the system $N_{\rm step}/2$ steps, and then adjust the timestep
    by sending
    \begin{equation}
        \Delta t \to \begin{cond}
            0.5 \Delta t & \mathcal{E} > 0.0005/N_{\rm snap}\\
            0.9 \Delta t & 0.0005/N_{\rm snap} \geq \mathcal{E} > 0.0003/N_{\rm snap}\\
            0.95 \Delta t & 0.0003/N_{\rm snap} \geq \mathcal{E} > 0.0002/N_{\rm snap}\\
            1.2 \Delta t & \mathcal{E} < 0.000001/N_{\rm snap}\\
            1.05 \Delta t & 0.000001/N_{\rm snap} \leq \mathcal{E} < 0.00005/N_{\rm snap}\\
            \Delta t & {\rm otherwise},
        \end{cond}
    \end{equation}
    where
    \begin{equation}
        \mathcal{E} = \left| \frac{2\ln (E_i/E_{i+1})}{N_{\rm step}} \right|.
    \end{equation}
    Then, one repeats the above $N_{\rm eq}$ times. Afterwards,
    one evolves the system for $N_{\rm step}/2$ using an Andersen thermostat
    and $N_{\rm step}/2$ without an Andersen thermostat, and takes a snapshot
    at the end with the L-BFGS algorithm (for ground states). One repeats this
    $N_{\rm snap}$ times.

    We have justified this change through a blocking analysis, where we have observed
    that upon splitting each trajectory into five equal portions sequentially, the
    value of $g_2(r)$ observed in each sub-trajectory is similar.
    Whenever an uncertainty for $g_2(r)$ was necessary (e.g., in the
    extrapolation to obtain the numerical $a$ coefficient) it is estimated by
    assuming each snapshot contributes independently to the final value. This
    independence assumption was corroborated by a standard block uncertainty
    analysis \cite{daan_frenkel_understanding_2002}. In three dimensions,
    we occasionally drop the first bin of $g_2(r)$ because no counts are recorded, even
    though the likely value of $g_2(r)$ is not zero. This is likely a finite size effect.

    The second change is that we conduct our simulations at $\rho=1$ rather
    than $K=1$. This has important implications for the choice of temperature
    used to equilibrate the system before taking snapshots. While we use the
    same values as Ref. \cite{zhang_ground_2015} ($T=2\times10^{-4}, 2\times10^{-6},$ and
    $1\times 10^{-6}$ for one, two, and three dimensions, respectively), it
    should be noted that this choice actually corresponds to physically
    distinct systems, since changing the density changes how far the particles
    need to move to obtain the same difference in energy. We have justified
    this choice by also simulating at temperatures one order of magnitude below
    those stated above. We observe that $g_2(r)$ does not change when
    simulated with this lower equilibration temperature.

    In general, we work with larger systems that have been
    equilibrated for a shorter amount of time and with fewer snapshots than the
    corresponding work in Refs. \cite{torquato_ensemble_2015,
    zhang_ground_2015, zhang_ground_2015-1, zhang_can_2017,
    zhang_transport_2016}. The values of the system size $N$, $N_{\rm snap}$,
    $N_{\rm step}$, and $N_{\rm eq}$ along with the shape of the unit cell of
    each system and a specification of which figures the data is used in, is
    given in Table \ref{simtable}.

    Throughout the article, we have used the rounded values of $\chi$ appearing in the figures to compute
    theoretical curves. Due to the finite size effects implicit in Eq. \ref{finitesizechi}, one cannot
    obtain exactly these values of $\chi$ with our chosen system sizes.
    Instead, we use a relatively close value of $\chi$, which rounds correctly
    to two significant figures. To give an idea of how much error is made when
    making this choice, we have reported the values of $\chi$ to the next non-trivial
    significant figure in Table \ref{simtable}. Figures where $\chi$ appears as
    the bottom axis use more precise estimates of $\chi$.

\subsection{Sampling the Nearest-Neighbor Functions}

    For the void quantities in one dimension, we use the fact that $E_V(r)$ is
    the ratio of uncovered space to total space in Fig. \ref{small_voronoi} and
    that $H_V(r)$ is the surface area of the covered space
    \cite{torquato_nearest-neighbor_1990}. This has been used
    previously to compute accurate results in two and three dimensions
    \cite{rintoul_algorithm_1995, sastry_statistical_1997,
    maiti_characterization_2013}. For the purposes of this article, we note that
    this interpretation gives rise to a simple
    method in 1D. In particular, we can sample the nearest neighbor function by simply
    compiling a list of all the gap sizes in the system, and computing the
    uncovered length of these gaps at each $r$. To ensure a meaningful estimate
    of the uncertainty in our calculation, we drop any $r$ for which fewer than
    10 individual gaps contribute. The uncertainty for
    $H_V(r)$ and $E_V(r)$ are then computed as the standard deviation
    of the mean with the value from each snapshot being treated as independent. $G_V(r)$
    is computed as the ratio, and the uncertainty propagated linearly.

    For the void functions in two and three dimensions and the particle
    functions in all dimensions, we use a sampling strategy. One computes the
    function $H_{V/P}(r)$ through binning nearest neighbor observations, the
    function $E_{V/P}(r)$ by recording every observation where the nearest
    neighbor is at least $r$ away, and $G_{V/P}(r)$ by taking their ratio.
    Since this method involves estimating a sensitive statistical quantity
    through a quotient, care must be taken to reduce systematic error. To this
    end, we compute these quantities using multiple bin sizes, and compare them
    to ensure that we have obtained a stationary estimate with respect to bin
    size. To ensure a meaningful estimate of the uncertainty in our calculation,
    we drop any bin for which fewer than 10 observations contribute to $H_V(r)$.
    Uncertainties for $H_V(r)$ and $E_V(r)$ are computed by assuming each snapshot
    contributes independently to the final value, and the uncertainty for $G_V(r)$ is
    propagated through the ratio linearly.

    We also sample $E_V(r)$ via the series (\ref{finiteseries}) in 1D for
    the purpose of determining how many terms in the series is needed. To do this,
    we use the fact that $v_n^{\rm int}({\bi r}^n)$ is just $v_2^{\rm int}(r)$, where
    $r$ is taken as the distance between the two points farthest apart. Thus, we
    can compute the series for $E_V(r)$ to arbitrary order as
    follows: first, compute all of the pair distances up to $r$ and the number
    of particles $m$ contained between the pair, and then compute the
    contribution of the pairs according to the formula
    \begin{equation}
        E_V^{ij}(r) = \sum_{k=0}^{m} (-1)^k {m \choose k} v_2^{\rm int}(r_{ij}).
    \end{equation}
    Finally, one sums the contribution of all pairs to obtain $E_V(r)$. The highest order
    needed is then $m_c + 2$, where $m_c$ is the highest $m$ value observed
    in the calculation for $E_V(r_c)$. The value of $r_c$ is determined by looking for
    a large drop in $E_V(r)$, as past $r_c$, the value of $E_V(r)$ is very
    close to zero. This drop is typically many orders of magnitude.
}

\section*{References}

\providecommand{\newblock}{}

% Taken from Jaeuk's arxiv version of hyperuniformity and window shapes paper
% arXiv:1610.03922

\pagebreak

\begin{center}
\large{\bf Supplementary Material}
\end{center}
\setcounter{equation}{0}
\setcounter{figure}{0}
\setcounter{table}{0}
\setcounter{page}{1}
% https://tex.stackexchange.com/questions/71162/reset-section-numbering-between-unnumbered-chapters
\setcounter{section}{0}
\makeatletter
\renewcommand{\theequation}{S\arabic{equation}}
\renewcommand{\thefigure}{S\arabic{figure}}
% For turning off linking citations:
% https://tex.stackexchange.com/questions/118182/selectively-turn-off-hyperref-links-for-citations

\section{Comparison of Nearest-Neighbor Functions for Selected Systems}

In developing intuition for the behavior of the nearest-neighbor functions, it is instructive
to compare their behaviors for a variety of systems of physical importance.
In Fig. \ref{comparevoid} below, we compare the void
nearest-neighbor functions for the Poisson point process, an equilibrium
hard-sphere fluid, and a representative disordered stealthy hyperuniform system
at intermediate $\chi$. If we focus on just $H_V(r)$ and $E_V(r)$, it may
appear that the stealthy hyperuniform process falls ``between'' that of
equilibrium hard spheres and the uncorrelated Poisson process, but inspection
of $G_V(r)$ makes it clear that the large-$r$ behavior of the stealthy system is
qualitatively different than that of the non-stealthy systems. Stealthy systems
have compact support for $H_V(r)$ and $E_V(r)$ \begin{NoHyper}\cite{zhang_can_2017,
ghosh_generalized_2018}\end{NoHyper}, which manifests itself clearly as a divergence of
$G_V(r)$ at a finite $r$. Thus, while in principle one can compute all three functions from
knowledge of just one, one can obtain a clearer understanding of the behavior
of a system by considering each in turn.

We plot the particle nearest-neighbor functions for the same systems in Fig.
\ref{compareparticle} of the Supplementary Material. Similar to the case of the void quantities, the
divergence of $G_P(r)$ is indicative of the bounded hole property of stealthy systems. It is also
important to note that the determination of the particle properties is in some sense
``harder'' for stealthy systems than for previously investigated systems. While
there are fundamental symmetries in the Poisson process and equilibrium
hard-sphere fluid systems that allow for the determination of the particle from
the void quantities \begin{NoHyper}\cite{torquato_nearest-neighbor_1990},\end{NoHyper} no such symmetry
exists in our stealthy hyperuniform systems.

\begin{figure}[h!]
    \centering
    \begin{subfigure}{0.33\textwidth}
        \includegraphics[width=\linewidth]{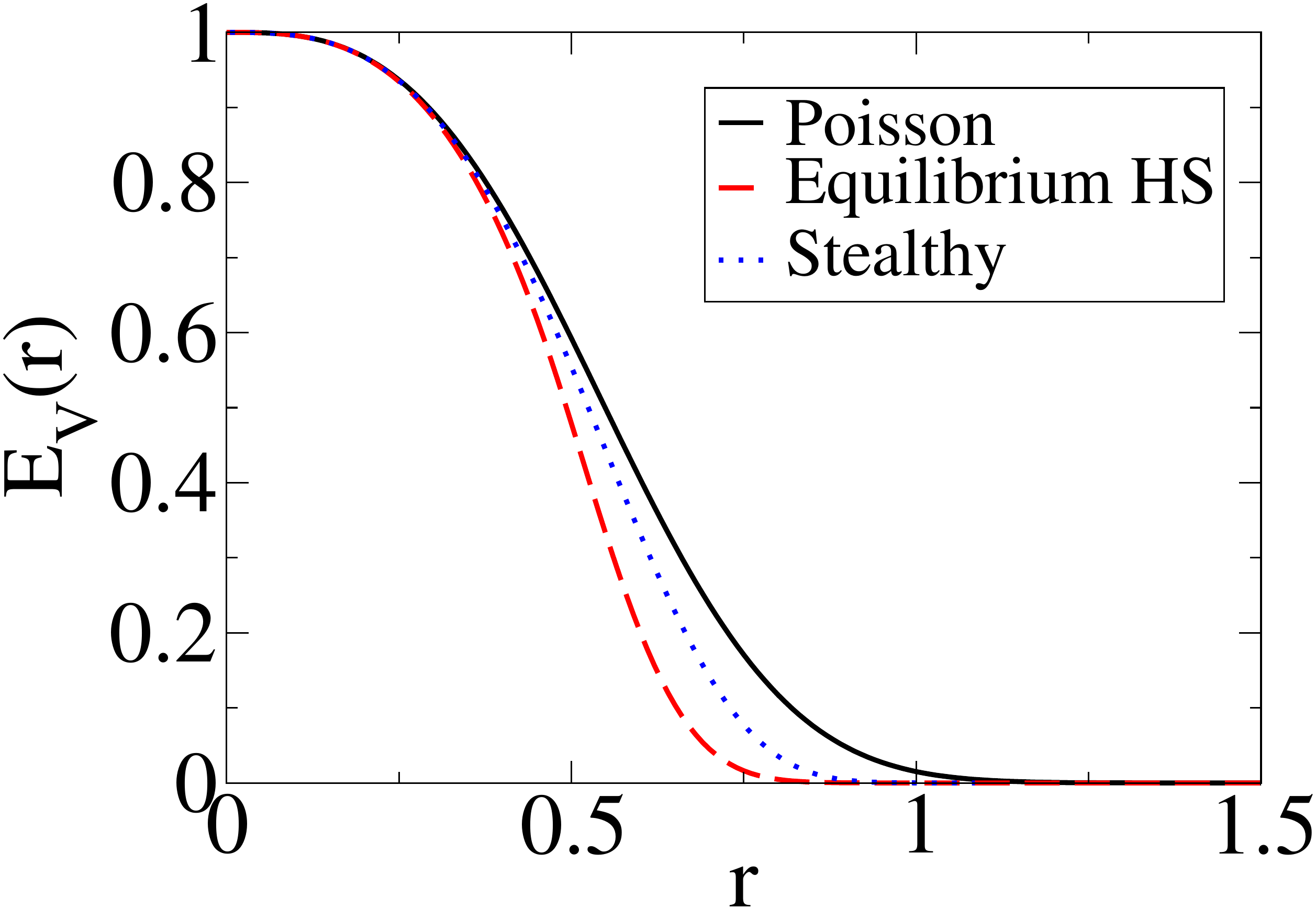}
        \caption{}
    \end{subfigure}%
    \begin{subfigure}{0.33\textwidth}
        \includegraphics[width=\linewidth]{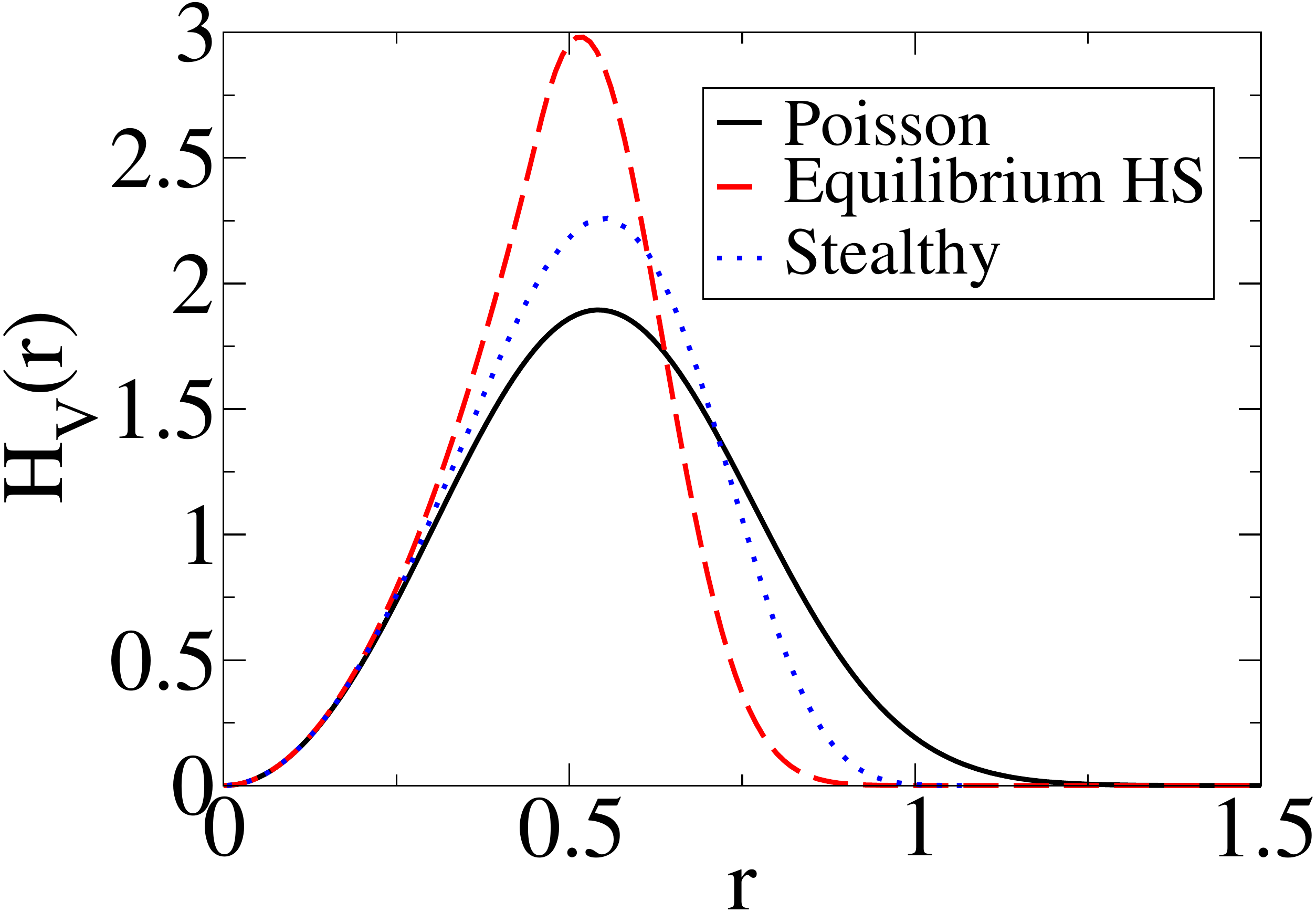}
        \caption{}
    \end{subfigure}%
    \begin{subfigure}{0.33\textwidth}
        \includegraphics[width=\linewidth]{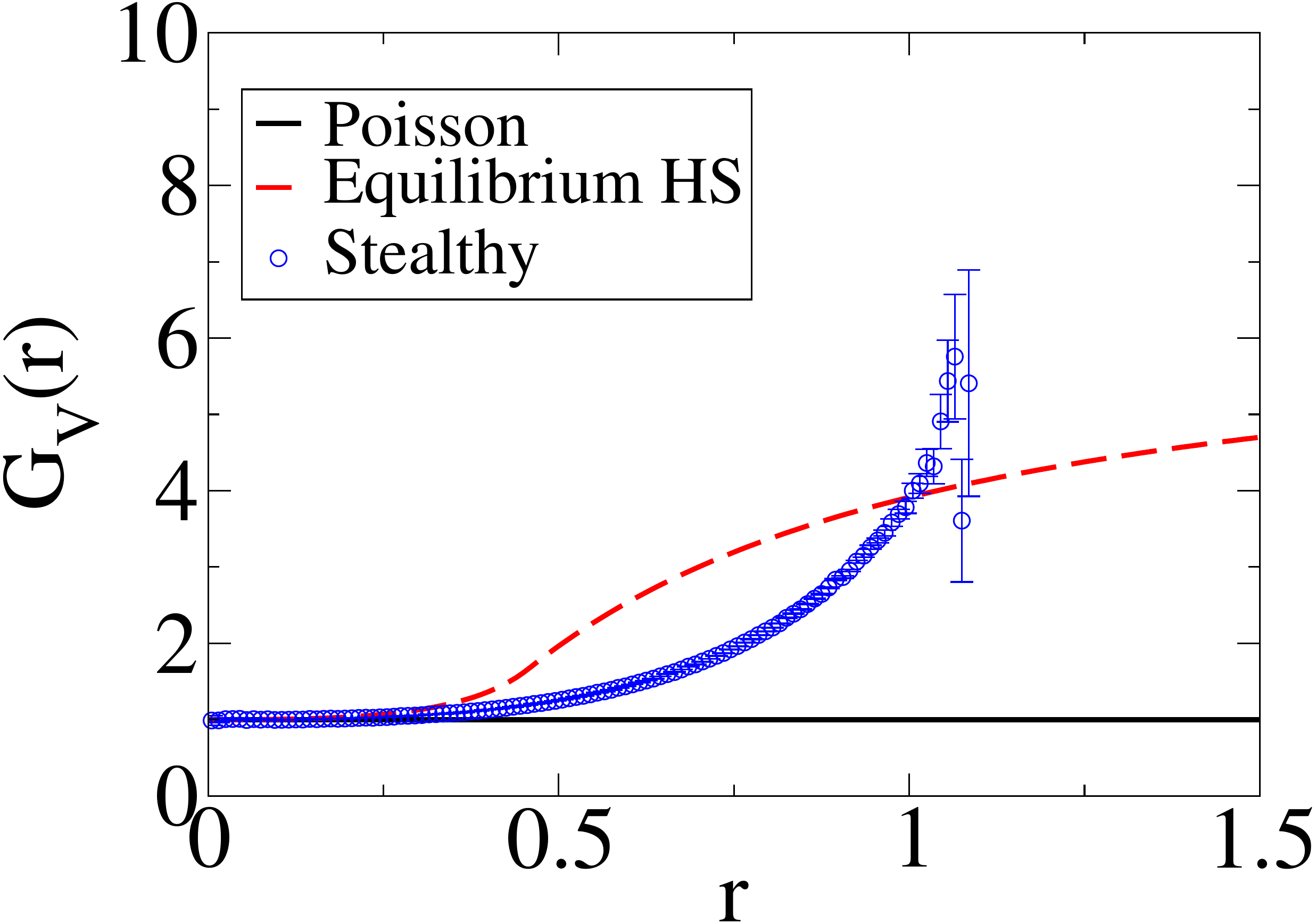}
        \caption{}
    \end{subfigure}
    \caption{The void nearest-neighbor functions $E_V(r)$, $H_V(r)$, and $G_V(r)$ for a
    Poisson point process at $\rho = 1$, an equilibrium hard-sphere system at
    $\rho = 1$ and $\phi = 0.4$, and a stealthy ensemble at $\rho = 1$ and $\chi
    = 0.10$. The Poisson results were computed with an exact formula
    % https://tex.stackexchange.com/questions/228973/argument-of-captionydblarg-has-an-extra
    \protect\begin{NoHyper}
    \cite{hertz_uber_1909}, the hard-sphere system was computed with an
    accurate approximation based on the Carnahan-Starling formula
    \cite{torquato_nearest-neighbor_1990}, and details concerning
    \protect\end{NoHyper}
    the stealthy ensemble can be found in the Appendix of the main article.
    (a) and (b) We can see that the Poisson process as the widest distribution
    of hole sizes and nearest-neighbor distances. (c) The bounded hole size
    property of the stealthy ensemble can be clearly seen in the divergence of
    $G_V(r)$.}\label{comparevoid}
\end{figure}

\begin{figure}
    \centering
    \begin{subfigure}{0.33\textwidth}
        \includegraphics[width=\linewidth]{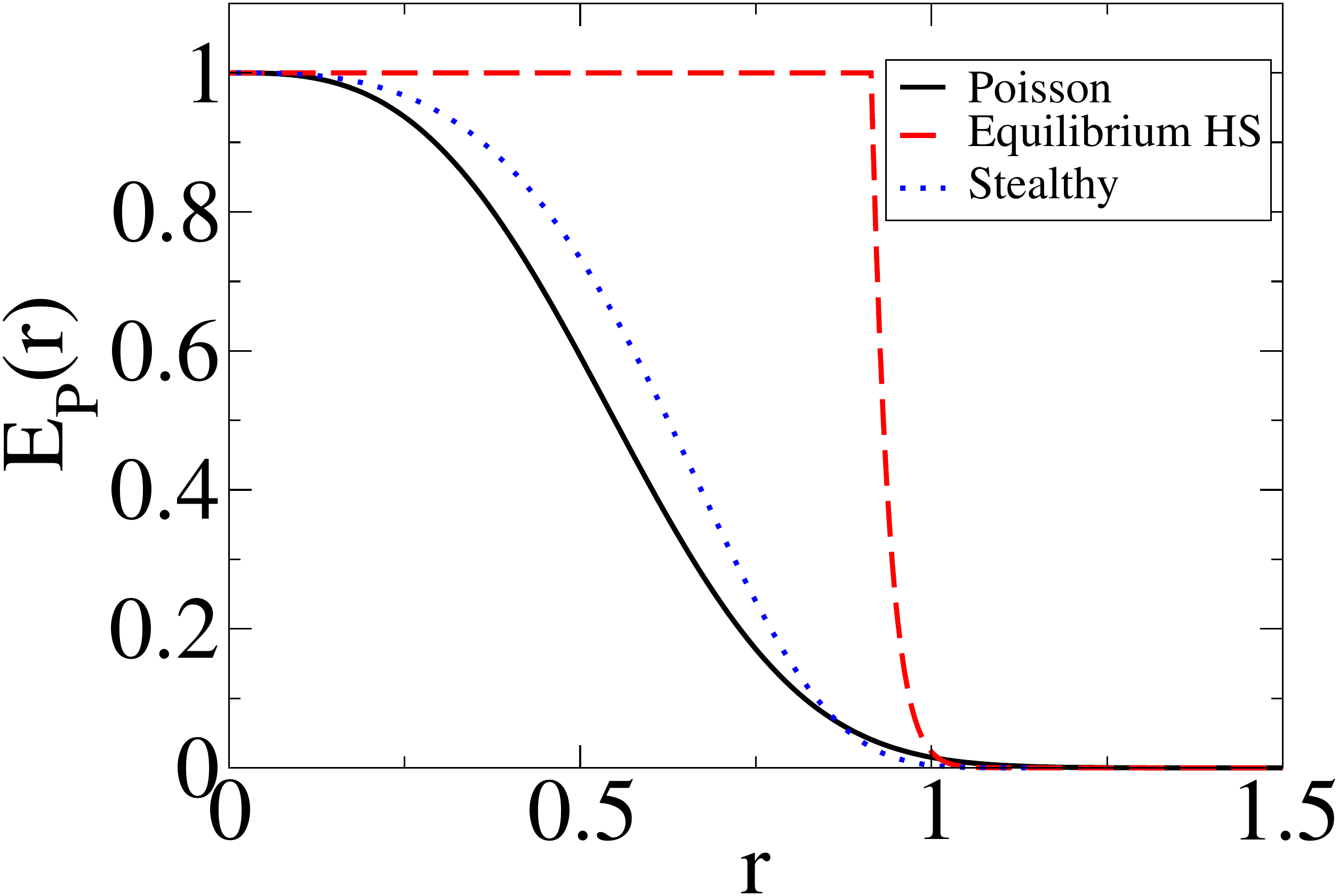}
        \caption{}
    \end{subfigure}%
    \begin{subfigure}{0.33\textwidth}
        \includegraphics[width=\linewidth]{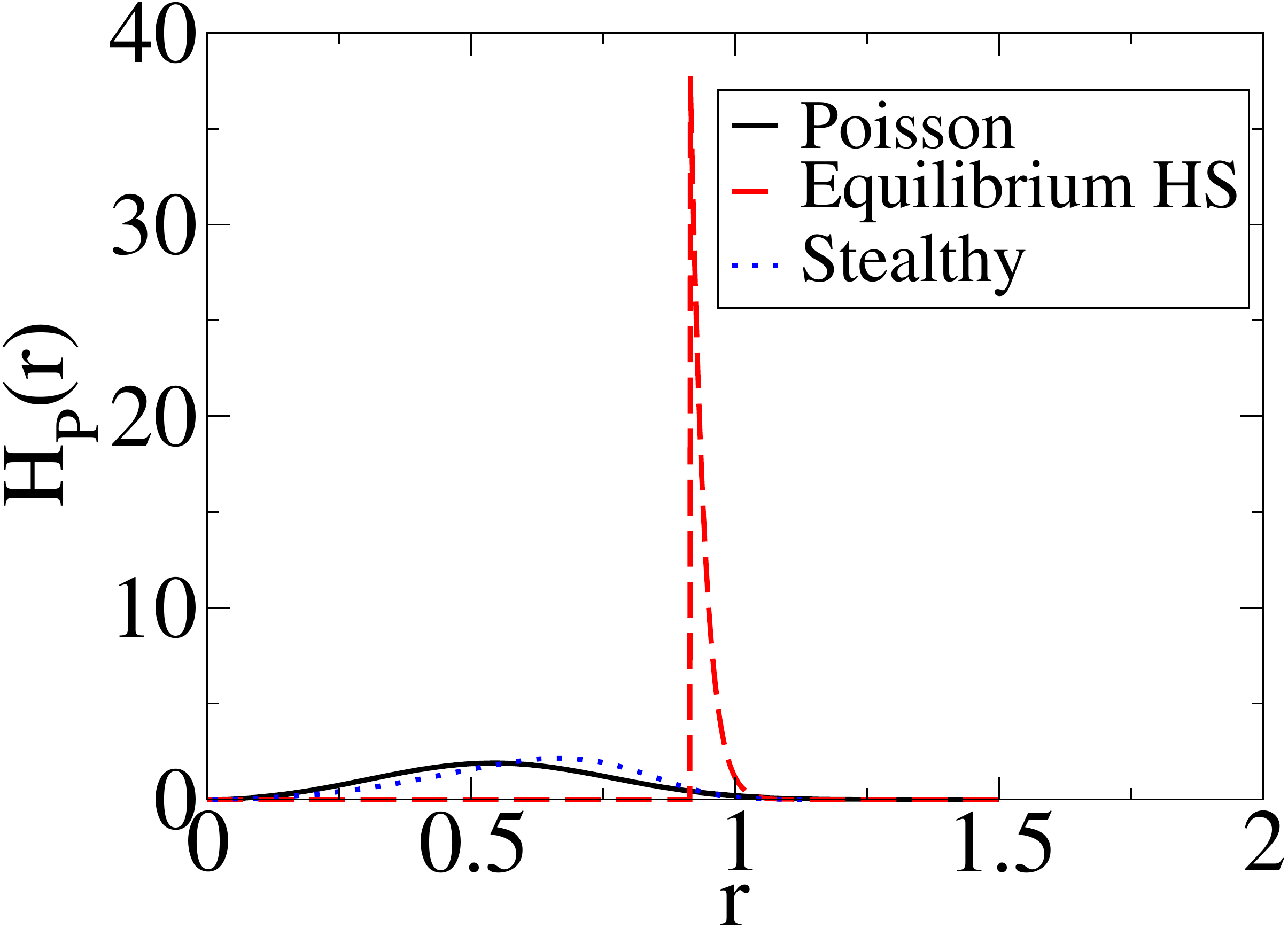}
        \caption{}
    \end{subfigure}%
    \begin{subfigure}{0.33\textwidth}
        \includegraphics[width=\linewidth]{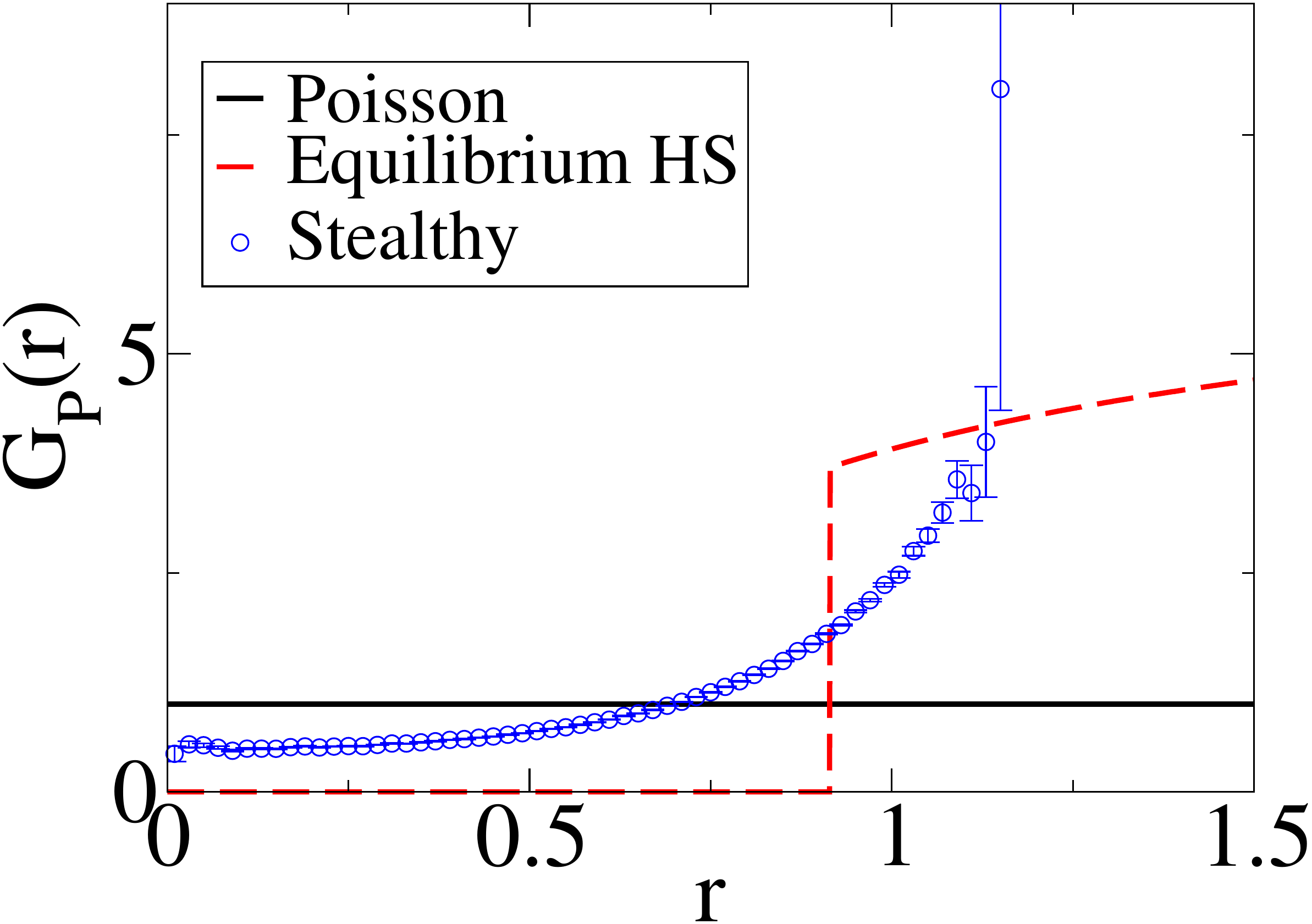}
        \caption{}
    \end{subfigure}
    \caption{The particle nearest-neighbor functions $E_P(r)$, $H_P(r)$, and $G_P(r)$ for a
    Poisson point process at $\rho = 1$, an equilibrium hard-sphere system at
    $\rho = 1$ and $\phi = 0.4$, and a stealthy ensemble at $\rho = 1$ and
    $\chi \approx 0.10$. The Poisson results were computed with an exact formula
    \protect\begin{NoHyper}
    \cite{hertz_uber_1909}, the hard-sphere system was computed with an
    accurate approximation based on the Carnahan-Starling formula
    \cite{torquato_nearest-neighbor_1990}, and details concerning
    \protect\end{NoHyper}
    the stealthy ensemble can be found in the Appendix of the main article. (a)
    and (b) We can see that the equilibrium hard sphere process has sharply
    localized nearest-neighbor statistics. (c) The
    presence of a bounded hole size for the stealthy ensemble can be
    clearly seen in the divergence of $G_P(r)$.}\label{compareparticle}
\end{figure}

\section{Pair Statistics of Stealthy Hyperuniform Point Processes}

\begin{figure}
    \centering
    \begin{subfigure}{0.4\textwidth}
        \includegraphics[width=\linewidth]{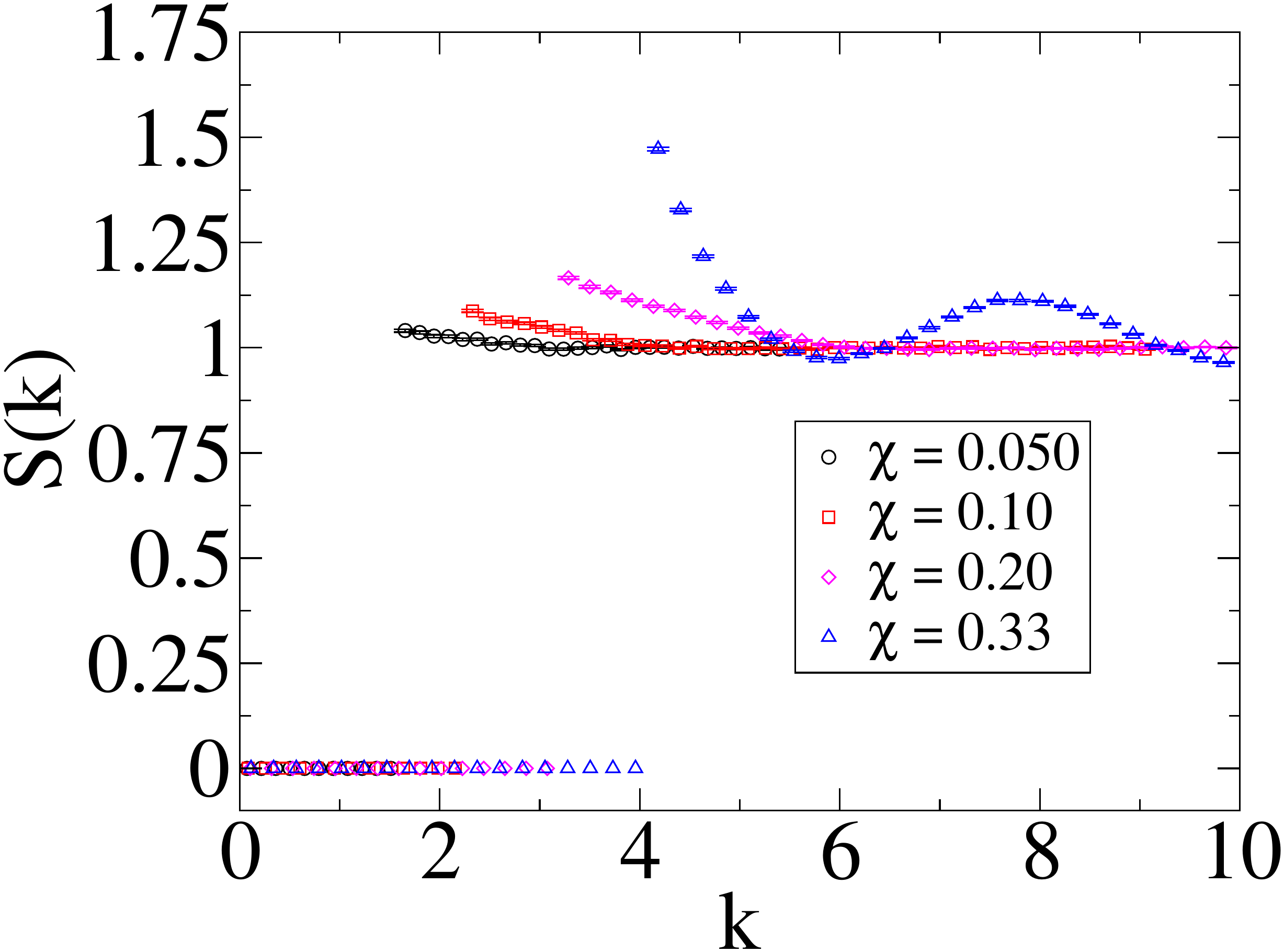}
        \caption{}
    \end{subfigure}%
    \begin{subfigure}{0.4\textwidth}
        \includegraphics[width=\linewidth]{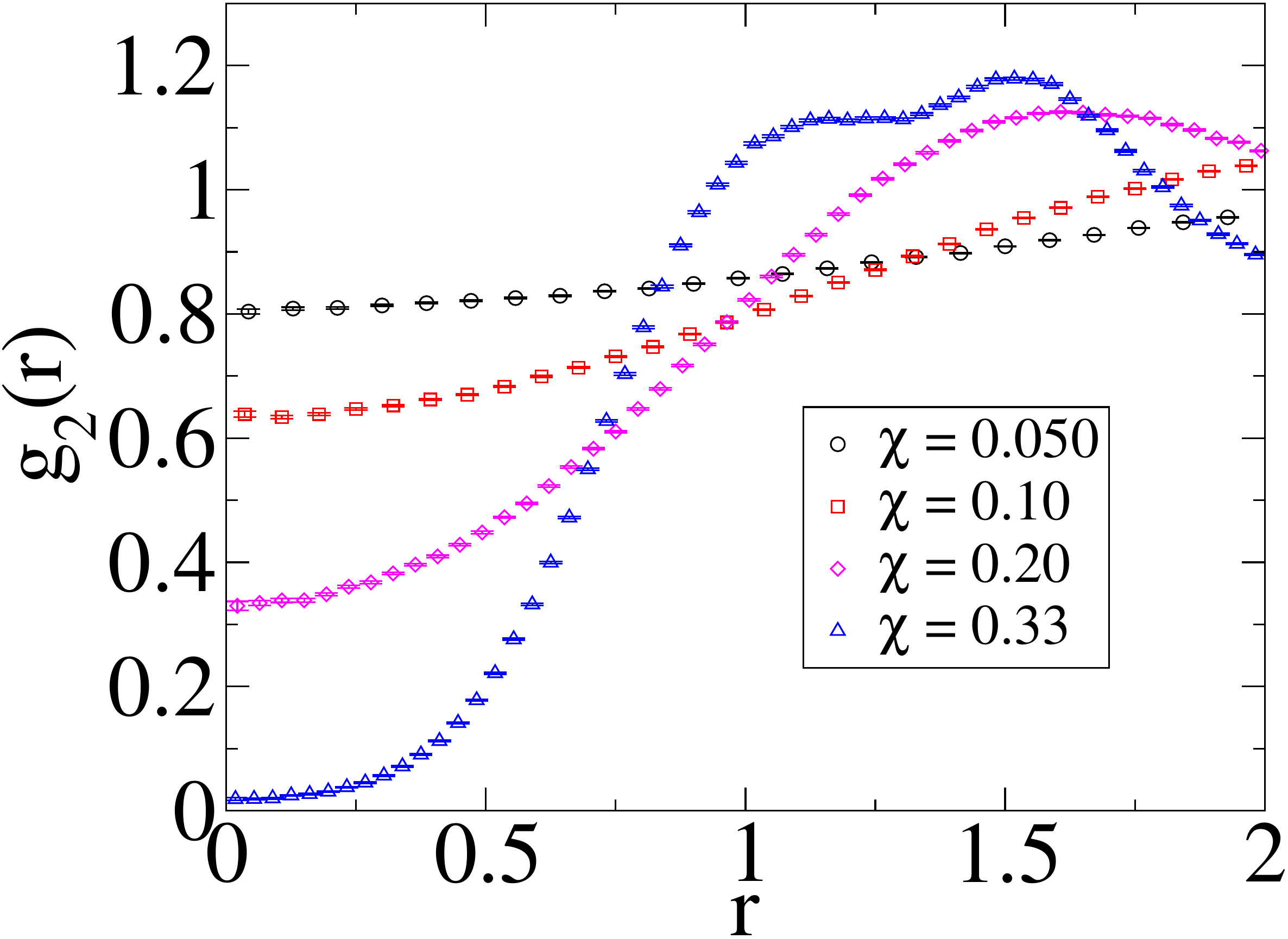}
        \caption{}
    \end{subfigure}
    \caption{A comparison of (a) $S(k)$ and (b) $g_2(r)$ for simulated 2D stealthy
    systems at $\rho=1$ and a variety of $\chi$. See the Appendix of the main
    article for the simulation details of the systems shown.}\label{skg2}
\end{figure}

\begin{figure}
    \centering
    \begin{subfigure}{0.33\textwidth}
        \includegraphics[width=\linewidth]{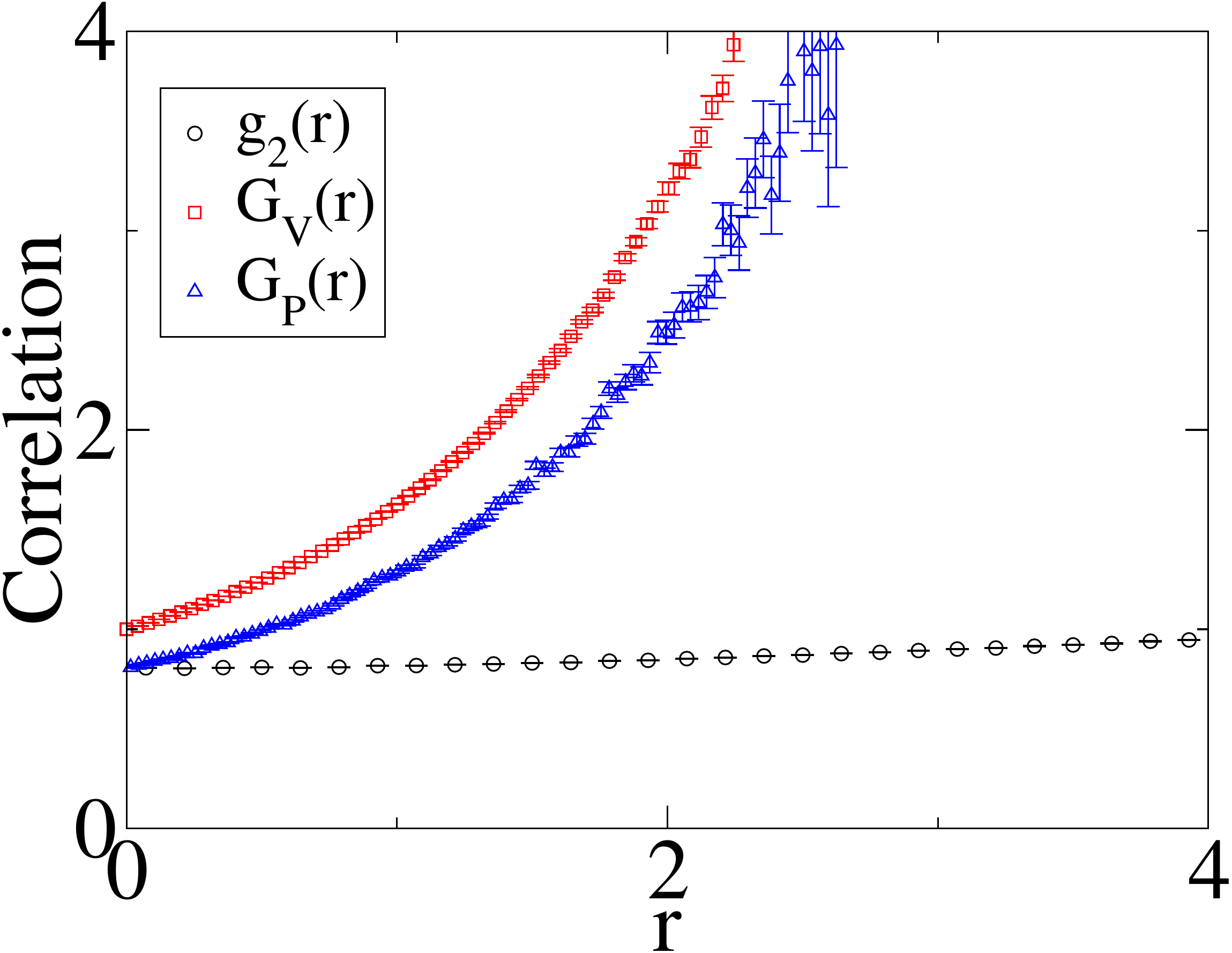}
        \caption{}
    \end{subfigure}%
    \begin{subfigure}{0.33\textwidth}
        \includegraphics[width=\linewidth]{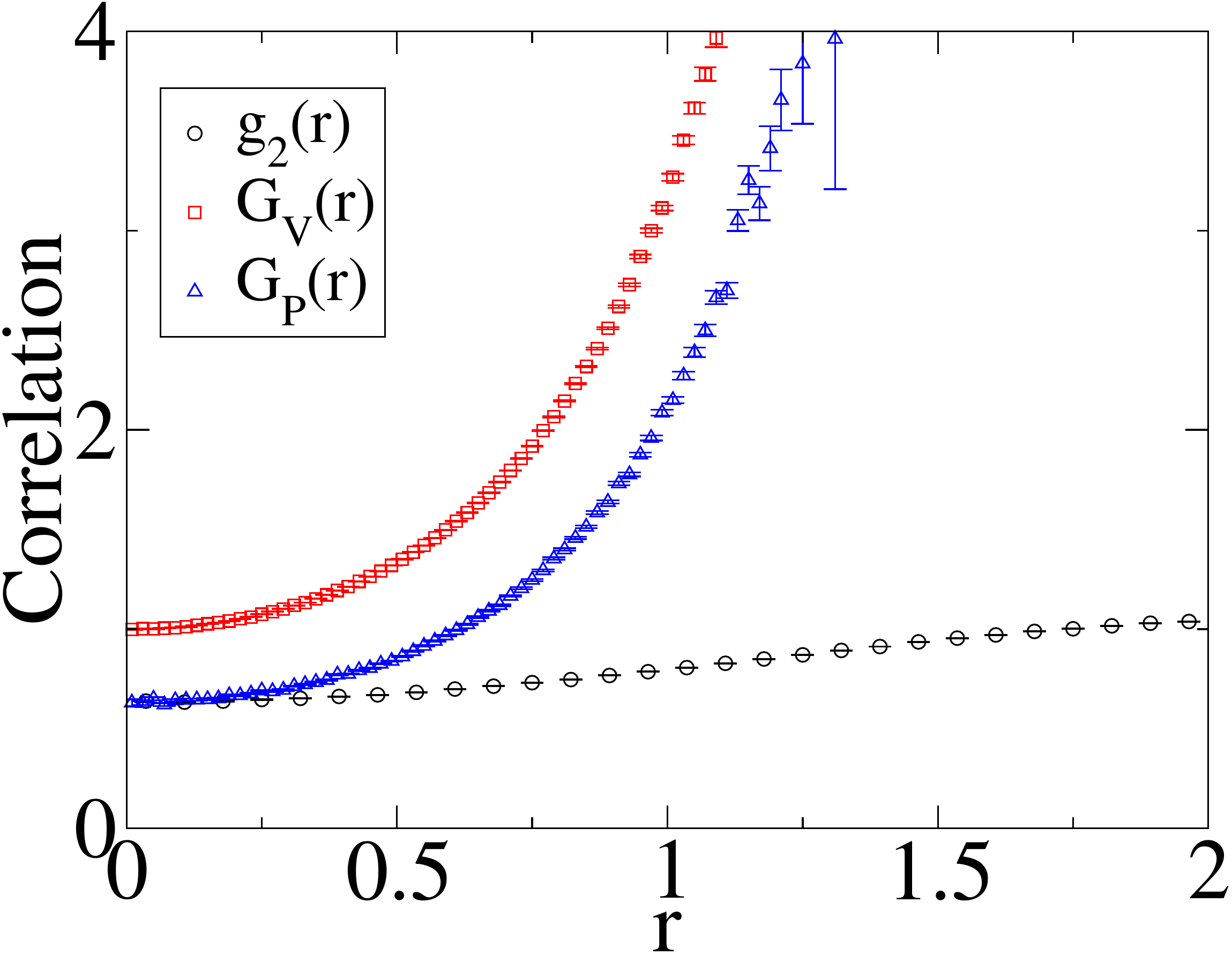}
        \caption{}
    \end{subfigure}%
    \begin{subfigure}{0.33\textwidth}
        \includegraphics[width=\linewidth]{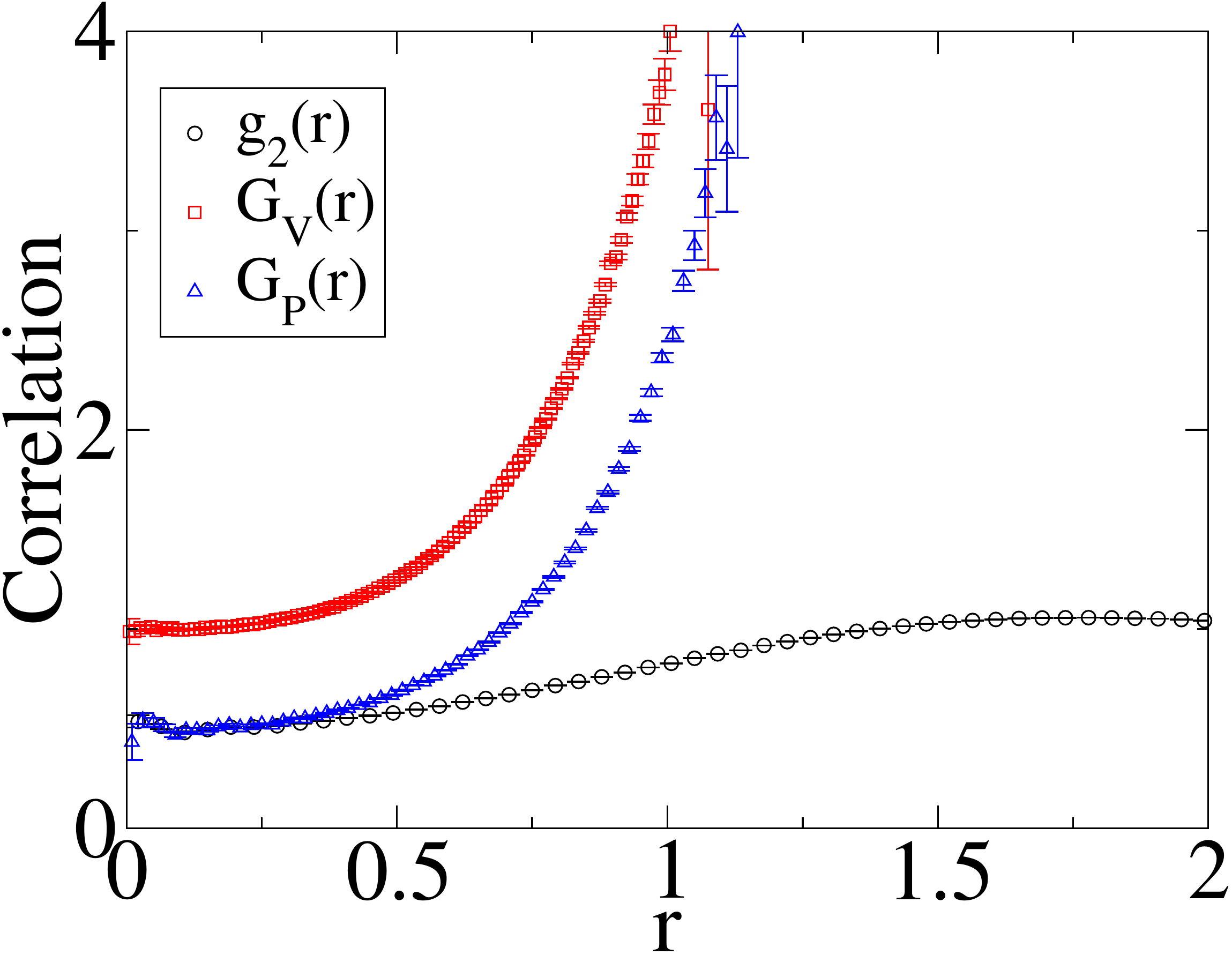}
        \caption{}
    \end{subfigure}\\
    \begin{subfigure}{0.33\textwidth}
        \includegraphics[width=\linewidth]{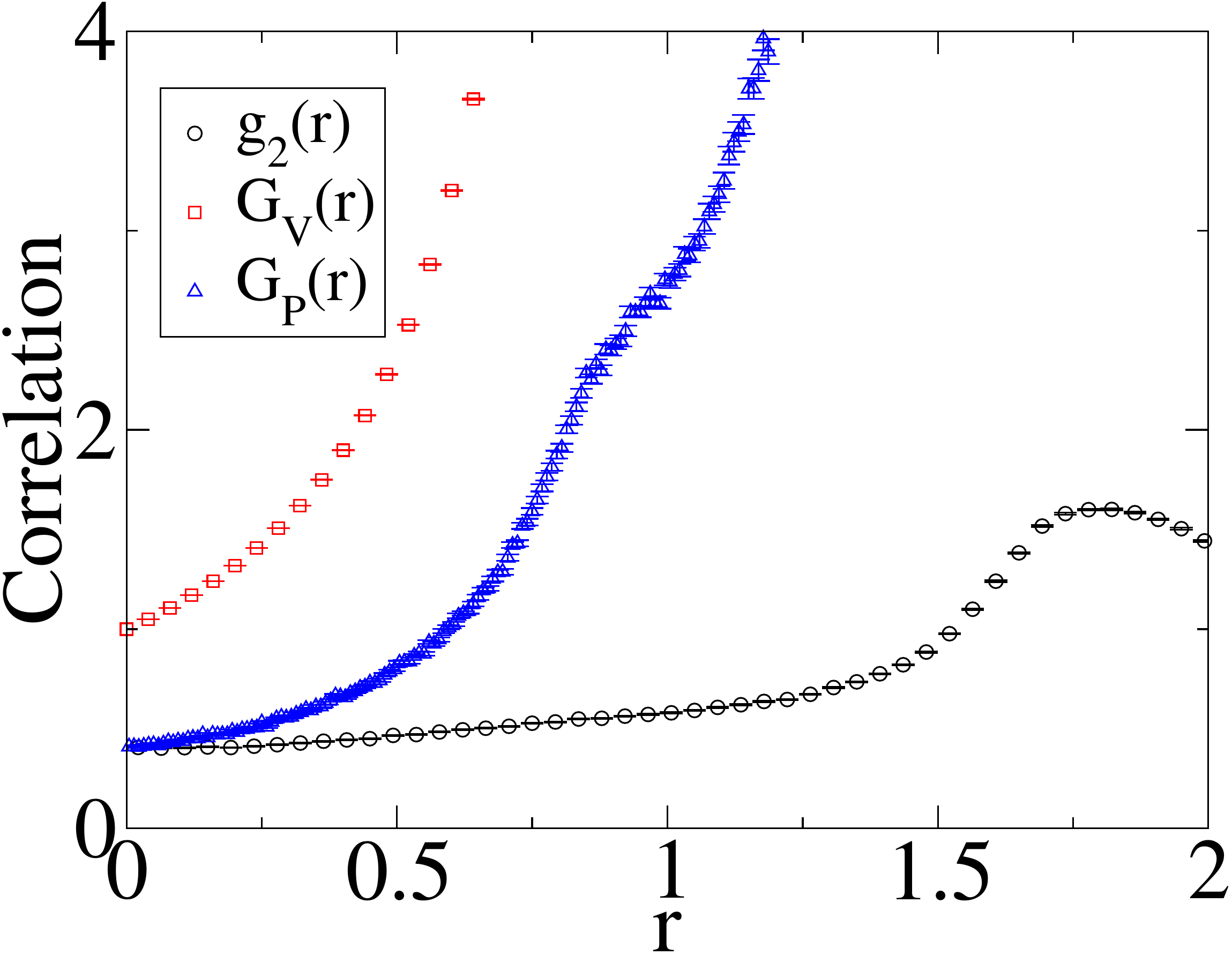}
        \caption{}
    \end{subfigure}%
    \begin{subfigure}{0.33\textwidth}
        \includegraphics[width=\linewidth]{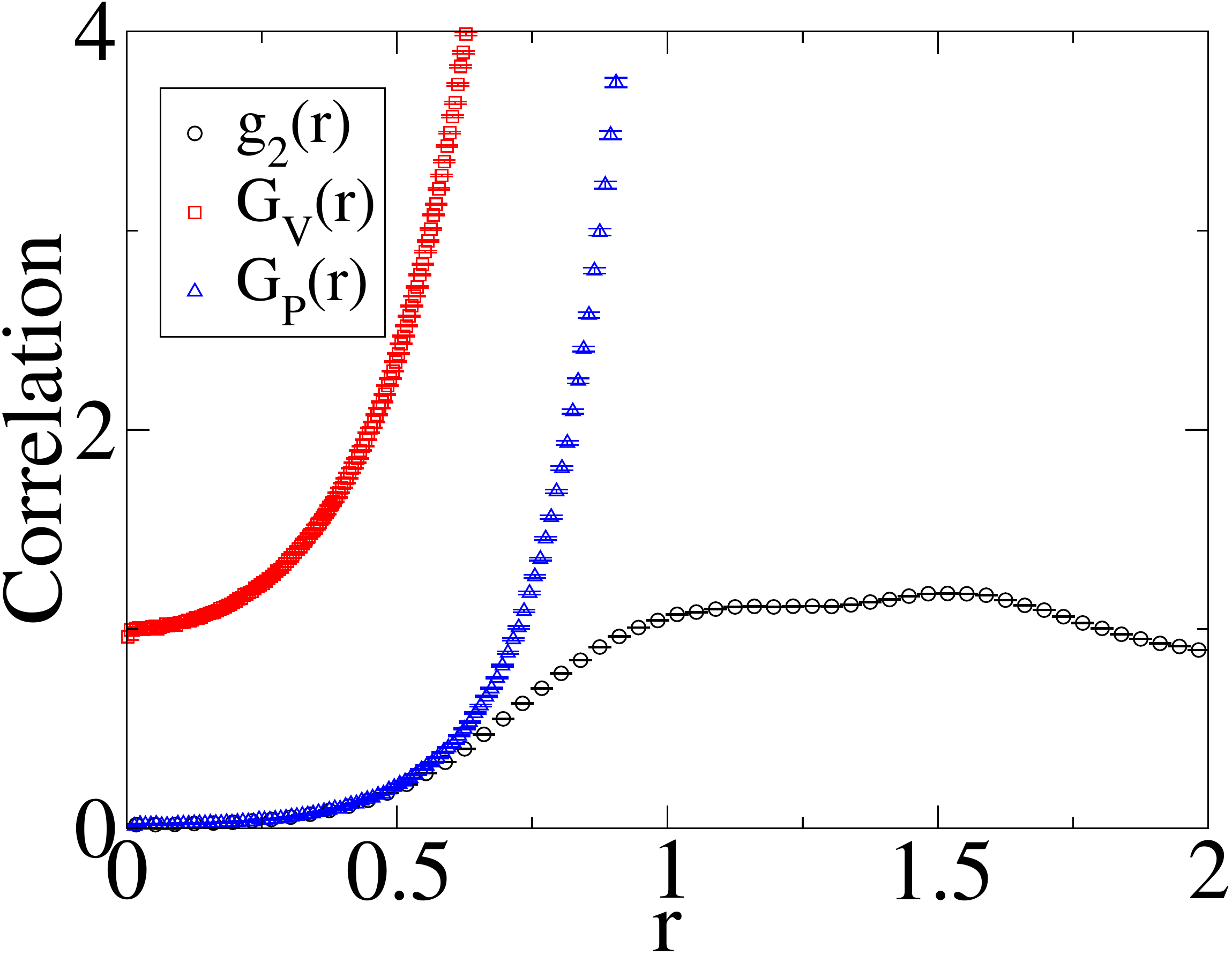}
        \caption{}
    \end{subfigure}%
    \begin{subfigure}{0.33\textwidth}
        \includegraphics[width=\linewidth]{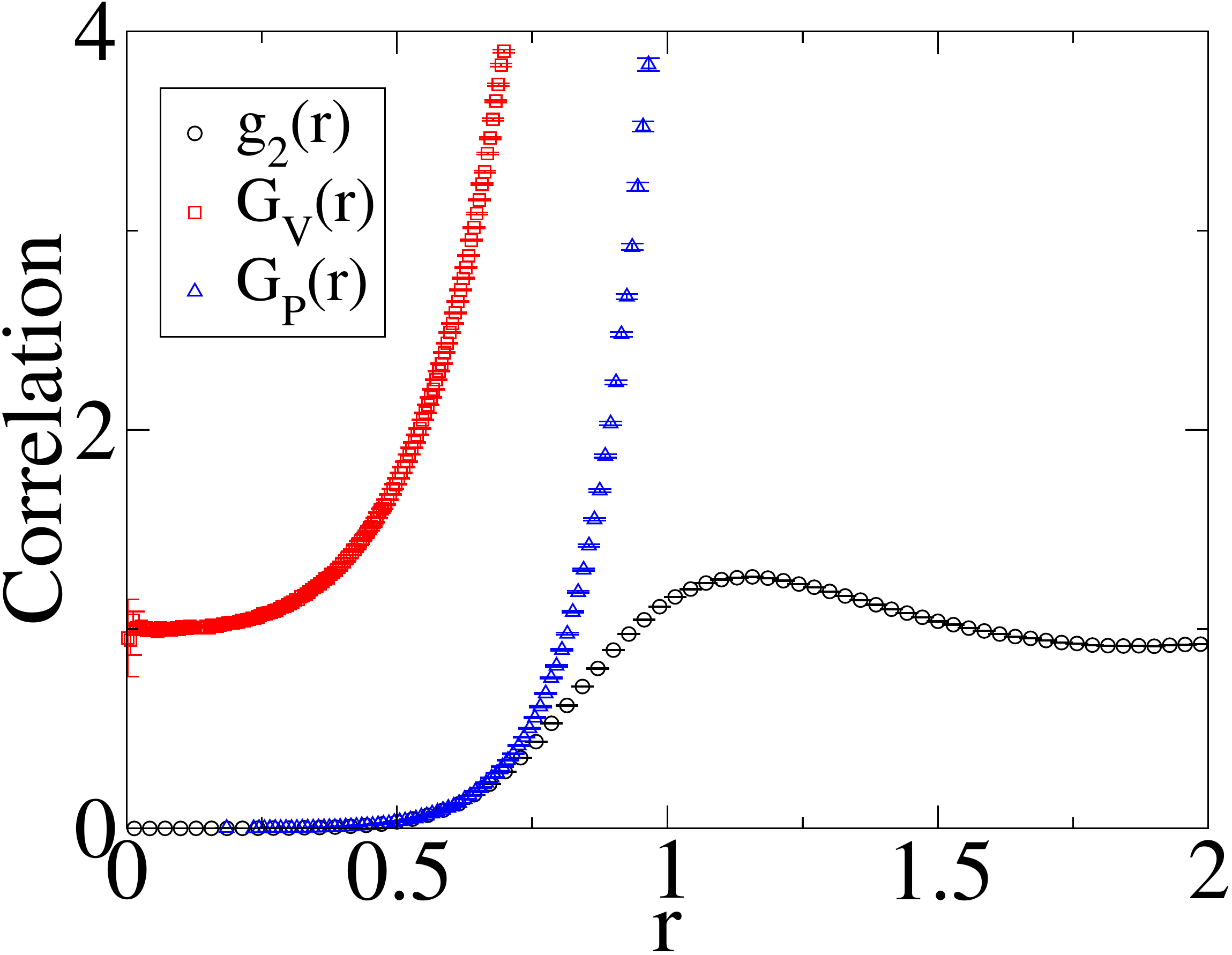}
        \caption{}
    \end{subfigure}
    \caption{A comparison of the correlation functions $g_2(r)$,
    $G_V(r)$, and $G_P(r)$ for a variety of stealthy systems. (a-c) Comparisons
    for $\chi\approx 0.10$ (see Appendix of main article for details) for one,
    two, and three dimensions, respectively. (d-f) Comparisons for $\chi\approx
    0.33$ (see Appendix of main article for details) for one, two, and three
    dimensions, respectively.}\label{comparepair}
\end{figure}

In this section, we investigate the pair statistics (the pair
correlation function $g_2(r)$ and the structure factor $S(k)$) of stealthy systems. While
theoretical expressions and simulation data on these two correlation functions have previously been reported in Refs. \begin{NoHyper}\cite{torquato_ensemble_2015} and
\cite{zhang_ground_2015}\end{NoHyper}, we make several observations
of key interest in the main article. In Fig. \ref{skg2} below, we show $S(k)$
and $g_2(r)$ for 2D stealthy system at various $\chi$. One sees that as $\chi$
is increased, the maximum constrained wavevector $K$ for which $S(k) = 0$
increases, and one has an increase in short-range order in the form of stronger
low-$r$ correlations in $g_2(r)$. In Fig.  \ref{comparepair} of the
Supplementary Material, we compare $G_V(r)$ and $G_P(r)$ to the pair correlation function
$g_2(r)$. In the case of hard sphere systems, these functions are directly
related at contact \begin{NoHyper}\cite{torquato_nearest-neighbor_1990}\end{NoHyper}. While no simple
relation exists in the case of stealthy systems, one can still observe a number
of useful generalities. One finds that for stealthy systems, there exists good
numerical evidence for the conjecture $G_V(r) \geq G_P(r) \geq g_2(r)$, at
least for small enough $r$. One can also see that while $g_2(r)$ and $G_P(r)$
tend to the same value as $r\to 0$, $G_V(r)$ always tends to unity.

\section*{References}
\providecommand{\newblock}{}

\end{document}